\documentclass[twocolumn,showpacs,superscriptaddress,prb,aps,floatfix]{revtex4-1}

\usepackage{graphicx}
\usepackage{rotating}
\usepackage{amsmath}
\usepackage[usenames,dvipsnames]{color}

\makeatletter

\newcommand{\Rmnum}[1]{\expandafter\@slowromancap\romannumeral #1@}
\makeatletter

\hfuzz=\maxdimen
\begin{document}

\title{Screened hybrid functional applied to 3$d^0$$\rightarrow$3$d^8$ transition-metal
perovskites La$M$O$_3$ ($M$=Sc-Cu):
influence of the exchange mixing parameter on the structural, electronic and magnetic properties}

\author{Jiangang He and Cesare Franchini}
\affiliation{University of Vienna, Faculty of Physics and Center for Computational
Materials Science, Vienna, Austria}

\date{\today}


\begin{abstract}

We assess the performance of the Heyd-Scuseria-Ernzerhof (HSE) screened hybrid
density functional scheme applied to the perovskite family La$M$O$_3$ ($M$=Sc-Cu)
and discuss the role of the mixing parameter $\alpha$ (which determines the fraction of
exact Hartree-Fock exchange included in the density functional theory (DFT) exchange-correlation
functional) on the structural, electronic, and magnetic properties.
The physical complexity of this class of compounds, manifested by the largely varying
electronic characters (band/Mott-Hubbard/charge-transfer insulators and metals),
magnetic orderings, structural distortions (cooperative Jahn-Teller like instabilities),
as well as by the strong competition between localization/delocalization effects
associated with the gradual filling of the $t_{2g}$ and $e_g$ orbitals,
symbolize a critical and challenging case for theory.
Our results indicates that HSE is able to provide a consistent picture of the complex
physical scenario encountered across the La$M$O$_3$ series
and significantly improve the standard DFT description. The only exceptions are the correlated
paramagnetic metals LaNiO$_3$ and LaCuO$_3$, which are found to be treated better within DFT.
By fitting the ground state properties with respect to $\alpha$ we have constructed a set of 'optimum'
values of $\alpha$ from LaScO$_3$ to LaCuO$_3$: it is found that the 'optimum'
mixing parameter decreases with increasing filling of the $d$ manifold
(LaScO$_3$: 0.25; LaTiO$_3$ \& LaVO$_3$: 0.10-0.15;
LaCrO$_3$, LaMnO$_3$, and LaFeO$_3$: 0.15; LaCoO$_3$: 0.05; LaNiO$_3$ \& LaCuO$_3$: 0).
This trend can be nicely correlated with the modulation of the screening and dielectric
properties across the La$M$O$_3$ series, thus providing a physical justification to the
empirical fitting procedure. Finally, we show that by using this set of optimum mixing parameter HSE 
predict dielectric constants in very good agreement with the experimental ones.
\end{abstract}

\maketitle

\section{INTRODUCTION}
The physics of transition metal perovskites with general chemical formula ABO$_3$
(where A is a large cation, similar in size to O$^{2-}$ and B is a small transition metal (TM) cation)
has attracted and challenged the interest and curiosity of the material science community for many decades,
due to huge variety of complex phenomena arising from the subtle coupling between structural, electronic
and magnetic degrees of freedom. The high degree of chemical flexibility and the localized (i.e. not spatially
homogeneous) character of the dominant TM partially filled $d$ states lead to the coexistence of several
physical interactions -- spin, charge, lattice and orbital -- which are all simultaneously active.
The occurrence of strong lattice-electron, electron-spin and spin-orbit couplings causes several fascinating
phenomena, including metal-insulator transitions\cite{Mott90, Imada98},
superconductivity\cite{bednorz}, colossal magnetoresistance\cite{helmolt,Salamon01}, multiferroicity\cite{Wang09},
bandgaps spanning the visible and ultraviolet\cite{arima}, and surface chemical reactivity from active to
inert\cite{Tanaka01,Suntivich11}. When the additional degrees of freedom afforded by the combinatorial
assemblage of perovskite building blocks in superlattices, heterointerfaces and thin films are introduced the
range of properties increases all the more, as demonstrated by the recent several remarkable discoveries
in the field of oxide heterostructures\cite{Zubko11}.
Tunability and control of these intermingled effects can be further achieved
by means of external stimuli such as doping\cite{Ramirez97,Coey99}, pressure\cite{Loa01,Zhou11},
temperature and magnetic or electric fields\cite{Asamitsu95,Varma96},
thereby enhancing the tailoring capability of perovskites for a wide range of functionalities.
This rich array of behaviors uniquely suit perovskites for
novel solutions in different sectors of modern technology (optoelectronics, spintronics, piezoelectric devices
and (photo)catalysis), for which conventional semiconductors cannot be used\cite{Ramesh,Chakhalian,Kudo09,adler}.

Theoretical studies of TM perovskites, aiming to describe and understand the underlying physical mechanisms
determining their complex electronic structures have been mainly developed within two
historically distinct solid state communities, model Hamiltonians\cite{Anderson61,Gunnarsson83} and
{\em first principles}\cite{Kohn99},
which in the recent years have initiated to fruitfully cross-connect each other methodologies towards more general
schemes such as DFT+DMFT (Density Functional Theory\cite{Kohn65} + Dynamical Mean Field
Theory\cite{Metzner89, Georges96, Kotliar06}), with the aim to overtake the individual limitations and to
improve the applicability and predictive power of electronic structure theory\cite{Solovyev08,Imada10}.
Model Hamiltonians approaches adopt simplified lattice fermion models, typically the celebrated {\em Hubbard model},
inspired by the seminal works of Anderson\cite{Anderson59}, Hubbard\cite{Hubbard} and Kanamori\cite{Kanamori63}
in which the many-body problem is solved using a small number of {\em relevant} bands and short-ranged electron
interactions. These effective models can solve the many-body problem very accurately,
also including ordering and quantum fluctuations, but critically depends on an large number of adjustable parameters
(which can be in principle derivable by {\em first principles} 
schemes\cite{Calzado02, Moreira07, Boilleau10, kovacik10,kovacik11,Franchini12}),
and its applicability is restricted to finite-size systems\cite{Imada98, Imada10}.
In DFT the intractable many-body problem of interacting electrons is mapped into a simplified problem of
non-interacting electrons moving in an effective potential throughout the Kohn-Sahm scheme\cite{Kohn65},
and electron exchange-correlation (XC) effects are accounted by the XC potential which is approximated
using XC functionals such as the Local Density Approximation (LDA), the Generalized Gradient Approximation
(GGA) {\em et similaria}\cite{Perdew05}. As the name suggests, in DFT the ground state properties are
obtained only from the charge density, and this makes DFT fundamentally different from wavefunction-based
approaches as the Hartee-Fock method, the simplest approximation to the many-body problem
which includes the exact exchange but no correlation\cite{Parr89}.
Though DFT has been widely and successfully used in the last 40 years
in solid-state physics and quantum-chemistry to calculate structural data, energetics and, to a lesser extent,
electronic and magnetic properties, it suffers of fundamental difficulties mostly due to the approximate
treatment of XC effects. This drawback is particularly severe when DFT is applied to the so called
strongly correlated systems (SCSs), whose prototypical examples are transition metal oxides (TMOs).
A systematic improvement of these XC-related deficiencies in DFT is essentially impossible, but
several 'beyond-DFT' approaches have been proposed which deliver much more satisfying results. The most
renewed ones are the DFT+U\cite{Anisimov91}, SIC\cite{Perdew81, Svane90, Szotek91, Filippetti03}, hybrid
functionals\cite{Becke93}, and GW\cite{Hedin65}.
For a recent review on DFT and beyond applied to transition metal oxides see \onlinecite{Cramer09}.

\begin{table*}[t]
\caption{Summary of the fundamental ground state properties of La$M$O$_3$: (i) Crystal structure: O=orthorhombic,
M=monoclinic, R=rhombohedral, and T=tetragonal; (ii) Transition metal (TM) spin-projected electronic configuration and (line below) corresponding oxidation state, (iii) electronic character:
I=insulator and M=metal; Magnetic ordering: NM=non-magnetic, different type of AFM arrangements
(see Fig.\ref{fig:ordering}), and P=paramagnetic.} \vspace{0.3cm}
\begin{ruledtabular}
\begin{tabular}{cccccccccc}
                           & LaScO$_3$ & LaTiO$_3$ & LaVO$_3$ & LaCrO$_3$ & LaMnO$_3$ & LaFeO$_3$ & LaCoO$_3$ & LaNiO$_3$ & LaCuO$_3$  \\
Crystal structure          & O-$P_{nma}$ & O-$P_{nma}$ & M-$P_{2_1/b}$ & O-$P_{nma}$ & O-$P_{nma}$ & O-$P_{nma}$ & R-$R_{\bar3c}$ &
R-$R_{\bar3c}$ & T-P$_{4/m}$\\
TM electronic configuration& $d^0$ &
                             ${t_{2g}}^{\uparrow}$ &
                             ${t_{2g}}^{\uparrow\uparrow}$ &
                             ${t_{2g}}^{\uparrow\uparrow\uparrow}$ &
                             ${t_{2g}}^{\uparrow\uparrow\uparrow}{e_{g}}^{\uparrow}$&
                             ${t_{2g}}^{\uparrow\uparrow\uparrow}{e_{g}}^{\uparrow\uparrow}$ &
                             ${t_{2g}}^{\uparrow\downarrow\uparrow\downarrow\uparrow\downarrow}$&
                             ${t_{2g}}^{\uparrow\downarrow\uparrow\downarrow\uparrow\downarrow}{e_{g}}^{\uparrow}$ &
                             ${t_{2g}}^{\uparrow\downarrow\uparrow\downarrow\uparrow\downarrow}{e_{g}}^{\uparrow\downarrow}$ \\
                           &  3+       &   3+      &     3+      &    3+     &  3+       &    3+     &     3+       &     3+       &    3+     \\
Electronic character       &  I        &   I       &     I       &    I      &  I        &    I      &     I        &     M        &    M      \\
Magnetic structure         &  NM       & G-AFM     &   C-AFM     &  G-AFM    & A-AFM     &  G-AFM    &     PM       &     PM       &    PM
\end{tabular}
\end{ruledtabular}
\label{tab:0}
\end{table*}

\begin{figure}
\includegraphics[clip=,width=0.5\textwidth]{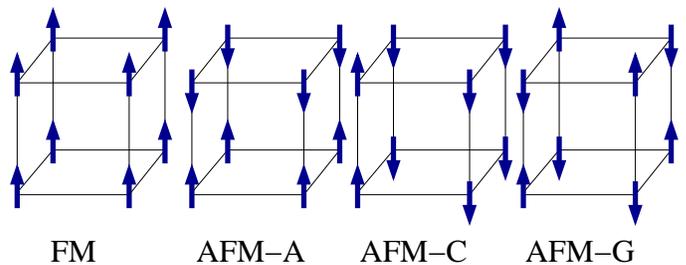}
\caption
{(Color online) Schematic representation of the typical magnetic orderings for the  perovskites.}
\label{fig:ordering}
\end{figure}

\begin{figure}
\includegraphics[clip=,width=0.5\textwidth]{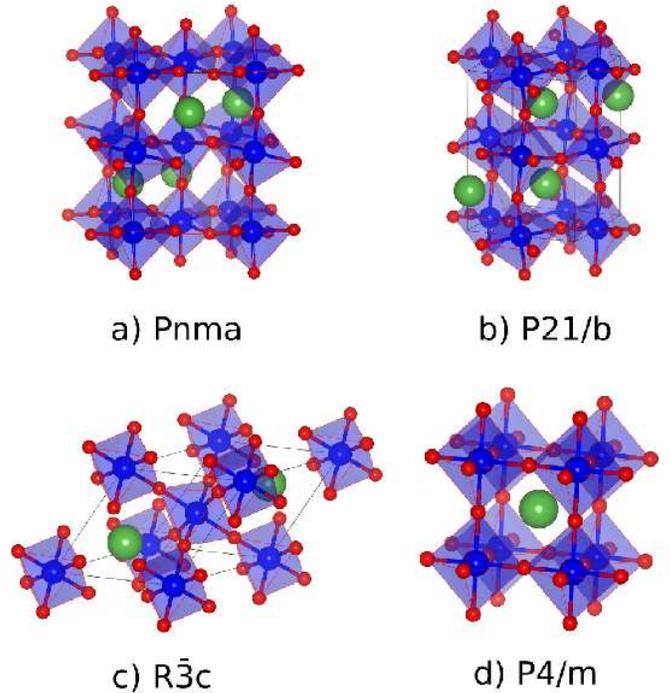}
\caption
{(Color online) The structures of perovskite oxides studied in this paper. P$_{nma}$ for LaScO$_3$,
LaTiO$_3$, LaCrO$_3$, LaMnO$_3$, and LaFeO$_3$, P$_{2_1/b}$ for LaVO$_3$, $R_{\bar3c}$
for LaCoO$_3$ and LaNiO$_3$, and P$_{4/m}$ for LaCuO$_3$, respectively. The large (green), medium-sized (blue), and
small (red) balls denote La, $M$, and O atoms, respectively. Plot done using the VESTA visualization
program~\cite{vesta}.}
\label{fig:models}
\end{figure}

In this article we applied the screened hybrid functional introduced by Heyd, Scuseria,
and Ernzerhof (HSE)~\cite{hse} to study the structural, electronic and magnetic properties of the
series of 3$d$ TMO perovskites La$M$O$_3$, with $M$ ranging from Sc to Cu. This is a rather challenging
family of compounds for electronic structure methods for several
reasons\cite{czyzyk94, sarma95, solovyev96, solovyev96b, Mizokawa96, korotin96, sawada96, sawada97, hamada97, yang99,
ravindran02, munoz04, ravindran04, fang04, evarestov05, sahnoun05, okatov05, trimarchi05, Kotomin05, solovyev06, knizek06,
ahn06, Ong08, nohara06, Hsu09, nohara09, zwanziger09, knizek09, Hsu10, Gryaznov10, hashimoto10, laref10, Ederer11, guo11, Ahmad11, Hong12, Franchini12, He12} (see Table \ref{tab:0}:
(i) it encompasses band, Mott-Hubbard (MH) and charge-transfer (CT)
insulators as well as correlated metals (the last two members of the series: LaNiO$_3$ and LaCuO$_3$)\cite{arima};
(ii) different type of antiferromagnetic (AFM) orderings are encountered
across the series (A-type, C-type and G-type, graphically represented  in Fig.\ref{fig:ordering}), but also
non-magnetic (NM, LaScO$_3$) and paramagnetic (PM, La(Co$\rightarrow$Cu)O$_3$ systems\cite{solovyev96};
(iii) the dominating electronic character varies from $d^0$ to $d^8$, and ranges from $t_{2g}$/$e_g$ localization
(with variable crystal filed splitting between $t_{2g}$ and $e_g$ states) to more spatially delocalized
$d$-orbitals\cite{solovyev96};
(iv) the crystal symmetry spans orthorhombic (O), monoclinic (M), rhombohedral (R) and tetragonal (sketches of the
crystal structures is given in Fig.\ref{fig:models}) characterized by a different level of structural distortions
(Jahn-Teller (JT: staggered disproportionation of the $M$-O bondlengths),
GdFeO$_3$-type (GFO: collective tiltings and rotations of the oxygen octahedra), monoclinic angle $\beta$).

Before describing the method and presenting the result we briefly recall previous {\em ab initio} investigations
of this set of compounds performed using conventional DFT and "beyond-DFT" methodologies. The most widely studied
member of this family is certainly the classical JT-GFO distorted Mutt-Hubbard AFM insulator LaMnO$_3$, but also
other compounds have received significant attention, in particular LaTiO$_3$ and LaVO$_3$, and to a lesser extent,
LaFeO$_3$, LaCoO$_3$, LaNiO$_3$, and LaCuO$_3$. Relatively scarce studies on the band insulator LaScO$_3$
are present in literature.

{\em DFT\cite{czyzyk94, solovyev96, solovyev96b, sawada96, sawada97, hamada97, ravindran02, sahnoun05, Ong08, hashimoto10,
guo11, Franchini12, He12}:} The seminal works of the Terakura group in the late
90s\cite{solovyev96, solovyev96b, sawada96, sawada97, hamada97} have extensively assess the performance of LDA for
the La$M$O$_3$ series ($M$=Ti-Cu), and revealed that LDA is unable to predict the observed insulating ground state for the first
members (LaTiO$_3$ and LaVO$_3$), wrongly favors a non magnetic solution for LaTiO$_3$ and severely underestimate the
insulating gap in LaCrO$_3$, LaMnO$_3$, LaFeO$_3$, and LaCoO$_3$. The situation does not improve using the
GGA\cite{hamada97}. However, the recent GGA-based re-exploration of the electronic properties of LaCrO$_3$
by Ong {\em et al.}\cite{Ong08} has reported that a good agreement with experiment can be achieved, upon a
proper (re)interpretation of the optical spectra.
It should be noticed that all these results were obtained using the experimental geometries. The very few structural
optimizations at DFT level, mostly focused on LaMnO$_3$, have shown that though LDA/GGA reproduce the experimental volume
within 1-3 \% \cite{hashimoto10, Kotomin05, He12}, the lattice distortions associated with the JT and GFO instabilities
are significantly underestimated. For compounds with more delocalized 3$d$ electrons like LaNiO$_3$
the LDA performance get better as recently reported by Guo {\em et al.}\cite{guo11}.

{\em DFT+U\cite{czyzyk94, solovyev96, korotin96, sawada97, hamada97, yang99, ravindran04, fang04, okatov05,
trimarchi05, Kotomin05, knizek06, ahn06, Ong08, Hsu09, nohara09, zwanziger09, knizek09, Hsu10, hashimoto10,
laref10, hu00, Ederer11, guo11, Franchini12}:}
In some cases the drawbacks of LDA and GGA in treating localized partially filled $d$ states can be adjusted by
introducing a strong Hartree-Fock like intra-atomic interaction U properly balanced by the so-called double-counting (dc) correction.
The resulting LDA(GGA)+U energy functional can be written as\cite{Anisimov91}
\begin{equation}
E_{\rm tot}(n,\hat n)=E_{\rm DFT}(n)+E_{\rm HF}(\hat n)-E_{\rm dc}(\hat n)
\label{eq:ldau}
\end{equation}
where $\hat n$ is the operator for the number of electrons occupying a particular site and  $n$ is its expectation
value. This expression can be written in terms of the direct (U) and exchange (J) contributions, which lead to a set
of {\em slightly} different LDA(GGA)+U energy functionals depending on the way the dc term is constructed\cite{Ylvisaker09}.
Among the numerous applications of DFT+U to La$M$O$_3$ the study of I. Solovyev and coworkers represents the most
comprehensive and systematic one\cite{solovyev96}. There it is found that LDA+U conveys a substantially improved
description of the bandstructure of La$M$O$_3$ from LaTiO$_3$ to LaCuO$_3$ with respect to conventional DFT, though the
results critically depend on the specific treatment of localization effects in the 3$d$ manifold. By applying U-correction
to $t_{2g}$ electrons only the authors show that LaTiO$_3$ and LaVO$_3$ are correctly predicted to be insulating, thus
curing the deficient LDA picture. At variance with DFT LaTiO$_3$ is found to be magnetic, but with a magnetic moment
twice larger than the
experimental one. The bandgap of early (LaTiO$_3$, LaVO$_3$) and late (LaCoO$_3$) La$M$O$_3$ members which
have a predominant  $t_{2g}$ character are better described than the $e_g$ compounds LaMnO$_3$ and LaFeO$_3$, for which
an on site U applied to the entire 3$d$ is needed to improve the agreement with experiment. The values of the gap clearly
depends on the value of the U parameter, as discussed by Z. Yang {\em et al.}\cite{yang99}. By fitting the U using the measured gap as
reference quantity, these authors have shown that the best agreement with experiment is achieved for U progressively
increasing from 5 eV (LaCrO$_3$) to 7 eV (LaNiO$_3$), about 2 eV smaller than those computed by I. Solovyev using
constrained LDA\cite{solovyev96}. Similarly to the standard LDA case,
few attempts have been made to optimize the structural parameters
at DFT+U level\cite{sawada97, ahn06, hashimoto10, laref10, guo11}:
(i) {\em LaTiO$_3$}. H.S. Ahn\cite{ahn06} and coworkers have shown that the application of LDA+U (U=3.2 eV) systematically
increase the (underestimated) LDA lattice parameters of LaTiO$_3$ and the internal distortions, thus
improving the overall agreement with experiment.
(ii) {\em LaMnO$_3$}. using the Perdew-Burke-Ernzerhof\cite{pbe} (PBE) approximation with an on-site effective U=2 eV
T. Hashimoto {\em et al.}\cite{hashimoto10} have performed a full (volume and internal coordinates) structural
optimization in LaMnO$_3$ and demonstrated that, unlike GGA, GGA+U accounts well for the experimental JT and tilting
distortions;
(iii) {\em LaCoO$_3$}.
H. Hsu {\em et al.}\cite{Hsu09} and A. Laref {\em et al.}\cite{laref10} have shown that LDA+U
describes well the lattice parameter, rhombohedral angle and atomic coordinates of LaCoO$_3$;
better agreement with experiment is obtained using a self-consistent U\cite{Hsu09} rather than a fixed U value
of $\approx$7-8 eV\cite{laref10}.
(iv) {\em LaNiO$_3$}.
The work of G. Guo {\em et al.}\cite{guo11} on LaNiO$_3$ reported that for this correlated
metal LDA+U (U=6 eV) delivers geometrical data very similar to the already satisfactory LDA ones
(though, as already pointed out, LDA does a better job in predicting the electronic properties).

{\em HF\cite{Mizokawa96, solovyev06, su}:}
The application of a purely Hartee-Fock (HF) procedure, i.e. including an exact treatment of the exchange interaction
and neglecting electron correlation, has been extensively investigated by T. Mizokawa and A. Fujimori\cite{Mizokawa96}
and by I. Solovyev\cite{solovyev06}. Though the HF method suffers for the absence of electron correlation which is
reflected by its tendency to overestimate the magnitude of band gaps (which can be cured by including the correlation
effects beyond the HF approximation), these studies show that HF can qualitatively explain the ground state electronic
and magnetic properties of this class of magnetic oxides. Important exceptions are LaNiO$_3$ and LaCuO$_3$,
which are found to be FM insulator (LaNiO$_3$) and G-type AFM insulator (LaCuO$_3$), in contrast with the observed
PM metallic ground state. Another critical case for HF and in general for electronic structure methods is the
origin of the type-G AFM ordering in LaTiO$_3$\cite{Mizokawa96, Mochizuki04, Pavarini05, solovyev06}: in Ref.\onlinecite{Mizokawa96} the authors report that the stabilization
of the G-type arrangement can be achieved by fixing the $\widehat{{\rm Ti}-{\rm O}-{\rm Ti}}$ angle to approximately
the experimental value. The resulting magnetic moment, downsized by spin orbit interaction effect, results in good
agreement with the measured value, but the calculated bandgap is dramatically wrong, about 2.7 eV, against the measured
value of 0.1 eV\cite{arima}. The results of Ref.\onlinecite{solovyev06} go to the opposite direction:
the magnetic ground state remains wrong even upon inclusion of correlation effects, but the bandgap, 0.6 eV, is in
much better agreement with experiment. A similar trend is also observed for LaVO$_3$.

{\em Hybrid Functionals\cite{munoz04, evarestov05, Gryaznov10, guo11, Ahmad11, Hong12, Franchini12, He12, Xao12,
Rivero09, Rivero10}:}
An alternative methodology to DFT and HF which has attracted a considerable attention
in the solid state physics and chemistry communities in the last two decades is the so called hybrid functional approach.
Originally introduced by A.D.J. Becke in 1993\cite{Becke93}, the hybrid functional scheme relays on a suitable mixing
between HF and $local/semilocal$ (LDA/GGA) DFT theories, in which a portion of the exact $non-local$ HF exchange
\begin{equation}
E_X^{\rm HF}(r,r') = -\frac{1}{2} \sum_{i.j} \int\int d^3{\bf r} d^3{\bf r}'
\frac{\phi_{i}^{*}({\bf r})\phi_{j}({\bf r})\phi_{j}^{*}({\bf r}')\phi_{i}({\bf r}')}
{|{\bf r}-{\bf r}'|}
\end{equation}
is mixed with the complementary LDA/GGA $local/semilocal$ approximated exchange $E_X^{\rm DFT}(r)$. The resulting general hybrid
XC kernel $E_{XC}^{\rm Hybrid}$ (decomposed over its exchange (X) and correlation (C) terms) can be written in the form:
\begin{equation}
E_{XC}^{\rm Hybrid} = {\alpha}E_X^{HF} + (1-\alpha)E_X^{LDA/GGA} + E_C^{LDA/GGA}
\end{equation}
where the mixing factor $\alpha$ controls the amount of exact $E_X^{HF}$ incorporated in the hybrid functional.
Similarly to DFT+U (which makes use of the HF-like intra-atomic interaction U, as recalled above) hybrid functionals tends
to correct the LDA/GGA delocalization error and to provide a better description of TMO with partially filled $d$ and $f$ states.
The advantages with respect to DFT+U is that hybrid functionals (i) do not suffer from the double counting term (see Eq. \ref{eq:ldau})
and, even most importantly, (ii) use an orbital dependent functional acting on all states, extended as well as localized
(in the DFT+U method, the improved treatment of exchange effects is limited to states localized inside the atomic spheres,
and usually limited to the partially filled TM shell). Though both schemes problematically depends on
$semiempirical$ parameters such as U and J in DFT+U and the mixing factor $\alpha$ in hybrid functionals, many attempts
have been made to overcome these
difficulties\cite{Gunnarsson89, Anisimov91b, Imada04, Cococcioni05, Solovyev06, Karlsson10, Marques11}.

Though sparse in literature, hybrid functionals studies of La$M$O$_3$ are increasing in the last few years
\cite{munoz04, evarestov05, Paier05, Gryaznov10, guo11, Ahmad11, archer, Franchini12, He12, Iori12}.
Applications of the renowned Becke, 3-parameter, Lee-Yang-Parr B3LYP functional\cite{Becke93} to LaMnO$_3$\cite{munoz04,
evarestov05, Ahmad11} have shown that this method properly favors the type-A AFM ground state and provides an accurate description
of the band gap, magnetic coupling constants, and Gibbs formation energies. The only structural optimization of the
JT-distorted structure, however, deliver lattices constants which deviate by 5\% from experiment\cite{Ahmad11}.
We have recently reported that HSE performs very well in predicting the ground state properties of LaMnO$_3$,
including the optimized structural parameters, and that the data are slightly dependent on the actual value
of the mixing factor\cite{Franchini12, He12}. D. Gryaznov {\em et al.} have successfully studied the structural and
phonon properties of LaCoO$_3$ using the PBE0 (Perdew-Ernzerhof-Burke)\cite{pbe0} hybrid functional and reported
a substantial improvement with respect to conventional DFT. The application of HSE and PBE0 functionals to LaNiO$_3$,
conversely, turned out to give poor agreement with the experimental photoemission spectroscopy (PES);
this is in line with precedent unsatisfactory HSE/PBE0 results obtained for other itinerant magnetic metals\cite{paier06}.
The influence of the non-local exchange on the electronic properties of LaTiO$_3$ has been investigated recently
by F. Iori and coworkers\cite{Iori12}. By adopting the experimental structure
these authors clarified that the improved description of HSE over DFT+U is due to a correct re-positioning of the O $p$ states,
and show that by fixing the mixing parameter $\alpha$ to its "standard" value, 0.25, the bandgap and the magnetic moment
are significantly overestimated with respect to measurements.

{\em SIC\cite{Zenia05,Filippetti11}:}
Another approach to correct the self-interaction (SI) LDA/GGA problem is the self-interaction correction
method\cite{Perdew81, Svane90, Szotek91, Filippetti03}, in which an approximated (atomic-like and orbitally averaged)
self-interaction is subtracted from the LDA XC functional. Though conceptually different from LDA+U
(in LDA+U an additional effective Coulomb term is added to the LDA/GGA functional), the SIC method is often
pragmatically viewed as a generalized LDA+U approach in which the atomic SI plays the role of the U\cite{Filippetti11}.
Several implementations of the SIC scheme have been proposed, characterized by a different level of complexity
in treating the SI term and from the different underlying computational
framework\cite{Perdew81,Svane90, Szotek91, Filippetti03}, but all demonstrated an appreciable accuracy in
predicting and interpreting the electronic structure of a vast range of systems, including SCSs and TMOs\cite{Filippetti09}.

A valid illustration of the performance of the SIC method is supplied by the results obtained for
LaTiO$_3$ recently discussed by Filippetti {\em et al.}\cite{Filippetti11}. By assuming the experimental cell parameters
SIC finds the correct AFM type-G insulating ordering and delivers internal structural distortions close to
the experimental ones. As a downside, however, the magnitudes of the bandgap (1.6 eV) and magnetic moment
(0.89 $\mu_{\rm B}$) are substantially larger than the corresponding measured values
($\approx$0.2 eV and $\approx$0.5 $\mu_{\rm B}$, respectively).
Other SIC applications to the La$M$O$_3$ series are limited, to our knowledge, to the ideal undistorted cubic phase
of LaMnO$_3$\cite{Zenia05}, for which a stringent comparison with experiment is difficult to do.

{\em GW\cite{nohara06, nohara09, Franchini12}:}
We finally recall the main achievements on La$M$O$_3$ acquired using the GW approximation, a computational
method fundamentally different from both DFT and HF. GW is configured to reflect and to treat the quasi-particle
nature of electrons on the basis of Green's function many-body perturbation theory\cite{Hedin65}, by
explicitly accounting for the non-local and frequency-dependent self-energy ($\Sigma$) in a suitably rewritten
Schr\"odinger-like equation. In the GW approximation $\Sigma$ is approximated to the lowest order term of the
Hedin's equation, and can be written as:
\begin{equation}
\Sigma = iGW
\end{equation}
where G is the Green's function and W is the dynamically screened Coulomb kernel.
In the most widely used single-shot G$_0$W$_0$ approximation both G and W are treated in an
unperturbed manner, but with increasing computer power self-consistent or partially self-consistent
GW schemes are becoming more and more possible\cite{Shishkin07, Franchini10, Franchini12}.
Due to the extensive computing time required to perform GW-like calculations,
only few GW data are available in literature for complex systems.
Among these, the works of Y. Nohara {\em et al.}\cite{nohara06, nohara09}
represent a very comprehensive example of a systematic application of GW to La$M$O$_3$ starting from preconvergent
LDA+U wavefunctions. These authors have obtained excellent agreement with experimental spectra, but
probably due to the uncertainties connected to the choice of U in preparing the initial wavefuncitons the
values of the computed bandgaps deviate significantly from the experimental estimations, especially for
LaTiO$_3$, LaVO$_3$, and LaCoO$_3$. Good agreement with experiment has been also obtained for LaMnO$_3$ using
a partially self-consistent GW$_0$ approach, in this case starting from GGA wavefunction\cite{Franchini12}.

The paper is organized as it follows. In Section \ref{sec:comp} we illustrate the computational method
and its technical aspects; in Section \ref{sec:results} we report the results on the structural optimization
(Section \ref{sec:struc}) and electronic \& magnetic properties (Section \ref{sec:elec}).
A more general discussion on the observed trends and behaviors is developed in
Section \ref{sec:discussion}, and finally in Section \ref{sec:con} we draw our summary and conclusions.

\section{COMPUTATIONAL ASPECTS}
\label{sec:comp}
All calculations were performed using the Vienna ab initio Simulation Package (VASP)\cite{gk1,gk2}
employing DFT and hybrid-DFT approaches within the projector augmented wave method\cite{blochl,gk4} and
the PBE parametrization scheme\cite{pbe} for the XC functional. 
In the screened hybrid-DFT HSE approach adopted in this study, part of the short-range (sr) PBE exchange functional
is replaced by an equal portion of exact HF exchange, according to the general prescription:
\begin{equation}
E_{XC}^{HSE} = {\alpha}E_X^{HF,sr,\mu} + (1-\alpha)E_X^{PBE,sr,\mu} + E_X^{PBE,lr,\mu} + E_C^{PBE}
\end{equation}
where $\mu$, controls the range separation between the sr and long-range (lr) part of the Coulomb kernel
(1/$r$, with $r= | {\bf r}-{\bf r'} |$), decomposed over long (L) and short (S) terms:
\begin{equation}
\frac{1}{r}=S_{\mu}(r) + L_{\mu}(r)=\frac{\rm erfc(\mu{r})}{r} + \frac{\rm erf(\mu{r})}{r}
\end{equation}
The reason to include a screening parameter $\mu$ is motivated by the computational effort required in
computing the spatial decay of the HF X interaction. In the refined HSE06 hybrid functional, $\mu$ is set
equal to 0.20 $\AA^{-1}$ which corresponds to the distance 2/$\mu$ at which the HF X interactions
starts to become negligible. For $\mu$=0 the PBE0 functional is recovered\cite{pbe0}, whereas for
$\mu\rightarrow\infty$ HSE becomes identical to PBE. Beside the computational cost, the main beneficial
consequence of the inclusion of a screening strategy in PBE0 is that screened hybrids can give access to
the metallic state, which is unaffordable by unscreened PBE0-like hybrids. The HSE method has proven to 
to improve the quantitative and qualitative prediction of a large variety of materials, incluiding 
conventional semiconductors\cite{Heyd05,Gerber07}, transition metal oxides\cite{marsman,Franchini07,Franchinitc}, 
ferroeletrics\cite{Stroppa10}, and surfaces\cite{Stroppa08, Franchinis10}.
The mixing parameter $\alpha$, determining the amount of exact non-local HF X included in the
hybrid XC functional is usually set to 0.25\cite{hse}. In this HSE case the PBE functional is recovered for $\alpha$=0.

Thus, the HSE06 depends by construction on two parameters, $\mu$ and $\alpha$. Though their
standard values are routinely used in solid state calculations, it is to be expected that they may vary
from material to material\cite{Marques11,Varley12}, or that they may be property-dependent\cite{Moreira02,Feng04,munoz04}.
Unfortunately, a rigorous first principle procedure to determine the choice of these parameters does not exist.
The conventional value $\alpha$=1/4 is determined by perturbation theory\cite{pbe0}.
The choice $\mu=0.20$ $\AA^{-1}$ has proven to be a practical compromise between
computational cost and quality of the results\cite{Krukau06}.
Considering that most of the tests and fitting procedures have been performed taking as a
reference atomic or molecular energetical and structural properties, the direct acquisition of these standard values
in extended solid state system is not straightforward\cite{pbe0,Krukau06}.

By linking hybrid density functional theory with many-electron XC self-energy $\Sigma$ within a GW framework
it has been proposed that the mixing factor $\alpha$ can be interpreted as the inverse of the dielectric
constant $\epsilon_\infty$\cite{Gygi89, Clark10, Alkauskas10, Varley12}. Based on this idea, an approximated
recipe to determine the {\em optimum} value of $\alpha$ can be obtained:
\begin{equation}
\alpha_{\rm opt} \approx \frac{1}{\epsilon_\infty}
\label{eq:alpha}
\end{equation}
which depends solely on the dielectric constant and on the "unknown" factor of proportionality.
It is important to emphasize that this relation should be interpreted as
an {\em a posteriori} justification of the choice of the optimum value
of $\alpha$, and not as a fundamental quantum mechanical definition of the
mixing factor.
 It follows straightforwardly that for metal ($\epsilon_\infty=\infty$)
$\alpha_{\rm opt}$ is equal to zero. Several limitations affect this practical rule and degrade its {\em ab initio}
nature\cite{Alkauskas10}, above all an accurate calculation of the dielectric constant, which is presently very difficult
in particular for complex TMOs.

Following this line of thought other strategies have been introduced to overcome
this problem invoking density-functional estimators\cite{Gutle99} in the spirit of the Tran and Blaha
functional\cite{Tran09}, which furnishes parametric expressions inevitably dependent on the specific material dataset,
usually limited to monoatomic and binary semiconductors\cite{Marques11}. To complicate the situation even further, there is some amount of arbitrariness
in transferring the $\alpha_{\rm opt} \approx \frac{1}{\epsilon_\infty}$ relation from unscreened PBE0-like hybrids to
screened ones like HSE, where screening is already present in some form in the range separation controlled by the screening
factor $\mu$. These complications become particularly cumbersome when one moves from 'standard' monoatomic and binary
semiconductors to the more complex ternary TM perovskites. As a matter of fact, due to the absence of a systematic study on
the influence of $\alpha$ in this class of compounds, the large majority of hybrid functionals
studies on ternary TMOs have been performed using the standard 1/4 compromise, though there are neither fundamental
nor practical justifications for this choice.

Thus, in order to shed some light on the role of $\alpha$ in a representative class of ternary TMOs with a
largely varying degree of screening and competition between localization/delocalization effects we have performed
our HSE calculations using 4 different values of $\alpha$: (i) low mixing (strong screening): 0.10 (HSE-10), 0.15 (HSE-15),
(ii) standard mixing: 0.25 (HSE-25), and (iii) high mixing (low screening): 0.35 (HSE-35).
The careful analysis of structural, electronic and magnetic properties will allow us to draw some general trends
which should serve as a guidance for future HSE applications.

{\em Technical setup:}
The plane wave cutoff energy was set to 300 eV. A 4$\times$4$\times$4, 6$\times$6$\times$6, 8$\times$8$\times$8
Monkhorst-Pack $k$-points grids were used to sample the Brillouin zones for P$_{nma}$/P$_{2_1/b}$, R$_{\bar3c}$,
and P$_{4/m}$ structures, respectively. Structural optimization was achieved by relaxing the volume, lattice parameters,
lattice angles and internal atomic positions throughout the minimization of the stress tensor and forces using standard
convergence criteria. Finally, the dielectric constant ${\epsilon_\infty}$ were computed adopting the perturbation
expansion after discretization (PEAD) method\cite{pead,Franchini10}.

\section{RESULTS}
\label{sec:results}

This section is subdivided into two parts which are devoted to the presentation of the structural
(Sec. \ref{sec:struc}), and electronic and magnetic (Sec. \ref{sec:elec}) properties, respectively.
In each subsection we will summarize the specific results obtained for each member of the La$M$O$_3$ series,
and in the next section (Sec. \ref{sec:discussion}) we will provide a more reasoned discussion on the
general trends observed across the series.

As anticipated in the introduction hybrid functionals can be simplistically viewed as an orbital-dependent
DFT+U approach in which the on-site electron-electron interaction parameter U is replaced by the parametric
inclusion of a portion of the exact HF exchange quantified by the mixing factor $\alpha$.
In DFT+U calculations, the U is usually either tuned to fit some specific physical property (i.e. bandgap,
magnetic moment, volume, etc.), or calculated within constrained-LDA procedures\cite{Gunnarsson89, Anisimov91b,
Imada04, Cococcioni05, Solovyev06}. In contrast, most of the available HSE-based calculations present
in literature are done at fixed mixing parameter $\alpha$=0.25. This might erroneously convey the idea
of a minor role played by the mixing factor or, even more fundamentally misleading, that HSE is a purely
{\em ab initio} (i.e. parameter free) scheme. As already discussed previously, in the last few years
the modeling community has started to address this issue\cite{munoz04, Marques11, Varley12} but the amount of
available data are still very limited, in particular for complex oxide. It is therefore instructive to briefly
recall a few results on the choice of the U in DFT+U studies of transition metal oxides in order to
possibly formulate some expectations on the behavior of the mixing parameter $\alpha$ in HSE.
A good example to start with are transition metal monoxides (TMOs: MnO, FeO, CoO and NiO), where the
TM possesses the oxidation state 2+ ($M^{2+}$). Several LDA+U studies have shown that a U between 6 and 8 eV
can provide an accurate enough prediction of band gaps for all TMOs\cite{Han06, Anisimov91}.
Going from $M^{2+}$ to $M^{3+}$  the number of the localized electrons decreases.
Thus, it might be expected that the magnitude of the Coulomb interaction increases due to the
contraction of the spatial extension of the of the 3$d$ ($M^{3+}$) wave functions.\cite{Solovyev94}
However, by comparing $M^{2+}$O and La$M^{3+}$O$_3$ photoemission data it can be unambiguously concluded that the effective
Coulomb interaction decreases in $M^{3+}$ compounds\cite{Solovyev94, Mizokawa95, Sarma94,saitoh,chainani,Abbate93}.

Under the assumption that in La$M^{3+}$O$_3$ the $t_{2g}$ electrons are localized and the $e_g$ electrons are itinerant,
Solovyev\cite{solovyev96} has explained this apparent contradiction by invoking the strong screening
associated with the $e_g$ electrons. Indeed, the computed value of U for the $t_{2g}$ shell in La$M^{3+}$O$_3$
are significantly reduced with respect to U for the $d$ states in $M^{2+}$O. The strength of the screening
depends on the filling of the $e_g$ orbital: it is strong at half-filling and less efficient when the $e_g$ are
nearly empty of occupied.  The results of Solovyev indicate that this $t_{2g}$-U approach reproduces sufficiently well
the main features of early (Ti-V-Cr) and late (Co-Ni) La$M$O$_3$ compounds but fails for LaFeO$_3$ (much too small
bandgap and magnetic moment) and LaMnO$_3$ (small gap). Clearly, effects other than $e_g$ itinerancy contribute
to the strength of the U, such as the screening from non-3$d$ electrons, $M$(3$d$/4$s$)-O(2$p$) hybridization, and
lattice relaxation which can explain the discrepancy between self-consistent +U methods and experiments.

A fitting-U approach can selectively adjust the comparison with the experimental gap (not for LaCrO$_3$) at the expense
of a rigorous description of the position of the $e_g$, $t_{2g}$ and O($p$) sub-bands (i.e. the 'correct' value of the
bandgap can arise from an fundamentally incorrect artificial electronic structure). This failure prevents any
physically sound specification/understanding of the (MH or CT) character of the gap:
in Ref.\onlinecite{yang99}, for instance, the gap of LaMnO$_3$ is found to be predominantly CT like,
in discrepancy with the actual situation (LaMnO$_3$ is a MH insulator with a gap opened between occupied and
empty $e_g$ states, partially hybridized with O $p$ states).
Furthermore, the 'optimum' U's resulting from fitting-U schemes do not seem to reflect the observed
$M^{2+}$ to $M^{3+}$ U-reduction, which is an additional sign of the inadequacy of such a procedure.

Considering that standard HSE ($\alpha$=0.25) performs quite well for TM monoxides\cite{Franchini05,marsman} we
can expect that a smaller value will turn out to be more appropriate for reproducing the ground state properties
of La$M$O$_3$. Furthermore, given the full-orbital character of HSE we may expect that hybridization effects
and screening from non 3$d$ electrons will be better described as compared to DFT+U.
Finally, we should point out that the choice to perform a complete structural optimization at each considered
value of $\alpha$  allows for a more genuine account of the structural contribution
to the screening which is disguised in frozen lattice (atomic positions fixed to experimental ones) calculations.

\subsection{Structural Properties}
\label{sec:struc}

As already mentioned in the introduction four different crystal symmetries are encountered across the
La$M$O$_3$ series (see Fig. \ref{fig:models}):
(i) Orthorhombic P$_{nma}$ for LaScO$_3$, LaTiO$_3$, LaCrO$_3$, LaMnO$_3$, and LaFeO$_3$;
(ii) Monoclinic P$_{2_1/b}$ for LaVO$_3$;
(iii) Rhombohedral $R_{\bar3c}$ for LaCoO$_3$ and LaNiO$_3$; and
(iv) Tetragonal P$_{4/m}$for LaCuO$_3$.
All these different structures share the same octahedral perovskitic building block $M$O$_6$, characterized
by one central TM metal surrounded by two apical (O$_1$) oxygen atoms and four planar (O$_2$) oxygen atoms.
Depending on the specific compound,  the $M$O$_6$ octahedra can undergo two kinds of structural distortions:
the JT distortion, manifested by a short (s) and long (l) $M$-O$_2$ in-plane distances and
medium (m) $M$-O$_1$ vertical ones (along the octahedral axis), and the GFO tilting of the
$\widehat{{M}-{\rm O_1}-{M}}$ and $\widehat{{M}-{\rm O_2}-{M}}$ 180$^\circ$ angles.
The cooperative JT distortion is usually measured in
terms of the JT modes Q$_2$=2($l$-$s$)/$\sqrt{2}$ and Q$_3$=2(2$m$-$l$-$s$)/$\sqrt{6}$.
In our full structural relaxation we have optimized the volume (V), lattice parameters $a$, $b$, and $c$,
the monoclinic/rhombohedral angle $\beta$, as well as all internal atomic coordinates (this clearly includes
all relevant GFO and JT structural parameters $\widehat{{\rm M}-{\rm O_1}-{\rm M}}$ (${\theta}_1$),
$\widehat{{M}-{\rm O_2}-{M}}$ (${\theta}_2$), Q$_2$, and Q$_3$).

\begin{figure}
\includegraphics[clip,width=0.50\textwidth]{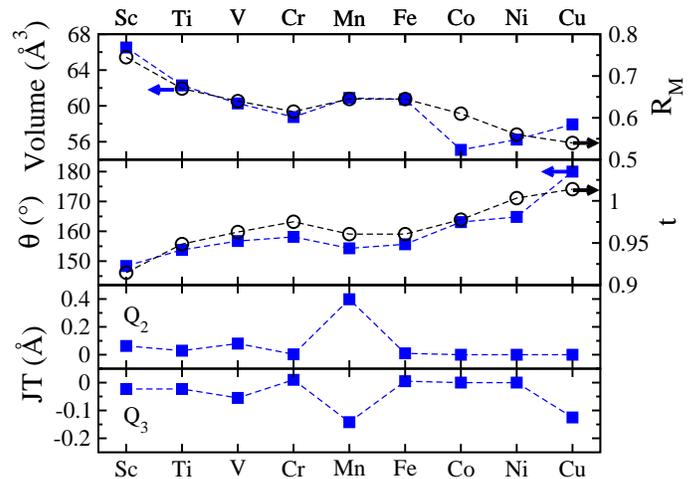}
\caption
{Experimental trend of Volume (V),  average tilting angle ($\theta$), and JT distortions (Q$_2$ and  Q$_3$),
for the La$M$O$_3$ series from $M$=Sc to $M$=Cu. The corresponding trend of the tolerance factor
($t$=($\rm R_A$+$\rm R_O$)/$\sqrt{2}$($\rm R_M$+$\rm R_O$), and $\rm R_M$ is also shown.}
\label{fig:trend_exp}
\end{figure}

\begin{table}
\caption{
Structural data for LaScO$_3$.
Comparison between the optimized parameters calculated using PBE and HSE (with different values of the mixing factor)
and the available (room temperature) experimental data taken from Ref. \onlinecite{geller}.
The relative error (in brackets, in \%) and the mean absolute relative error (MARE, \%) is also supplied.
} \vspace{0.3cm}
\begin{ruledtabular}
\begin{tabular}{ccccccc}
                       & Expt.   &    HSE-35   &   HSE-25    &   HSE-15    &   HSE-10    &  PBE        \\
V (\AA$^3$)       & 266.09  & 262.02      & 263.48      & 265.12      & 265.99      & 267.90      \\
                       &         &       (1.5) &       (1.0) &       (0.4) &       (0.0) &       (0.7) \\
$a$ (\AA)              &  5.787  &  5.764      &  5.780      &  5.794      &  5.798      &  5.810      \\
                       &         &       (0.4) &       (0.1) &       (0.1) &       (0.2) &       (0.4) \\
$b$ (\AA)              &  8.098  &  8.050      &  8.061      &  8.076      &  8.088      &  8.108      \\
                       &         &       (0.6) &       (0.5) &       (0.3) &       (0.1) &       (0.1) \\
$c$ (\AA)              &  5.678  &  5.647      &  5.655      &  5.666      &  5.672      &  5.686      \\
                       &         &       (0.5) &       (0.4) &       (0.2) &       (0.1) &       (0.1) \\
Sc-O$_m$ (\AA)         &  2.104  &  2.091      &  2.096      &  2.100      &  2.103      &  2.108      \\
                       &         &       (0.6) &       (0.4) &       (0.2) &       (0.0) &       (0.2) \\
Sc-O$_l$ (\AA)         &  2.140  &  2.095      &  2.101      &  2.108      &  2.109      &  2.115      \\
                       &         &       (2.1) &       (1.8) &       (1.5) &       (1.4) &       (1.2) \\
Sc-O$_s$ (\AA)         &  2.096  &  2.082      &  2.086      &  2.091      &  2.093      &  2.098      \\
                       &         &       (0.7) &       (0.5) &       (0.2) &       (0.1) &       (0.1) \\
${\theta}_1$ ($^\circ$)
                       & 148.39  & 148.42      & 148.08      & 148.14      & 148.19      & 148.18      \\
                       &         &       (0.0) &       (0.2) &       (0.2) &       (0.1) &       (0.1) \\
${\theta}_2$ ($^\circ$)
                       & 146.29  & 149.98      & 149.95      & 149.55      & 149.68      & 149.53      \\
                       &         &       (2.5) &       (2.5) &       (2.2) &       (2.3) &       (2.2) \\
\hline
MARE                   &         &        1.0  &    0.8      &    0.6      &    0.5      &     0.6     \\
\hline
Q$_2$                  &  0.063  &  0.018      &  0.021      &  0.024      &  0.023      &  0.023      \\
Q$_3$                  & -0.023  &  0.004      &  0.005      &  0.000      &  0.003      &  0.002      \\
\end{tabular}
\end{ruledtabular}
\label{tab:1}
\end{table}

\begin{table}
\caption{
Structural data for LaTiO$_3$.
Comparison between the optimized parameters calculated using PBE and HSE (with different values of the mixing factor)
and the available experimental data taken from Ref. \onlinecite{cwik}.
The relative error (in brackets, in \%) and the mean absolute relative error (MARE, \%) is also supplied.
} \vspace{0.3cm}
\begin{ruledtabular}
\begin{tabular}{ccccccc}
                       & Expt.   &   HSE-35   &   HSE-25    &    HSE-15   &   HSE-10   &  PBE       \\
V (\AA$^3$)       & 249.17  & 250.03     & 250.00      & 250.98      & 251.17     & 250.61     \\
                       &         &       (0.3)&       (0.3) &       (0.7) &       (0.8)&       (0.6)\\
$a$ (\AA)              & 5.589   & 5.599      & 5.597       & 5.612       & 5.617      &  5.646     \\
                       &         &      (0.2) &      (0.1)  &      (0.4)  &      (0.5) &       (1.0)\\
$b$ (\AA)              & 7.901   & 7.931      & 7.909       & 7.915       & 7.913      &  7.929     \\
                       &         &      (0.4) &      (0.1)  &      (0.2)  &      (0.2) &       (0.4)\\
$c$ (\AA)              & 5.643   & 5.631      & 5.648       & 5.651       & 5.650      &  5.598     \\
                       &         &      (0.2) &      (0.1)  &      (0.1)  &      (0.1) &       (0.8)\\
Ti-O$_s$ (\AA)         & 2.028   & 2.038      & 2.034       & 2.033       & 2.029      &  2.018     \\
                       &         &      (0.5) &      (0.3)  &      (0.2)  &      (0.0) &       (0.5)\\
Ti-O$_l$ (\AA)         & 2.053   & 2.059      & 2.069       & 2.071       & 2.068      &  2.033     \\
                       &         &      (0.3) &      (0.8)  &      (0.9)  &      (0.7) &       (1.0)\\
Ti-O$_m$ (\AA)         & 2.032   & 2.027      & 2.023       & 2.028       & 2.030      &  2.029     \\
                       &         &      (0.2) &      (0.4)  &      (0.2)  &      (0.1) &       (0.1)\\
${\theta}_1$ ($^\circ$)& 153.78  & 153.12     & 152.98      & 153.54      & 154.25     & 158.47     \\
                       &         &      (0.4)&       (0.5) &       (0.2) &       (0.3)&       (3.0)\\
${\theta}_2$ ($^\circ$)& 152.93  & 152.64     & 152.54      & 152.59      & 152.97     & 156.37     \\
                       &         &      (0.2)&       (0.3) &       (0.2) &       (0.0)&       (2.2)\\
\hline
MARE                   &         &     0.31   &    0.33     &    0.35     &    0.31    &    1.07    \\
\hline
Q$_2$                  &  0.029  &  0.046     &  0.065      &  0.062      &  0.054     &  0.006     \\
Q$_3$                  & -0.023  & -0.008     & -0.021      & -0.027      & -0.031     & -0.021     \\
\end{tabular}
\end{ruledtabular}
\label{tab:2}
\end{table}

\begin{table}
\caption{
Structural data for LaVO$_3$.
Comparison between the optimized parameters calculated using PBE and HSE (with different values of the mixing factor)
and the available (T=10 K) experimental data taken from Ref. \onlinecite{bordet}.
The relative error (in brackets, in \%) and the mean absolute relative error (MARE, \%) is also supplied.
} \vspace{0.3cm}
\begin{ruledtabular}
\begin{tabular}{cccccccc}
                            & Expt.   &   HSE-35    &   HSE-25    &   HSE-15   &   HSE-10   &    PBE     \\
V (\AA$^3$)            & 241.10  & 240.31      & 241.33      & 242.20     & 242.45     & 241.64     \\
                            &         &       (0.3) &       (0.1) &       (0.5)&       (0.6)&       (0.2)\\
a (\AA)                     & 5.5917  &  5.562      &  5.582      &  5.622     &  5.637     &  5.613     \\
                            &         &       (0.5) &       (0.2) &       (0.5)&       (0.8)&       (0.4)\\
b (\AA)                     & 7.7516  &  7.801      &  7.787      &  7.729     &  7.713     &  7.729     \\
                            &         &       (0.6) &       (0.5) &       (0.3)&       (0.5)&       (0.3)\\
c (\AA)                     & 5.5623  &  5.538      &  5.552      &  5.574     &  5.577     &  5.570     \\
                            &         &       (0.4) &       (0.2) &       (0.2)&       (0.3)&       (0.1)\\
$\beta$ ($^\circ$)          &  90.13  &  89.93      &  90.16      &  90.16     &  90.18     &  90.03     \\
                            &         &       (0.2) &       (0.0) &       (0.0)&       (0.1)&       (0.1)\\
V$_2$-O$_s$ (\AA)           &  1.979  &  2.019      &  1.993      &  1.974     &  1.966     &  1.962     \\
                            &         &       (2.0) &       (0.7) &       (0.3)&       (0.7)&       (0.9)\\
V$_2$-O$_{l}$ (\AA)         &  2.039  &  2.007      &  2.019      &  2.055     &  2.054     &  2.012     \\
                            &         &       (1.6) &       (1.0) &       (0.8)&       (0.7)&       (1.3)\\
V$_2$-O$_s$ (\AA)           &  1.979  &  2.004      &  1.999      &  1.984     &  1.991     &  2.011     \\
                            &         &       (1.3) &       (1.0) &       (0.3)&       (0.6)&       (1.6)\\
V$_1$-O$_s$ (\AA)           &  1.978  &  1.969      &  1.989      &  1.972     &  1.965     &  1.961     \\
                            &         &       (0.5) &       (0.6) &       (0.3)&       (0.7)&       (0.9)\\
V$_1$-O$_{l}$ (\AA)         &  2.042  &  2.007      &  2.026      &  2.063     &  2.059     &  2.013     \\
                            &         &       (1.7) &       (0.8) &       (1.0)&       (0.8)&       (1.4)\\
V$_1$-O$_{m}$ (\AA)         &  1.989  &  1.997      &  1.997      &  1.982     &  1.990     &  2.013     \\
                            &         &       (0.4) &       (0.4) &       (0.4)&       (0.1)&       (1.2)\\
${\theta}_1$ ($^\circ$)     & 156.74  & 155.83      & 155.88      & 156.65     & 157.70     & 160.15     \\
                            &         &       (0.6) &       (0.5) &       (0.1)&       (0.6)&       (2.2)\\
${\theta}_{21}$ ($^\circ$)     & 157.83  & 156.23      & 156.90      & 157.05     & 157.06     & 158.76     \\
                            &         &       (1.0) &       (0.6) &       (0.5)&       (0.5)&       (0.6)\\
${\theta}_{21}$ ($^\circ$)     & 156.12  & 157.08      & 156.23      & 156.28     & 156.51     & 158.26     \\
                            &         &       (0.6) &       (0.1) &       (0.1)&       (0.2)&       (1.4)\\
\hline
MARE                        &         &     0.82    &    0.48     &     0.35   &      0.51  &     0.98   \\
\hline
Q$_{21}$                    &  0.085  &  0.003      &  0.028      &  0.101     &  0.088     &  0.001     \\
Q$_{31}$                    & -0.050  &  0.023      & -0.027      & -0.075     & -0.093     & -0.080     \\
Q$_{22}$                    &  0.074  &  0.015      &  0.042      &  0.114     &  0.098     &  0.002     \\
Q$_{32}$                    & -0.060  & -0.054      & -0.037      & -0.082     & -0.097     & -0.084     \\
\end{tabular}
\end{ruledtabular}
\label{tab:3}
\end{table}

\begin{table}
\caption{
Structural data for LaCrO$_3$.
Comparison between the optimized parameters calculated using PBE and HSE (with different values of the mixing factor)
and the available (room temperature)
experimental data taken from Ref. \onlinecite{khattak} (similar structural data can be found in
Refs.\onlinecite{tezuka} and \onlinecite{li}).
The relative error (in brackets, in \%) and the mean absolute relative error (MARE, \%) is also supplied.
} \vspace{0.3cm}
\begin{ruledtabular}
\begin{tabular}{ccccccccc}
                       & Expt.   &   HSE-35   &   HSE-25   &    HSE-15   &  HSE-10   &  PBE        \\
V (\AA$^3$)       & 235.02  & 233.45     & 234.74     & 236.13      & 236.69    & 237.70      \\
                       &         &       (0.7)&       (0.1)&       (0.5) &       (0.7)&       (1.1) \\
$a$ (\AA)              & 5.483   & 5.478      &  5.509     &  5.494      &  5.531    &  5.512      \\
                       &         &      (0.1) &       (0.5)&       (0.2) &       (0.9)&       (0.5) \\
$b$ (\AA)              & 7.765   & 7.752      &  7.766     &  7.776      &  7.785    &  7.795      \\
                       &         &      (0.2) &       (0.0)&       (0.1) &       (0.3)&       (0.4) \\
$c$ (\AA)              & 5.520   & 5.498      &  5.487     &  5.527      &  5.531    &  5.533      \\
                       &         &      (0.4) &       (0.6)&       (0.1) &       (0.2)&       (0.2) \\
Cr-O$_l$ (\AA)         & 1.977   & 1.973      &  1.977     &  1.982      &  1.984    &  1.983      \\
                       &         &      (0.2) &       (0.0)&       (0.3) &       (0.4)&       (0.3) \\
Cr-O$_{m}$ (\AA)       & 1.972   & 1.971      &  1.975     &  1.979      &  1.981    &  1.985      \\
                       &         &      (0.1) &       (0.2)&       (0.4) &       (0.5)&       (0.7) \\
Cr-O$_{s}$ (\AA)       & 1.970   & 1.970      &  1.975     &  1.979      &  1.980    &  1.984      \\
                       &         &      (0.0)&       (0.3)&       (0.5) &       (0.5)&       (0.7) \\
${\theta}_1$ ($^\circ$)& 158.14  & 158.29     & 158.10     & 159.76      & 157.71    & 158.51      \\
                       &         &       (0.1)&       (0.0)&       (1.0) &       (0.3)&       (0.2) \\
${\theta}_2$ ($^\circ$)& 161.32  & 159.72     & 159.60     & 157.66      & 159.73    & 159.30      \\
                       &         &       (1.0)&       (1.1)&       (2.3) &       (1.0)&       (1.3) \\
\hline
MARE                   &         &    0.30    &     0.30   &    0.60     &    0.51   &    0.61     \\
\hline
Q$_2$                  & 0.003   & 0.001      &  0.001     &  0.000      & 0.001     &  0.001      \\
Q$_3$                  & 0.010   & 0.004      &  0.004     &  0.004      & 0.005     & -0.002      \\
\end{tabular}
\end{ruledtabular}
\label{tab:4}
\end{table}

\begin{table}
\caption{
Structural data for LaMnO$_3$.
Comparison between the optimized parameters calculated using PBE and HSE (with different values of the mixing factor)
and the available (T=4.2 K) experimental data taken from Ref. \onlinecite{exp1}.
The relative error (in brackets, in \%) and the mean absolute relative error (MARE, \%) is also supplied.
} \vspace{0.3cm}
\begin{ruledtabular}
\begin{tabular}{cccccccc}
                      & Expt.  &   HSE-35   &   HSE-25  &   HSE-15   &   HSE-10   &  PBE       \\
V (\AA$^3$)      & 243.57 & 243.98     & 245.82    & 247.36     & 244.24     & 244.21     \\
                      &        &       (0.2)&       (0.9)&       (1.6)&       (0.3)&       (0.3)\\
\emph{a} (\AA)        & 5.532  &  5.526     & 5.537     &  5.553     &  5.661     &  5.569     \\
                      &        &       (0.1)&      (0.1)&       (0.4)&       (2.3)&       (0.7)\\
\emph{b} (\AA)        & 5.742  &  5.789     & 5.817     &  5.820     &  5.594     &  5.627     \\
                      &        &       (0.8)&      (1.3)&       (1.4)&       (2.6)&       (2.0)\\
\emph{c} (\AA)        & 7.668  &  7.628     & 7.633     &  7.653     &  7.712     &  7.793     \\
                      &        &       (0.5)&      (0.5)&       (0.2)&       (0.6)&       (1.6)\\
Mn-O$_m$ (\AA)        & 1.957  &  1.954     & 1.957     &  1.962     &  1.979     &  1.992     \\
                      &        &       (0.2)&      (0.0)&       (0.3)&       (1.1)&       (1.8)\\
Mn-O$_l$ (\AA)        & 2.184  &  2.204     & 2.214     &  2.213     &  2.134     &  2.063     \\
                      &        &       (0.9)&      (1.4)&       (1.3)&       (2.3)&       (5.5)\\
Mn-O$_s$ (\AA)        & 1.903  &  1.899     & 1.905     &  1.914     &  1.923     &  1.971     \\
                      &        &       (0.2)&      (0.1)&       (0.6)&       (1.1)&       (3.6)\\
${\theta}_1$ ($^\circ$)&154.3  & 154.78     & 154.35    & 154.36     & 153.96     & 155.85     \\
                      &        &       (0.3)&       (0.0)&       (0.0)&       (0.2)&       (1.0)\\
${\theta}_2$ ($^\circ$)&156.7  & 154.38     & 154.08    & 154.17     & 157.59     & 157.71     \\
                      &        &       (1.5)&       (1.7)&       (1.6)&       (0.6)&       (0.6)\\
\hline
MARE                  &        &   0.52     &   0.66    &    0.81    &    1.22    &  1.90      \\
\hline
Q$_2$                 & 0.398  &  0.431     & 0.437     &  0.423     &  0.298     &  0.131     \\
Q$_3$                 &-0.142  & -0.159     &-0.167     & -0.165     & -0.080     & -0.041     \\
\end{tabular}
\end{ruledtabular}
\label{tab:5}
\end{table}

\begin{table}
\caption{
Structural data for LaFeO$_3$.
Comparison between the optimized parameters calculated using PBE and HSE (with different values of the mixing factor)
and the available (room temperature)
experimental data taken from Ref. \onlinecite{dann}.
The relative error (in brackets, in \%) and the mean absolute relative error (MARE, \%) is also supplied.
} \vspace{0.3cm}
\begin{ruledtabular}
\begin{tabular}{ccccccc}
                      & Expt.   &   HSE-35   &   HSE-25    &   HSE-15    &    HSE-10   &  PBE        \\
V (\AA$^3$)      & 242.90  & 240.39     & 242.08      & 244.02      & 245.09      & 246.47      \\
                      &         &       (1.0)&       (0.3) &       (0.5) &       (0.9) &       (1.5) \\
$a$ (\AA)             &  5.565  &  5.530     &  5.569      &  5.587      &  5.557      &  5.618      \\
                      &         &       (0.6)&       (0.1) &       (0.4) &       (0.1) &       (1.0) \\
$b$ (\AA)             &  7.854  &  7.829     &  7.842      &  7.861      &  7.868      &  7.878      \\
                      &         &       (0.3)&       (0.2) &       (0.1) &       (0.2) &       (0.3) \\
$c$ (\AA)             &  5.557  &  5.553     &  5.543      &  5.556      &  5.605      &  5.568      \\
                      &         &       (0.1)&       (0.3) &       (0.0) &       (0.9) &       (0.2) \\
Fe-O$_l$ (\AA)        &  2.009  &  2.001     &  2.006      &  2.012      &  2.015      &  2.018      \\
                      &         &       (0.4)&       (0.1) &       (0.1) &       (0.3) &       (0.4) \\
Fe-O$_l$ (\AA)        &  2.009  &  2.002     &  2.010      &  2.017      &  2.024      &  2.032      \\
                      &         &       (0.3)&       (0.0) &       (0.4) &       (0.7) &       (1.1) \\
Fe-O$_s$ (\AA)        &  2.002  &  1.995     &  2.002      &  2.008      &  2.013      &  2.018      \\
                      &         &       (0.3)&       (0.0) &       (0.3) &       (0.5) &       (0.8) \\
${\theta}_1$ ($^\circ$)&155.66  & 155.95     & 155.52      & 155.16      & 155.07      & 154.83      \\
                      &         &       (0.2)&       (0.1) &       (0.3) &       (0.4) &       (0.5) \\
${\theta}_2$ ($^\circ$)&157.26  & 157.10     & 156.69      & 156.29      & 155.78      & 155.21      \\
                      &         &       (0.1)&       (0.4) &       (0.6) &       (0.9) &       (1.3) \\
\hline
MARE                  &         &   0.38     &    0.16     &   0.30      &   0.56      &    0.79     \\
\hline
Q$_2$                 &  0.010  &  0.010     &  0.011      &  0.013      &  0.016      &  0.020      \\
Q$_3$                 &  0.005  &  0.004     &  0.000      & -0.001      & -0.006      & -0.011      \\
\end{tabular}
\end{ruledtabular}
\label{tab:6}
\end{table}

\begin{table}
\caption{
Structural data for LaCoO$_3$.
Comparison between the optimized parameters calculated using PBE and HSE (with different values of the mixing factor)
and the available (T=4 K) experimental data taken from Ref. \onlinecite{thornton} (room temperature data can be found
in Ref. \onlinecite{haas}).
The relative error (in brackets, in \%) and the mean absolute relative error (MARE, \%) is also supplied.
Here ${\theta}_1$=$\widehat{{\rm O}-{\rm Co}-{\rm O}}$, and ${\theta}_2$=$\widehat{{\rm Co}-{\rm O}-{\rm Co}}$.
For LaCoO$_3$ we have optimized the structure using a reduced 0.05 mixing parameter and obtained the following
data: V = 111.87 (\AA$^3$) (1.5), $a$ = 5.354 (\AA) (0.2), Co-O$_1$ = 1.940 (\AA) (0.8), O-Co-O =
88.07 ($^\circ$) (0.5), and Co-O-Co = 161.00 ($^\circ$) (1.3); MARE = 0.82.
}
\vspace{0.3cm}
\begin{ruledtabular}
\begin{tabular}{ccccccc}
                      & Expt.   &   HSE-35   &   HSE-25   & HSE-15     &   HSE-10   &     PBE    \\
V (\AA$^3$)      & 110.17  & 107.78     & 109.02     & 110.39     & 111.09     &  114.11    \\
                      &         &       (2.2)&       (1.0)&       (0.2)&       (0.8)&   (3.6)\\
$a$ (\AA)             &  5.342  &  5.314     &  5.328     &  5.343     &  5.348     &   5.405    \\
                      &         &       (0.5)&       (0.3)&       (0.0)&       (0.1)&   (1.2)\\
$\beta$ ($^\circ$)   &  60.99  &  60.70     &  60.87     &  61.05     &  61.20     &   60.99    \\
                      &         &       (0.5)&       (0.2)&       (0.1)&       (0.3)&   (0.0)\\
Co-O     (\AA)        &  1.924  &  1.904     &  1.915     &  1.925     &  1.932     &   1.948    \\
                      &         &       (1.0)&       (0.5)&       (0.1)&       (0.4)&   (1.2)\\
${\theta}_1$ ($^\circ$) & 88.55  &  88.96     &  88.72     &  88.49     &  88.27     &   88.50    \\
                      &         &       (0.5)&       (0.2)&       (0.1)&       (0.3)&   (0.1)\\
${\theta}_2$ ($^\circ$) & 163.08  & 165.56     & 164.06     & 162.88     & 161.82     &  162.45    \\
                      &         &       (1.5)&       (0.6)&       (0.1)&       (0.8)&   (0.4)\\
\hline
MARE                  &         &   0.95     &     0.42   &      0.09  &     0.44   &   0.92  \\
\end{tabular}
\end{ruledtabular}
\label{tab:7}
\end{table}

\begin{table}
\caption{
Structural data for LaNiO$_3$.
Comparison between the optimized parameters calculated using PBE and HSE (with different values of the mixing factor)
and the available (T=1.5 K) experimental data taken from Ref. \onlinecite{munoz92}.
Here ${\theta}_1$=$\widehat{{\rm O}-{\rm Ni}-{\rm O}}$, and ${\theta}_2$=$\widehat{{\rm Ni}-{\rm O}-{\rm Ni}}$.
The relative error (in brackets, in \%) and the mean absolute relative error (MARE, \%) is also supplied.
} \vspace{0.3cm}
\begin{ruledtabular}
\begin{tabular}{ccccccc}
                      & Expt.   &   HSE-35  &   HSE-25  &   HSE-15  &   HSE-10  &    PBE    \\
V (\AA$^3$)      & 112.48  & 112.02    & 112.47    & 113.42    & 113.83    & 115.20    \\
                      &         &       (0.4)&       (0.0)&       (0.8)&       (1.2)&       (2.4)\\
$a$ (\AA)             &  5.384  &  5.377    &  5.380    &  5.393    &  5.392    &  5.415    \\
                      &         &       (0.1)&       (0.1)&       (0.2)&       (0.1)&       (0.6)\\
$\beta$ ($^\circ$)   &  60.86  &  60.85    &  60.95    &  61.01    &  60.21    &  61.16    \\
                      &         &       (0.0)&       (0.1)&       (0.2)&       (1.1)&       (0.5)\\
Ni-O            (\AA) &  1.933  &  1.930    &  1.935    &  1.941    &  1.947    &  1.953    \\
                      &         &       (0.2)&       (0.1)&       (0.4)&       (0.7)&       (1.0)\\
${\theta}_1$ ($^\circ$) &88.78  &  88.78    &  88.63    &  88.55    &  88.30    &  88.41    \\
                      &         &       (0.0)&       (0.2)&       (0.3)&       (0.5)&       (0.4)\\
${\theta}_2$ ($^\circ$) &64.82  & 164.79    & 163.77    & 163.43    & 162.12    & 163.02    \\
                      &         &       (0.0)&       (0.6)&       (0.8)&       (1.6)&       (1.1)\\
\hline
MARE                  &         &    0.10   &   0.19    &   0.43    &    0.84   &    0.92   \\
\end{tabular}
\end{ruledtabular}
\label{tab:8}
\end{table}

\begin{table}
\caption{
Structural data for LaCuO$_3$.
Comparison between the optimized parameters calculated using PBE and HSE (with different values of the mixing factor)
and available (room temperature) experimental data taken from Ref. \onlinecite{bringley}.
The relative error (in brackets, in \%) and the mean absolute relative error (MARE, \%) is also supplied.
} \vspace{0.3cm}
\begin{ruledtabular}
\begin{tabular}{cccccccc}
                      & Expt.   &    HSE-35  &    HSE-25  &   HSE-15  &   HSE-10  &  PBE       \\
V (\AA$^3$)      &  57.94  &  56.07     &  56.38     & 57.05     & 57.28     &  57.85     \\
                      &         &       (3.2)&       (2.7)&      (1.5)&      (1.1)&       (0.2)\\
$a$ (\AA)             &  3.819  &  3.821     &  3.832     & 3.844     & 3.850     &  3.867     \\
                      &         &       (0.1)&       (0.3)&      (0.7)&      (0.8)&       (1.3)\\
$c$ (\AA)             &  3.973  &  3.840     &  3.840     & 3.861     & 3.865     &  3.869     \\
                      &         &       (3.3)&       (3.3)&      (2.8)&      (2.7)&       (2.6)\\
Cu-O$_l$ (\AA)        &  1.986  &  1.920     &  1.920     & 1.930     & 1.933     &  1.934     \\
                      &         &       (3.3)&       (3.3)&      (2.8)&      (2.7)&       (2.6)\\
Cu-O$_s$ (\AA)        &  1.909  &  1.911     &  1.916     & 1.922     & 1.925     &  1.934     \\
                      &         &       (0.1)&       (0.4)&      (0.7)&      (0.8)&       (1.3)\\
\hline
MARE                  &         &    2.01    &    2.01    &   1.70    &   1.64    &     1.59   \\
\hline
Q$_2$                 &  --     &   --       &    --      &   --      &   --      &    --      \\
Q$_3$                 &  0.125  &  0.016     &  0.007     & 0.014     & 0.013     &  0.002     \\
\end{tabular}
\end{ruledtabular}
\label{tab:9}
\end{table}

A graphical summary of the observed trend of the most relevant structural parameters is given
in Fig. \ref{fig:trend_exp}.  The progressive reduction of the Volume from Sc to Cu is clearly associated
with the almost monotonically decrease of the $M$ ionic radius $\rm R_M$, whose size is determined by the
competition between the size of the 4$s$ shell (where extra protons are pulled in) and the
additional screening due to the increasing number of 3$d$ electrons: adding protons should lead to a
decreased atom size, but this effect is hindered by repulsion of the 3$d$ and, to a lesser extent,  4$s$
electrons. The V/$\rm R_M$ curves show a plateau at about half filling (Cr-Mn-Fe) indicating that for
this trio of elements these two effects are essentially balanced and atom size does not change much.
The volume contraction is associated with a rectification of the average
($\widehat{{M}-{\rm O_1}-{M}}$+$\widehat{{M}-{\rm O_2}-{M}}$)/2 tilting angle $\theta$, which follows very well the
evolution of the tolerance factor $t$=($\rm R_A$+$\rm R_0$)/$\sqrt{2}$($\rm R_M$+$\rm R_O$)
(where $\rm R_A$, $\rm R_M$ and $\rm R_O$ indicate the ionic radius for La, $M$=Sc-Cu and O, respectively).
This indicates that the tolerance factor is indeed a good measure of the overall stability and
degree of distortion of perovskite compounds. Clearly, the value of $t$ is well within the range of
stability  set to $0.78 < t < 1.05$.
The bottom panel of Fig. \ref{fig:trend_exp} conveys the message that Q$_2$ and Q$_3$ assume non negligible values for
LaMnO$_3$ only, confirming that JT distortions are predominant in perovskites containing cations such as Cu$^{2+}$
and Mn$^{3+}$ in their octahedral cation site.

In the following subsections we will report on the full structural optimization of La$M$O$_3$ at PBE and HSE
(for different values of $\alpha$) and will provide a one-to-one comparison with available experimental data,
also in terms of the mean absolute relative error (MARE, not given for the very small quantities Q$_2$ and Q$_3$).

\subsubsection{LaScO$_3$}
LaScO$_3$ crystallizes with a P$_{nma}$ orthorhombic structure, and shows the largest tilting instabilities of
all La$M$O$_3$ series (147.3$^{\circ}$)\cite{geller}. The JT parameter Q$_2$ and Q$_3$ are almost zero (0.063 and -0.023,
respectively) and as a consequence the Sc-O bondlength disproportionation is negligible: both planar and vertical
Sc-O bondlengths are  all $\approx$ 2.1~\AA.
The computed structural data are collected in Table \ref{tab:1}. All methods deliver a quite satisfactory
description with an overall MARE less than 1\%. PBE supplies the best agreement with measurements (MARE = 0.5 \%).
The most critical quantities for theory are Sc-O$_l$ and $\theta_2$, for which relative errors larger than
1\%  and 2\% are found, respectively.
We can thus conclude that for an accurate account of the structural properties of LaScO$_3$ it is not required to
apply beyond-DFT methods. As we will see in the next section, this is not the case for the electronic properties.

\subsubsection{LaTiO$_3$}
Similarly to LaScO$_3$, the low temperature space group of LaTiO$_3$ is P$_{nma}$ with small JT distortions due
to the low JT activity of the single $t^{\uparrow}_{2g}$ orbital and large GFO distortions caused by the
large size difference between Ti and La ions.\cite{cwik,maclean}
Though the overall PBE MARE is only 1\%, the relaxed structure parameters given by HSE functionals are
appreciably better (MARE $\approx$ 0.3\%) than PBE, regardless the amount of exact HF exchange, as summarized in
Table \ref{tab:2}. The PBE errors mostly arise from an incorrect description of the tilting angles,
which are by far ($\approx$ 3\%) overestimated with respect to the low-temperature experimental data\cite{cwik}.
As for the volume, PBE furnishes a nice optimized value, which is improved by going to large $\alpha$
HSE setups (HSE-35 HSE-25).

\subsubsection{LaVO$_3$}
LaVO$_3$ is the only member of the La$M$O$_3$ series displaying a monoclinic structure with
P$_{2_1/b}$ space group\cite{bordet}. The unit cell contains two inequivalent V sites (V$_1$ and V$_2$),
which sit in the center of GFO distorted octahedra not subjected to significant JT distortions.
Due to the occurrence of two different V atoms in the unit cell two different sets of V-O bondlengths and
tilting angle $\theta_2$ (${\theta}_{21}$ and ${\theta}_{22}$) are identified.
The comparison between the low-temperature experimental data and the theoretical values
are collected in Table \ref{tab:3}. The general situation is similar to LaTiO$_3$: the PBE MARE, 1\%,
is about twice larger than the average HSE MARE. The PBE relative errors are large for the tilting angles
and V-O bondlengths, but rather small for volume and lattice constants. HSE leads to slightly better data, in
particular in the range $ 0.10 < \alpha < 0.25 $, but the volume and lattice constants are
reproduced less accurately than at PBE level.

\subsubsection{LaCrO$_3$}
The structural data for P$_{nma}$ LaCrO$_3$ are collected in Table \ref{tab:4}.
The full three-fold degenerate $t_{2g}$ shell inhibits completely any tendency to JT
distortions but the size difference between La and Cr drives a substantial GFO-like tilting
of the CrO$_6$ octahedra. Also in this case PBE performs as well as HSE-10 and HSE-15 (MARE = 0.6 \%).
The overall MARE is further reduced to 0.3 \% for larger values of $\alpha$ (HSE-25 and HSE-35).

\subsubsection{LaMnO$_3$}
LaMnO$_3$ is the most critical case across the La$M$O$_3$ series, due to the concomitant occurrence of
both GFO and JT structural distortions, the latter originating by the intrinsic instabilities associated
with the orbital degeneracy in the $e_g$ channel of the Mn$^{3+}$ cation.
The lattice constants and atom positions of O-P$_{nma}$ LaMnO$_3$ were fully optimized
at both PBE and HSE level within an AFM-A magnetic ordering, though, as we will discuss in the next section,
PBE is not able to catch the correct magnetic ground state and rather favors an FM
arrangement. The results are listed in Table \ref{tab:5}. In this case PBE does not supply a satisfactory
account of the structural properties, reflected by a quite large MARE ($\approx$ 2\%) significantly larger than
the corresponding HSE-10 (1.22), HSE-15 (0.81), HSE-25 (0.66), and HSE-35 (0.52). The major obstacle for PBE is the
correct prediction of the JT distortions: (i) the relative error for the M-O bondlength disproportionation
is as high as 5.5 \%, and  (ii)  Q$_2$ and Q$_3$ are found to be one third of the measured values.
The serious underestimation of Q$_2$ and Q$_3$ at PBE level has important consequences on the electronic
properties; we will discuss this issue in the next section. We can anticipate that the deficient treatment
of Q$_2$ and Q$_3$ prevents the opening of the band gap thereby leading to a metallic solution.
HSE-10 improves the estimations of Q$_2$ and Q$_3$ with respect to PBE, and with increasing $\alpha$
the MARE get progressively reduced down to 0.52 for $\alpha$=0.35.
The inaccuracy of local functional in reproducing the JT distortions was recently overviewed
by Hashimoto {\em {et al.}}\cite{hashimoto10}. In particular, these authors  have
pointed out that DFT+U can only supply a semiquantitative account of JT changes if the structure
is fully relaxed (including volume). This is also valid for purely unrestricted HF approaches.
All other non-JT related quantities are equally well described by both methodologies, with relative error
generally smaller than 1 \%, apart from the tilting angles which suffers of slightly larger deviations (1-1.5 \%).

\subsubsection{LaFeO$_3$}
The crystal of LaFeO$_3$ is orthorhombic with $P_{nma}$ space group\cite{dann}.
In the high-spin Fe$^{3+}$ configuration ${t_{2g}}^{\uparrow\uparrow\uparrow}{e_{g}}^{\uparrow\uparrow}$
the JT distortions are completely suppressed.
The optimized structural data, collected in Table \ref{tab:6}, show that PBE overestimates the volume by 1.5 \%
but describes well all other parameters, leading to a relatively small MARE of 0.79 \%.
HSE predicts a better volume, especially for $\alpha$ equal to 0.15 and 0.25, but in general HSE improves
only marginally the PBE results.

\subsubsection{LaCoO$_3$}
At low-temperature, LaCoO$_3$ possesses a slightly GFO-distorted perovskite structure with rhombohedral
symmetry ($R_{\bar3c}$ space group)\cite{thornton,radaelli}, characterized by a rhombohedral angle of 60.99$^\circ$ (see Fig. \ref{fig:models}).
The structural data are given in Table \ref{tab:7}.
The best agreement with experiment is achieved by HSE-15, but also HSE-10 and HSE-25 lead to relative
errors $\leq$ 1\%. The HSE-25 set of data is in good agreement with previous PBE0 results.\cite{Gryaznov10}
PBE performs not bad (MARE below 1\%), but overestimates too much the volume (+3.5\% with respect to experiment).

\subsubsection{LaNiO$_3$}
LaNiO$_3$ crystallizes with a rhombohedral structure with moderate GFO-like
distortions\cite{munoz92}.
The fully optimized structural parameters are listed in Table \ref{tab:8}.
Similarly to the isostructural LaCoO$_3$, also in this case
PBE gives a large volume (+2.4\%) but all other structural quantities are well reproduced
(our data are in line with the previous calculation of Guo {\em et al.}\cite{guo11}).
Within HSE, larger the amount of HF exchange is included, the more the MARE is reduced: from
0.84 \% (HSE-10) down to 0.1 \% (HSE-35).

\subsubsection{LaCuO$_3$}
LaCuO$_3$ is the only member of the La$M$O$_3$ family displaying a tetragonal structure (P$_{4/m}$),
which can be suitably tuned to a rhombohedral one under different oxygen pressure conditions.\cite{darracq}
In this paper we only examine the tetragonal phase.
The small elongation of the CuO$_6$ octahedron associated with the tetragonal form
induces a local JT-type distortion, manifested by  four equatorial
Cu-O bonds close to 1.909 $\AA$ and two apical bonds to 1.986 $\AA$.
The relaxed structure parameters are shown in Table \ref{tab:9}. From the structural data it is clear that LaCuO$_3$
represents the most challenging compound of the whole series for both level of theory, with MARE well above
1\%. PBE provides the overall best agreement with experiment (MARE = 1.59 \%) but produces
an almost cubic structure, dissimilar from the observed tetragonal one.
Hybrid functionals opens up a small structural disproportionation between long and short Cu-O bondlengths
which is however insufficient to stabilize a well defined tetragonal form: the lattice parameter $c$ is still
very badly accounted for (relative error of about 3\%).

\subsubsection{Concluding remarks}

Summing up the results presented in this section we can draw the following conclusions.
In general the structural properties of the La$M$O$_3$
series are sufficiently well described by standard PBE, which gives an overall MARE smaller than 1 \%, 
with the exceptions of:
(i) LaMnO$_3$: HSE is essential to treat correctly the JT distortions which are a crucial ingredient to
find and explain the A-AFM ordered insulating orbitally ordered state.
(ii) LaCuO$_3$: neither PBE nor HSE are capable to deliver MARE smaller than 1.5\%.
The amount of non-local HF exchange does not have a decisive and univocal effect on the structural properties
apart for the d$^0$ band insulator LaScO$_3$ for which the results get worse with increasing $\alpha$.
In all other cases, an improvement over PBE results is obtained for all values of $\alpha$ tested in the present study,
and the standard 0.25 compromise seems to appear a reasonable choice.  This was already noted for the case of actinide dioxides
for which the standard value of $\alpha$ yield to excellent volumes\cite{Prodan07}.
However, as we will discuss in the next section
with this value of $\alpha$ the band gaps are found to be exceedingly overestimate with respect to the
measured ones.

\subsection{Electronic Structures and Magnetic Properties}
\label{sec:elec}

The focus of the present section is the presentation of the electronic (density of states (DOS),
bandstructures and band gaps) and magnetic (magnetic moment $m$ and magnetic energies for different
spin orderings) results for the entire La$M$O$_3$ series, given for both the experimental and the fully
optimized structures. We note that from the magnetic energies it is possible to extract an estimation
of the magnetic coupling constants by means of a mapping onto an Heisenberg-like spin Hamiltonian
\cite{Moreira06,Cramer09,Bayer07,Franchini05,archer,Franchini12,Franchinitc,fischer}.

Here we are particularly interested on the modifications induced in the calculated
quantities by the value of the mixing parameter $\alpha$, from 0 (PBE) to 0.35 (HSE-35).
To this aim, following the outline adopted in the previous section we will sequentially discuss the
electronic and  magnetic structure case by case. A more general discussion on the evolution of the chemical and
physical properties of La$M$O$_3$ perovskites from $M$=Sc to $M$=Cu will be provided in the next section.

\subsubsection{$d^0$: LaScO$_3$}

\begin{figure}
\includegraphics[clip,width=0.5\textwidth]{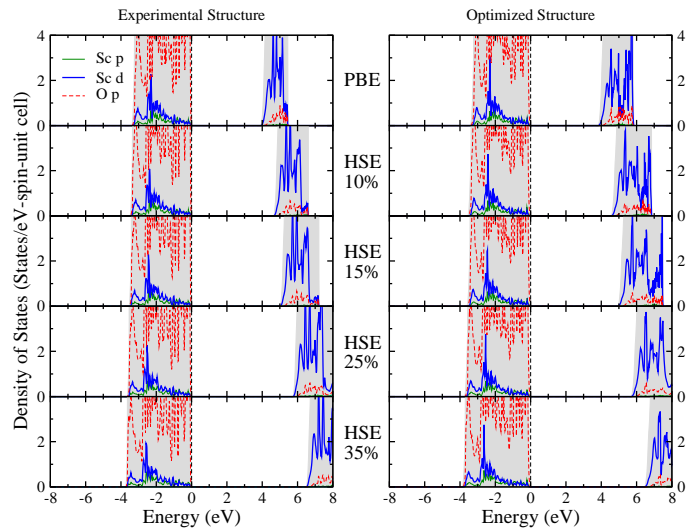}
\caption
{(Color online) $l$-projected DOS of non-magnetic LaScO$_3$ with experimental (left) and relaxed (right) structure
based on PBE and HSE (HSE-35, HSE-25, HSE-15, HSE-10) functionals. The shadow area indicates the total DOS.
}
\label{fig:1}
\end{figure}

\begin{figure}
\includegraphics[clip,width=0.45\textwidth]{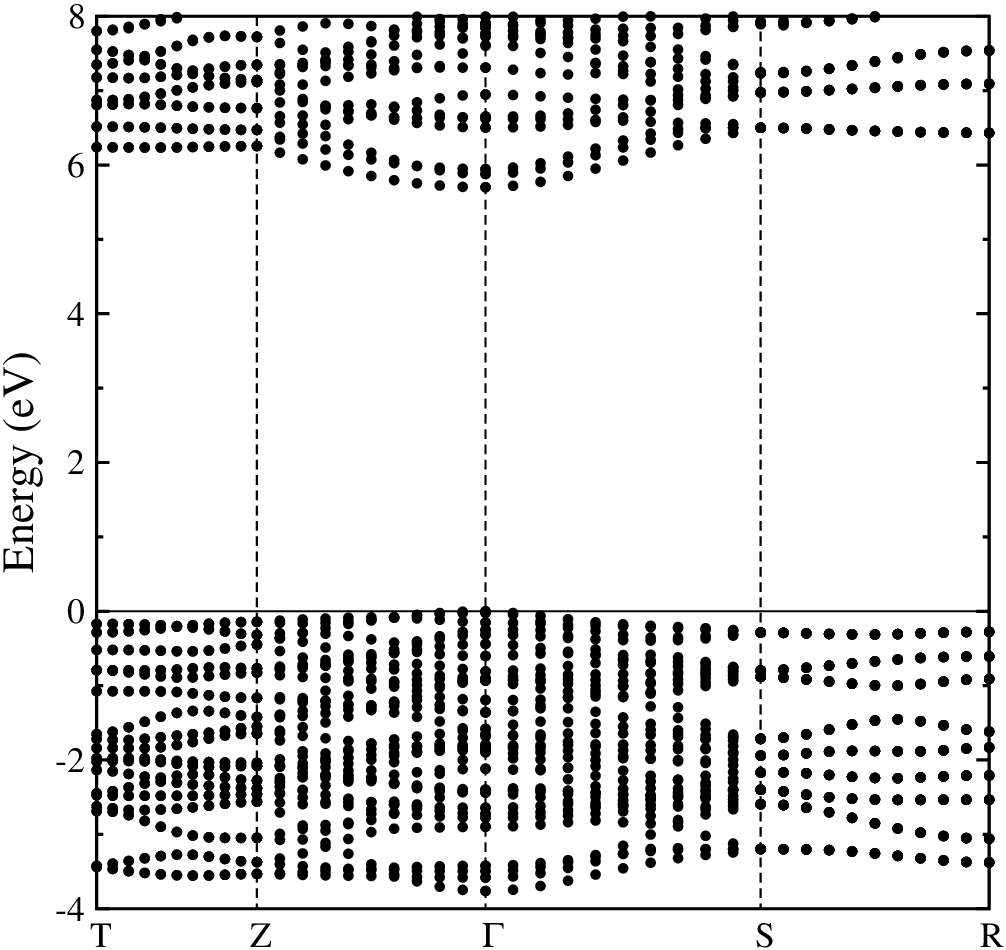}
\caption
{Bandstructure of LaScO$_3$ computed at HSE level ($\alpha$=0.25) using the optimized structure.
}
\label{fig:2}
\end{figure}

LaScO$_3$ is a non magnetic band insulator with the $d^0$ (Sc$^{3+}$) electronic configuration, and
an optically measured band gap of about 6.0 eV opened between the O-2$p$ valence band and the Sc
3$d$ unoccupied band.\cite{arima,afanas}. Our calculations confirm this picture as seen from the density of
states shown in Fig. \ref{fig:1}, but the band gap value predicted at PBE (3.81 eV and 3.92 eV for the experimental and
fully optimized structures, respectively, in agreement with previous calculations\cite{ravindran04}) seriously
underestimates the experimental value. The HSE data collected in Table \ref{tab:10} indicate that the correct value
of the gap is recovered by admixing 25\% of HF exchange. Clearly, the band gap increases with increasing
HF percentage, but the DOS (see Fig. \ref{fig:1}) always provide the same qualitative O-$p$/Sc-$d$ picture.
The band dispersion associated with the 25\% choice given in Fig. \ref{fig:2} shows that the band gap is
direct and located at $\Gamma$, but given the flatness of the topmost occupied bands
(O-$p$) and, to a lesser extent, the Sc-$d$ bands at about 6 eV, the value of the (direct) band gap does not change much
in the entire {\bf k}-space. This is in agreement with the experimental optical spectra which show
a sudden and very intense onset of the optical conductivity at 6 eV\cite{arima}.

\begin{table}
\caption{
The band gap $\Delta$ (eV) of LaScO$_3$ calculated within PBE and HSE (HSE-0.10, HSE-15,
HSE-25, and HSE-35) using both the Experimental and the relaxed structures (see Table \ref{tab:1}).
Other theoretical values are also listed for comparison, along with the experimental measurements.
} \vspace{0.3cm}
\begin{ruledtabular}
\begin{tabular}{ccccc}
 \multicolumn{5}{c}{Theory}                            \\
\hline
 \multicolumn{5}{c}{Optimized Structure}               \\
 HSE-35       & HSE-25   & HSE-15  & HSE-10   & PBE    \\
  6.495       &   5.730  &  4.995  &  4.635   & 3.915  \\
 \multicolumn{5}{c}{Experimental Structure}      \\
 HSE-35       & HSE-25   & HSE-15  & HSE-10   & PBE    \\
  6.435       &   5.685  &  4.920  &  4.560   & 3.810  \\
\multicolumn{5}{c}{Other works}                        \\
   LDA        &          &         &          &        \\
 3.98$^a$     &          &         &          &        \\
\hline\hline
 \multicolumn{5}{c}{Experiment}                        \\ \hline
 \multicolumn{5}{c}{$\sim$6.0$^b$, 5.7$^c$}
\end{tabular}
\end{ruledtabular}
\label{tab:10}
\begin{flushleft}
$^a$Ref.\cite{ravindran04}, $^b$Ref.\cite{arima}, $^c$Ref.\cite{afanas}.
\end{flushleft}
\end{table}

\begin{figure}
\includegraphics[clip,width=0.5\textwidth]{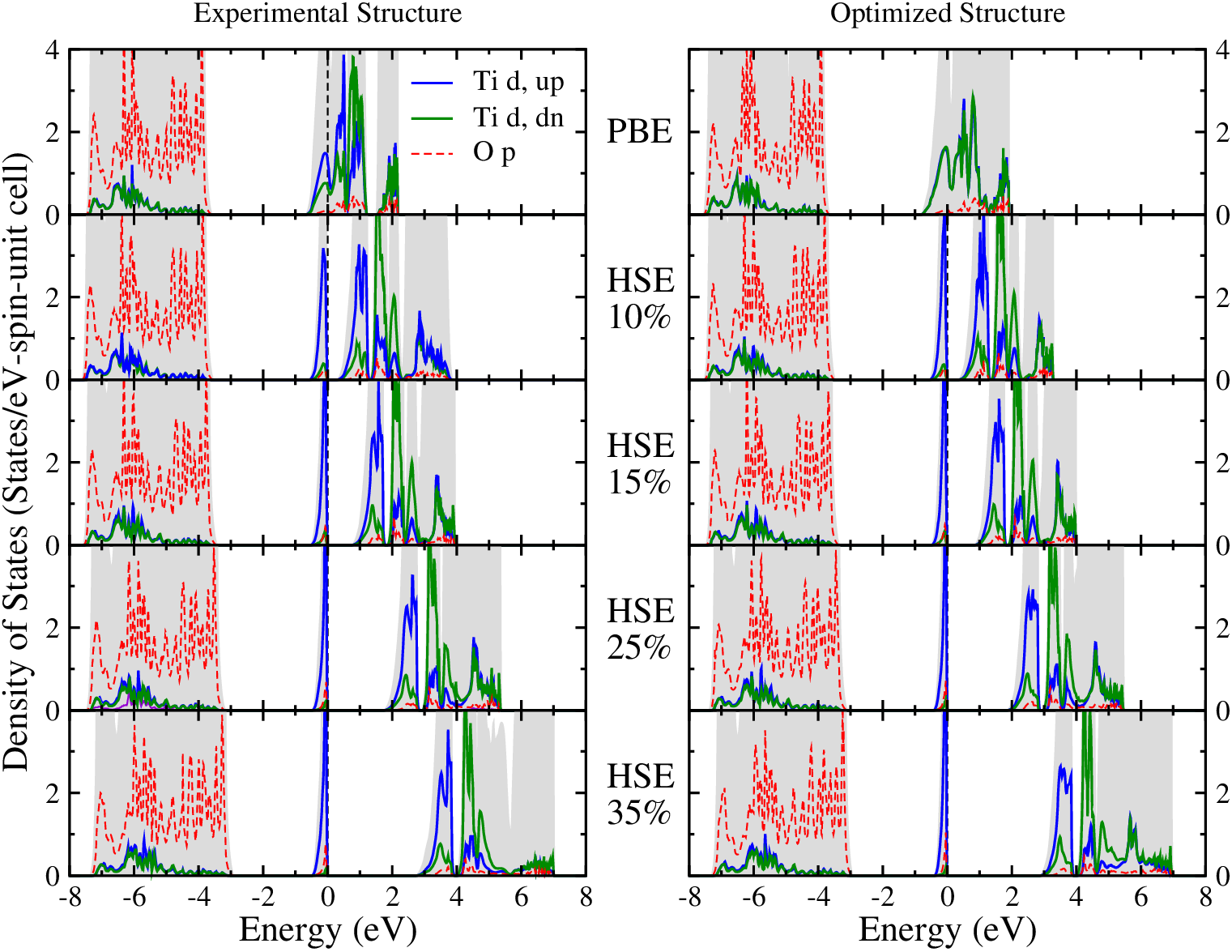}
\caption
{(Color online) $l$-projected DOS of AFM-G ordered LaTiO$_3$ with experimental (left) and relaxed (right) structure
based on PBE and HSE (HSE-35, HSE-25, HSE-15, HSE-10) functionals. The shadow area indicates the total DOS.
}
\label{fig:3}
\end{figure}

\begin{table}
\caption{
The band gap $\Delta$ (eV), magnetic moment $m$ ($\mu_{B}$/Ti), magnetic energy
(given with respect to the FM energy, in meV) of $\rm LaTiO_3$.
calculated by PBE and HSE (HSE-35, HSE-25, HSE-15, HSE-10) using both the experimental and
relaxed structures (Tab. \ref{tab:2}). The gaps in brackets are for the G-type which is not the
most favorable ordering for $\alpha$=0.10. Other theoretical values are also listed for comparison,
along with the experimental measurements.
} \vspace{0.3cm}
\begin{ruledtabular}
\begin{tabular}{cccccc}
          & \multicolumn{5}{c}{Theory}                            \\
          & \multicolumn{5}{c}{Optimized Structure}              \\
          & HSE-35   & HSE-25   & HSE-15  & HSE-10   & PBE       \\
$\Delta$  &  2.835   &  1.815   &  0.810  & 0.225           &  0.000    \\
          &          &          &         &       (0.315)   &           \\
$m$       &  0.908   &  0.868   &  0.790  &  0.707   &  0.497    \\
A-AFM     &  -26     &  -39     &  -57    &   -77    &   -23     \\
C-AFM     &   -3     &  -16     &    0    &    25    &   -13     \\
G-AFM     &  -33     &  -57     &  -62    &   -63    &   -17     \\
          & \multicolumn{5}{c}{Experimental Structure}     \\
          & HSE-35   & HSE-25   & HSE-15  & HSE-10   & PBE       \\
$\Delta$  &  2.700   &  1.710   &  0.720  &  0.165         & 0.0       \\
          &          &          &         &        (0.270) &           \\
$m$       &  0.905   &  0.868   &  0.789  &  0.702   & 0.621     \\
A-AFM     &  -29     &   -36    &   -53   &  -70     &  -49      \\
C-AFM     &  -35     &   -30    &    -7   &   32     &  -21      \\
G-AFM     &  -63     &   -68    &   -65   &  -52     &  -5       \\
          &\multicolumn{5}{c}{Other works}                       \\
          &   LDA    &   LDA+U                      &  GW      & HF              &  \\
Band gap  &          & 0.49$^a$, 0.5 $^b$           &1.00$^a$  & 2.7$^e$,0.6$^f$ &  \\
          &          & 0.4 $^c$, 0.16$^d$           &          &          &          \\
$m$       &          & 0.68$^a$, 0.92$^b$           & 0.68$^a$ & 0.55$^e$,0.76$^f$ & \\
          &          & 0.52$^g$, 0.7$^c$            &          &          &          \\
          &          &                    0.58$^d$  &          &          &          \\
\hline\hline
          & \multicolumn{5}{c}{Experiment}                        \\ \hline
$\Delta$  &  \multicolumn{5}{c}{0.1$^hc$, 0.2$^im$}                \\
$m$       &  \multicolumn{5}{c}{0.45$^la$, 0.57$^mb$}              \\
\end{tabular}
\end{ruledtabular}
\label{tab:11}
\begin{flushleft}
$^a$Ref.\cite{nohara09}, $^b$Ref.\cite{solovyev96}, $^c$Ref.\cite{okatov05}, $^d$Ref.\cite{ahn06},
$^e$Ref.\cite{Mizokawa96}, $^f$Ref.\cite{solovyev06}, $^g$Ref.\cite{zwanziger09},
$^h$Ref.\cite{arima}, $^i$Ref.\cite{okimoto}, $^l$Ref.\cite{goral}, $^m$Ref.\cite{cwik}
\end{flushleft}
\end{table}

\begin{figure}
\includegraphics[clip,width=0.45\textwidth]{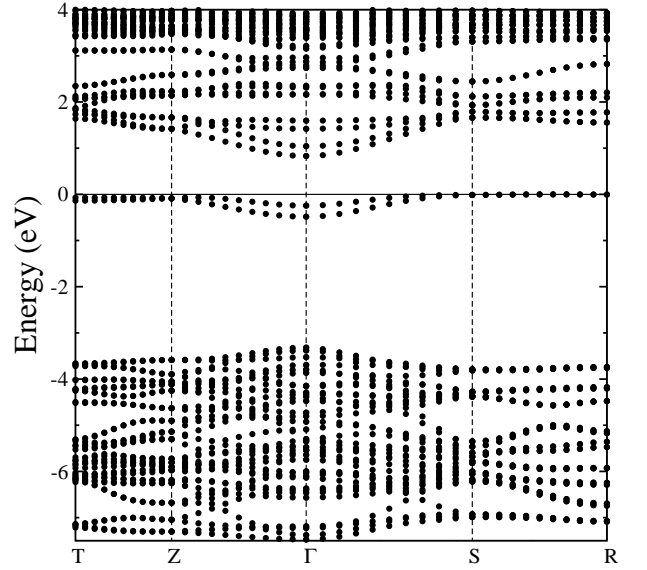}
\caption
{Bandstructure of LaTiO$_3$ computed at HSE level ($\alpha$=0.15) using the optimized structure.
}
\label{fig:4}
\end{figure}

\subsubsection{$d^1$: LaTiO$_3$}
\label{sec:elecTi}

LaTiO$_3$ is a G-AFM MH insulator with a magnetic moment of about 0.5 $\mu_{B}$\cite{cwik,hemberger},
in which the  single 3$d$ electron occupies one Ti $t_{2g}$ orbital. The physics of the orbital degree of
freedom has attracted considerable attention\cite{keimer,cwik}.
This nominal $t^{\uparrow}_{2g}$ configuration gives rise to a distinctive orbitally-ordered ground state
characterized by a very small band gap of 0.1-0.2 eV\cite{arima, okimoto} which has spurred a lot of
theoretical study aiming to clarify the physics underlying this peculiar behavior
\cite{Streltsov05, solovyev06, Fujitani95, Sawada98, Filippetti11, Pavarini04, solovyev04}.

In agreement with previous theoretical findings we find that local DFT, though it furnishes a very good
description of the structural properties, is incapable to reproduce the MH insulating state and
wrongly stabilize an AFM-A magnetic ordering. HSE delivers a coherent picture which is however
$\alpha$ dependent as summarized in Table \ref{tab:11} and Fig.\ref{fig:3}.
Regardless the value of the mixing parameter, HSE predicts an insulating ground state.
For $\alpha$=0.10 HSE conveys a band gap of about 0.1/0.2 eV (depending on whether the experimental or
the fully optimized structure is adopted), in excellent agreement with experiment. However, we found that
for $\alpha$ $\le$ 0.10 HSE finds the AFM-A as the most favorable magnetic ordering (like PBE), in contrast with
measurements. In order to stabilize the correct G-type AFM arrangement a larger value of $\alpha$
is required. But these larger portions of exact exchange lead to a band gap significantly larger than experiment.
The strong influence of the adjustable parameters in beyond-DFT methods such as U in DFT+U and $\alpha$ in HSE 
on the spin ordering which can lead to the stabilization of wrong or meta-stable magnetic arrangements  
is well known as recently discussed by Gryaznov {\em et al.} \cite{Gryaznov12}.
The 'optimum' choice is probably $\alpha$=0.15 for which HSE delivers an AFM-G insulating solution with a band gap of about
0.7-0.8 eV (depending on the structural details). For larger $\alpha$ the computed band gap is exceedingly
large: 1.8 and 2.8 for $\alpha$=0.25\cite{Iori12} and  $\alpha$=0.35, respectively.

The tendency of beyond-DFT
methods to overestimate the bandgap of LaTiO$_3$ was already reported in literature, based on
SIC (1.7 eV\cite{Filippetti11}) and other HSE\cite{Iori12} (1.7 eV using $\alpha$=0.25, in agreement with our
data) studies, and attributed to dynamical effects not included at this level of theory\cite{Pavarini04}.
Furthermore, HSE tends to overestimate the magnetic moment of  about 30 \%, again in analogy with previous beyond-DFT studies.

The MH like character of the band gap is evident by comparing the PBE and HSE DOS
given in Fig.\ref{fig:3}: the inclusion of non-local exchange split the $t_{2g}$ band near $E_F$,
thus opening a MH band gap between occupied and unoccupied $t_{2g}$ subbands. As expected the band gap
increases with increasing $\alpha$. The presence of an isolated peak on top of the valence band, well
separated from the states beneath has been also detected by X-ray photoemission spectroscopy (XPS) experiments\cite{Roth07}.
The CT gap, defined as the energy separation  between the O 2$p$ states and the upper $t_{2g}$
Hubbard band is also $\alpha$ dependent, and its value for the 'optimum' 0.15 choice, 4.7 eV, is in excellent
agreement with experiment, 4.5 eV\cite{arima}.

Finally, we underline that HSE is able to stabilize the correct orbitally-ordered state manifested
by a chessboard G-type arrangement of differently ordered $t_{2g}$ cigar-lobes.
We will come back to this point in the next section.

\begin{figure}
\includegraphics[clip,width=0.5\textwidth]{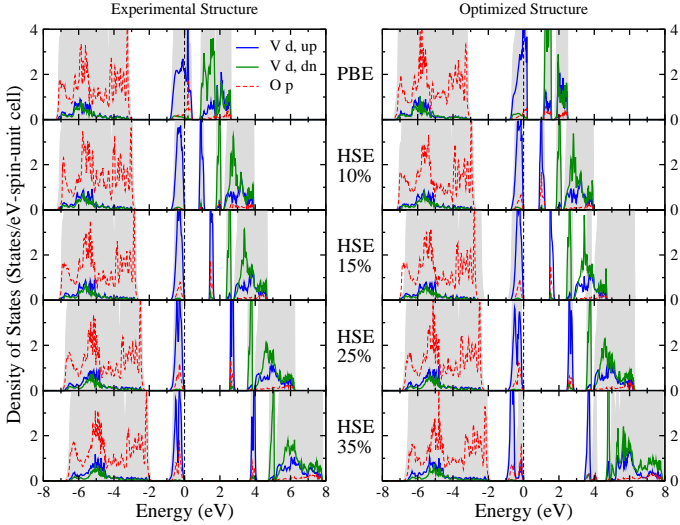}
\caption
{
(Color online) $l$-projected DOS of AFM-C ordered LaVO$_3$ with experimental (left) and relaxed (right) structure
based on PBE and HSE (HSE-35, HSE-25, HSE-15, HSE-10) functionals. The shadow area indicates the total DOS.
}
\label{fig:5}
\end{figure}

\begin{table}
\caption{
The band gap $\Delta$ (eV), magnetic moment $m$ ($\mu_{B}$/V), magnetic energy
(given with respect to the FM energy, in meV) of $\rm LaVO_3$.
calculated by PBE and HSE (HSE-35, HSE-25, HSE-15, HSE-10) using both the experimental and
relaxed structures (Table \ref{tab:3}).
Other theoretical values are also listed for comparison, along with the experimental measurements.
} \vspace{0.3cm}
\begin{ruledtabular}
\begin{tabular}{cccccc}
& \multicolumn{5}{c}{Theory}                            \\ \hline
          & \multicolumn{5}{c}{Optimized Structure}               \\
          & HSE-35   & HSE-25   & HSE-15  & HSE-10   & PBE        \\
$\Delta$  &   3.42   &   2.43   &  1.455  &  0.885   &  0.000     \\
$m$       &  1.876   &  1.855   &  1.819  &  1.782   &  1.625     \\
A-AFM     &    -73   &    -54   &    23   &    43    &   -77      \\
C-AFM     &   -124   &   -114   &  -144   &  -177    &  -216      \\
G-AFM     &    -96   &    -98   &   -30   &    33    &   137      \\
          & \multicolumn{5}{c}{Experimental  Structure}           \\
          & HSE-35   & HSE-25   & HSE-15  & HSE-10   & PBE        \\
$\Delta$  &  3.675   &  2.535   &  1.380  &   0.810  & 0.000      \\
$m$       &  1.882   &  1.858   &  1.813  &   1.774  & 1.629      \\
A-AFM     &    -2    &   33     &    16   &    11    &   -64      \\
C-AFM     &  -105    &  -119    &  -151   &  -179    &  -124      \\
G-AFM     &   -89    &   -80    &   -52   &   -11    &   203      \\
          &\multicolumn{5}{c}{Other works}                        \\
          & LDA                &  LDA+U                       &  GW       & HF       &            \\
$\Delta$  &  0.1$^a$           & 0.7 $^b$, 0.92$^c$           & 2.48$^c$  & 3.3$^e$, 0.9$^f$  &   \\
          &                    &                     1.2 $^d$ &           &                   &   \\
$m$       & 1.47$^a$, 1.85$^b$ & 1.98$^b$, 1.79$^c$           &  1.79$^c$ & 1.8$^e$,1.64$^f$  &   \\
          &                    &                     1.70$^d$ &           &                   &   \\
A-AFM     &     9$^a$          &   3.7$^d$                    &           &          &            \\
C-AFM     &   -35$^a$          &  -38.3$^d$                   &           &          &            \\
G-AFM     &    17$^a$          &  -14.8$^d$                   &           &          &            \\
\hline\hline
          & \multicolumn{5}{c}{Experiment}                        \\ \hline
$\Delta$  & \multicolumn{5}{c}{ 1.1$^g$ }                        \\
$m$       & \multicolumn{5}{c}{ 1.3$^h$ }                         \\
\end{tabular}
\end{ruledtabular}
\label{tab:12}
\begin{flushleft}
$^a$Ref. \cite{sawada96}, $^b$Ref.\cite{solovyev96}, $^c$Ref.\cite{nohara09},
$^d$Ref. \cite{fang04}, $^e$Ref.\cite{Mizokawa96}, $^f$Ref.\cite{solovyev06}
$^g$Ref.\cite{zubkov}, $^h$Ref.\cite{arima}.
\end{flushleft}
\end{table}

\begin{figure}
\includegraphics[clip,width=0.45\textwidth]{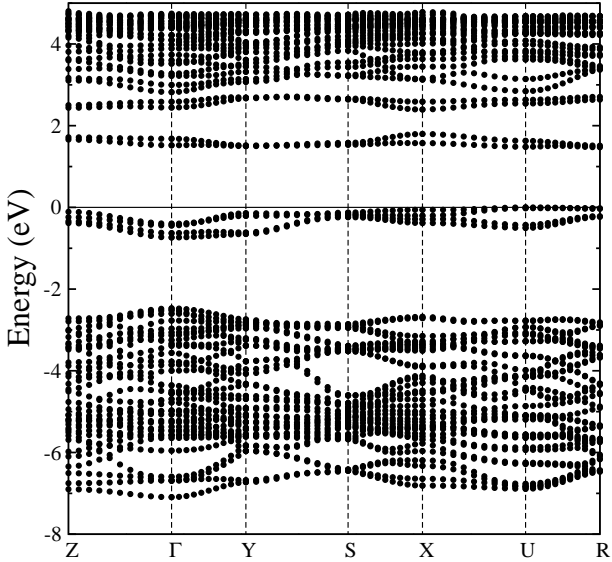}
\caption
{Bandstructure of LaVO$_3$ computed at HSE level ($\alpha$=0.15) using the optimized structure.
}
\label{fig:6}
\end{figure}

\subsubsection{$d^2$: LaVO$_3$}

LaVO$_3$ is another challenging material for conventional DFT: it is a $t^{\uparrow\uparrow}_{2g}$
AFM-C Mott insulator, but DFT finds an AFM-C metal. The C-type antiferromagnetic spin ordering
is stabilized by the JT induced bond length alternations in the $ab$ plane which cause the G-type
orderings of $d_{yz}$ and $d_{zx}$ orbitals\cite{sawada96}. The experimentally observed MH
and CT gaps are 1.1 and 4.0 eV, respectively\cite{arima}.

Regardless the fraction of non-local exchange, HSE correctly finds a AFM-C MH
insulating ground state, in which the gap is open between the lower and the upper MH $t_{2g}$
band, similarly to LaTiO$_3$ (in PBE the $t_{2g}$ band crosses the Fermi level, see Fig. \ref{fig:5}).
The best agreement with experiment is achieved for $\alpha$=0.10-0.15 for which HSE delivers satisfactory
values for both the MH ($\approx$ 0.8-1.4 eV for $\alpha$=0.10 and $\alpha$=0.15, respectively,
as summarized in Table \ref{tab:12})
and CT gaps ($\approx$ 4.4-4.9 eV for $\alpha$=0.10 and $\alpha$=0.15, respectively).
Similarly to all other theoretical DFT and beyond-DFT approaches, HSE
tends to overestimate the magnetic moment. It has been proposed that the origin of this discrepancy could
by an unquenched orbital magnetization or spin-orbit induced magnetic canting\cite{solovyev96}.

The bandstructure of LaVO$_3$ computed for the representative case $\alpha=0.15$ is displayed in
Fig.\ref{fig:6}. Also in this HSE is able to stabilize the correct G-type orbitally ordered state.
This will be discussed in more details in the next section.

\begin{figure}
\includegraphics[clip,width=0.5\textwidth]{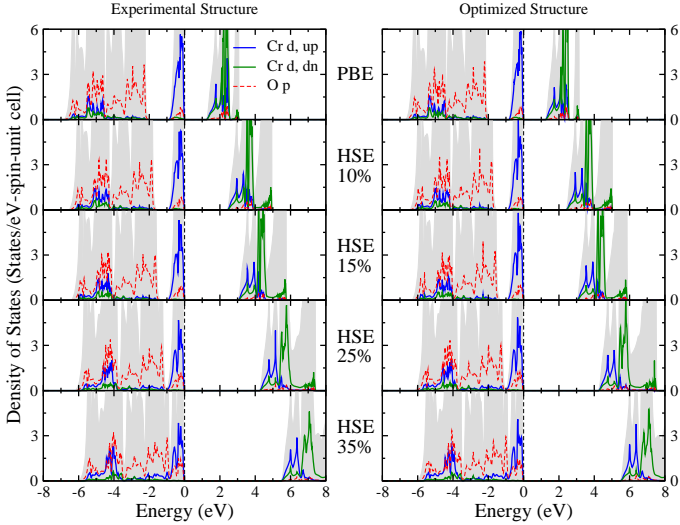}
\caption
{
(Color online) $l$-projected DOS of AFM-G ordered LaCrO$_3$ with experimental (left) and relaxed (right) structure
based on PBE and HSE (HSE-35, HSE-25, HSE-15, HSE-10) functionals. The shadow area indicates the total DOS.
}
\label{fig:7}
\end{figure}

\begin{table}
\caption{
The band gap $\Delta$ (eV), magnetic moment $m$ ($\mu_{B}$/Cr), magnetic energy
(given with respect to the FM energy, in meV) of $\rm LaCrO_3$.
calculated by PBE and HSE (HSE-35, HSE-25, HSE-15, HSE-10) using both the experimental and
relaxed structures (Table \ref{tab:4}).
Other theoretical values are also listed for comparison, along with the experimental measurements.
}
\vspace{0.3cm}
\begin{ruledtabular}
\begin{tabular}{cccccc}
          & \multicolumn{5}{c}{Theory}                            \\ \hline
          & \multicolumn{5}{c}{Optimized Structure}               \\
          & HSE-35   & HSE-25   & HSE-15  & HSE-10   & PBE   \\
$\Delta$  &  5.475   &  4.230   &  3.000  &  2.415   & 1.245 \\
$m$       &  2.866   &  2.836   &  2.790  &  2.756   & 2.643 \\
A-AFM     &   -79    &    -91   &  -108   &  -121    & -166  \\
C-AFM     &  -160    &   -184   &  -221   &  -245    & -309  \\
G-AFM     &  -226    &   -258   &  -305   &  -338    & -432  \\
          & \multicolumn{5}{c}{Experimental Structure}      \\
          & HSE-35   & HSE-25   & HSE-15  & HSE-10   & PBE   \\
$\Delta$  &  5.460   &  4.245   &  3.030  &  2.430   & 1.245 \\
$m$       &  2.868   &  2.835   &  2.784  &  2.748   & 2.626 \\
A-AFM     &   -76    &    -91   &   -113  &   -128   & -171  \\
C-AFM     &  -170    &   -203   &   -249  &   -281   & -375  \\
G-AFM     &  -233    &   -275   &   -335  &   -376   & -494  \\
          &\multicolumn{5}{c}{Other works}                        \\
          &   LDA        &  LDA+U             &  GW       &   HF     &       \\
$\Delta$  &1.40/3.4$^e$  &1.04$^f$, 1.40$^e$  &  3.28$^f$ &  4.5$^n$ &       \\
$m$       &2.56$^e$      & 2.58$^f$, 3.00$^m$ &  2.38$^f$ &  3.0$^n$ & \\
\hline\hline
          & \multicolumn{5}{c}{Experiment                  }      \\ \hline
$\Delta$  & \multicolumn{5}{c}{3.4$^c$}      \\
$m$       & \multicolumn{5}{c}{2.45$^a$, 2.8$^b$, 2.49$^d$, 2.63}      \\
\end{tabular}
\end{ruledtabular}
\label{tab:13}
\begin{flushleft}
$^a$Ref.\cite{koehler}, $^b$Ref.\cite{bertaut}, $^c$Ref.\cite{arima}, $^d$Ref. \cite{sakai}, $^e$Ref.\cite{Ong08}, $^f$Ref.\cite{nohara09}, $^m$Ref.\cite{yang99}, $^n$Ref.\cite{Mizokawa96}.
\end{flushleft}
\end{table}

\begin{figure}
\includegraphics[clip,width=0.45\textwidth]{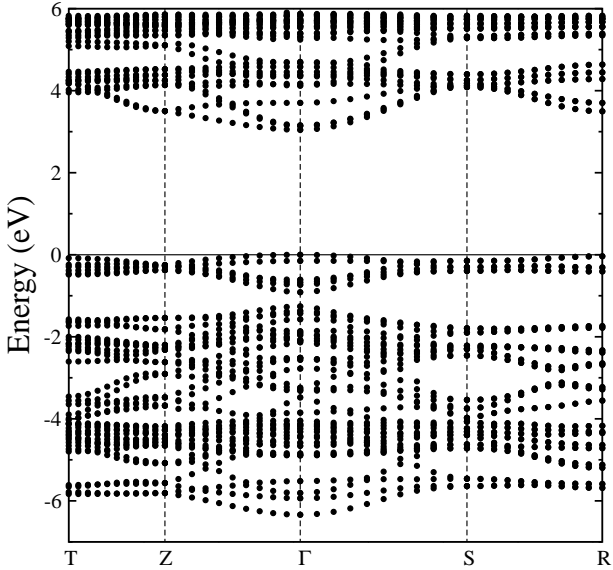}
\caption
{Bandstructure of LaCrO$_3$ computed at HSE level ($\alpha$=0.15) using the optimized structure.
}
\label{fig:8}
\end{figure}

\subsubsection{$d^3$: LaCrO$_3$}

Under equilibrium conditions $P_{nma}$-distorted LaCrO$_3$ exhibits a G-type AFM insulating
ground state with the Cr$^{+3}$ cation in the $d^3$ electron configuration $t^{\uparrow\uparrow\uparrow}_{2g}$.
The optical experiments by Arima {\em et al.} reported a coexistence of CT
and MH like excitations in LaCrO$_3$ at 3.4 eV\cite{arima}. These findings have been explained
by several theoretical HF\cite{Mizokawa96}, LDA+U\cite{yang99}, GW\cite{nohara09} studies in terms of a
significant mixing between Cr $t_{2g}$ and O $p$ states at the top of the valence band. In particular,
the LDA+U study of Yang and coworkers has shown that the CT/MH character of the band
gap is strongly U dependent: for small values of U (U$<$5 eV) the top of the valence band is mainly
formed by $t_{2g}$ states and the gap is predominantly MH, but for larger U (U$>$5 eV) the O
$p$ bands is progressively shifted towards higher energy thus reducing the size of the charge-transfer
gap which become indistinguishable from the MH one.
Our HSE calculations confirm this picture as shown in the DOS plotted in Fig.\ref{fig:7}:
the O$_p$-Cr$_d$ mixing at the top of the valence band increases with increasing $\alpha$.
As expected, $\alpha$ also influences the predicted band gap size which is found to be much smaller
than experiment at purely PBE level (1.2 eV) and reaches the value 3.0 eV for $\alpha$=0.15, in good agreement
with the reported optical gap. For larger $\alpha$ the gap starts to deviate substantially from
the measure value, and become exceedingly large for $\alpha$=0.35 (see Table \ref{tab:13}).
The bandstructure corresponding to the 'optimum' $\alpha$=0.15 choice is displayed in Fig. \ref{fig:8}.
The G-type spin ordering is very robust at any value of $\alpha$ and the magnetic moment changes by
only 0.2 $\mu_{B}$ going from $\alpha$=0 ($\approx$ 2.6 $\mu_{B}$) to $\alpha$=0.35 ($\approx$ 2.8 $\mu_{B}$).
Also in this case the electronic and magnetic properties obtained from the optimized structure are essentially
identical to those corresponding to the experimental structure.

A different interpretation of the bandstructure and optical properties of LaCrO$_3$ was proposed in 2008
by Ong and coworkers who suggested that LaCrO$_3$ should not be considered a strongly correlated material.\cite{Ong08}
These authors have attributed the 3.4 eV CT  gap as the excitation
from the top of the wide O $p$ band below the $t_{2g}$ states to the bottom of the Cr $d$ unoccupied band,
and called for a new optical experiment to confirm the presence of a smaller
MH gap of 2.38 eV open between Cr $t_{2g}$ and Cr $e_g$ bands, which would justify the
green-light color of LaCrO$_3$. We are not aware of more recent experimental data in support of
this interpretation.

\begin{figure}
\includegraphics[clip,width=0.45\textwidth]{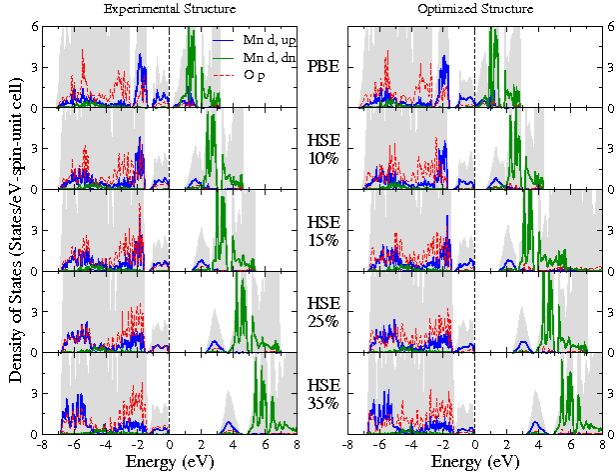}
\caption
{
(Color online) $l$-projected DOS of AFM-G ordered LaMnO$_3$ with experimental (left) and relaxed (right) structure
based on PBE and HSE (HSE-35, HSE-25, HSE-15, HSE-10) functionals. The shadow area indicates the total DOS.
}
\label{fig:9}
\end{figure}

\begin{table}
\caption{
The band gap $\Delta$ (eV), magnetic moment $m$ ($\mu_{B}$/Mn), magnetic energy
(given with respect to the FM energy, in meV) of $\rm LaMnO_3$.
calculated by PBE and HSE (HSE-35, HSE-25, HSE-15, HSE-10) using both the experimental and
relaxed structures (Table \ref{tab:5}).
Other theoretical values are also listed for comparison, along with the experimental measurements.
} \vspace{0.3cm}
\begin{ruledtabular}
\begin{tabular}{cccccc}
          & \multicolumn{5}{c}{Theory}                 \\ \hline
          & \multicolumn{5}{c}{Optimized Structure}               \\
          & HSE-35       & HSE-25   & HSE-15  & HSE-10   & PBE    \\
$\Delta$  &  3.41  &  2.47  &  1.63  &  0.75  &  0.00  \\
\emph{m}  &  3.78  &  3.74  &  3.67  &  3.65  &  3.52  \\
A-AFM     &  -7    &  -8    &   -24  &     3  &   171  \\
C-AFM     &  156   &  182   &   198  &   368  &   564  \\
G-AFM     &  161   &  192   &   208  &   428  &   899  \\
          & \multicolumn{5}{c}{Experimental Structure} \\
          & HSE-35 & HSE-25 & HSE-15 & HSE-10 &  PBE   \\
$\Delta$  &  3.30  &  2.40  &  1.52  &  1.10  &  0.23  \\
\emph{m}  &  3.78  &  3.73  &  3.67  &  3.62  &  3.50  \\
A-AFM     &   -4   &  -11   &  -28   &  -44   &  -63   \\
C-AFM     &  164   &  182   &  198   &  202   &  209   \\
G-AFM     &  175   &  195   &  212   &  216   &  228   \\
          &\multicolumn{5}{c}{Other works}             \\
          & GGA    & GGA+U  & B3LYP  &   HF   &   GW   \\
$\Delta$  &0.70$^a$&1.18$^a$&2.30$^b$& 3.0$^r$&1.6$^c$, 1.68$^d$\\
\emph{m}  &3.33$^a$           &3.46$^a$&3.80$^b$          &  3.9$^r$           &3.16$^c$          \\
          &          3.39$^e$ &        &          3.77$^f$&           3.96$^f$ &          3.51$^d$\\
\hline\hline
          & \multicolumn{5}{c}{Experiment}             \\
\hline
$\Delta$  & \multicolumn{5}{c}{1.1$^g$, 1.7$^{h}$, 1.9$^i$, 2.0$^{l,m}$}      \\
\emph{m}  & \multicolumn{5}{c}{3.87$^n$, 3.7$^o$, 3.42$^p$}\\
\end{tabular}
\end{ruledtabular}
\label{tab:14}
\begin{flushleft}
$^a$Ref. \cite{hashimoto10}, $^b$Ref. \cite{munoz04}, $^c$Ref. \cite{nohara09}, $^d$Ref. \cite{Franchini12},
$^e$Ref. \cite{ravindran02}, $^f$Ref. \cite{evarestov05},
$^g$ Ref.\cite{arima}, $^h$ Ref.\cite{saitoh},
$^i$ Ref.\cite{Jung97}, $^l$ Ref.\cite{Jung98}, $^m$ Ref.\cite{Kruger04},
$^n$ Ref.\cite{moussa}, $^o$ Ref.\cite{exp1}, $^p$ Ref.\cite{hauback}, $^r$Ref.\cite{Mizokawa96}.
\end{flushleft}
\end{table}

\begin{figure}
\includegraphics[clip,width=0.45\textwidth]{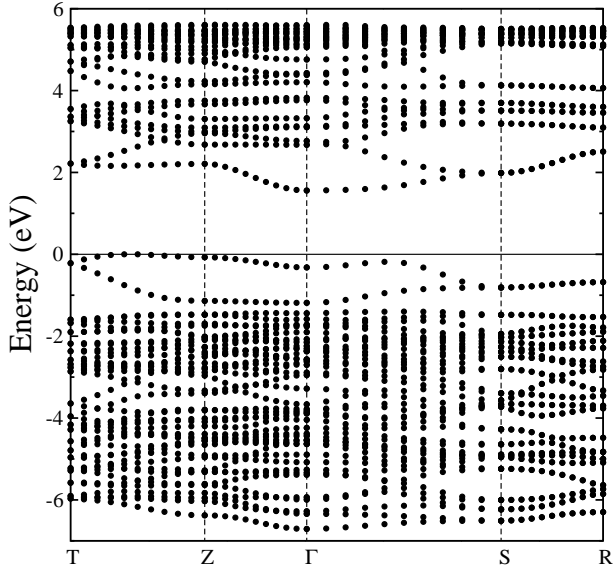}
\caption
{Bandstructure of LaMnO$_3$ computed at HSE level ($\alpha$=0.15) using the optimized structure.
}
\label{fig:11}
\end{figure}

\subsubsection{$d^4$: LaMnO$_3$}

LaMnO$_3$ is one of the most studied perovskite. Its properties have been widely studied both
experimentally and theoretically as mentioned in the introduction. The initial
tentative assignment of Arima and coworkers on the CT electronic nature of LaMnO$_3$
was successively disproved and nowadays it is widely accepted that LaMnO$_3$ represents the prototypical
example of a JT-distorted MH orbitally-ordered antiferromagnetic (type-A)
insulator\cite{Yin06, Pavarini10, Franchini12}.
In discussing the structural properties we have underlined that LaMnO$_3$ is a very critical
case for conventional band theory due to the small but crucial JT distortions which are only
marginally captured by PBE. The drawbacks of standard DFT are also reflected in the electronic and
magnetic properties summarized in Fig.\ref{fig:9} and Table \ref{tab:14}, especially for the
theoretically relaxed structure. Using the optimized geometry PBE favors the
wrong magnetic ordering (FM) and stabilizes a metallic solution, whereas by adopting the
experimental structure the correct AFM-A insulating ground state is stabilized, but the value of the
band gap, 0.23 eV, is significantly smaller than the experimental one, 1.1-2.0 eV (this is in agreement
with previous studies\cite{Kotomin05,hashimoto10}).
This indicate that the JT distortions alone are sufficient to open up a band gap in LaMnO$_3$,
but in order to predict a more accurate value it is necessary to go beyond DFT.
In fact, turning to HSE the situation improves significantly and the results achieved within the
theoretically optimized geometrical setup are essentially identical to those obtained for the
experimental structure. The only significant difference regards the relative stability of the AFM-A
ordering with respect to the FM one. For $\alpha$=0.10 the FM ordering is still more favored over
the AFM-A one using the optimized geometry, but by adopting the experimental
the AFM-A arrangement become the most stable one. For larger values of $\alpha$ both structural setups lead to essentially the same relative
stability among all considered spin arrangements.
As expected, the band gap increases linearly with increasing mixing parameter and the best agreement
with the measured values is reached again for $\alpha$=0.15 ($\approx$ 1.6 eV, well within the experimental
range of variation). The band gap is open between occupied and unoccupied Mn $e_g$ states which are
almost completely separated from the other bands, as clarified in the bandstructure plot provided
in Fig. \ref{fig:11}. The associated orbitally ordered state will be presented in the next section.
The HSE prediction for the Mn magnetic moment is in good agreement with low
temperature measurements, 3.7-3.87 $\mu_{B}$.\cite{exp1,exp2}, and previous B3LYP data
($\sim$ 3.8 $\mu_{B}$).\cite{hashimoto10,evarestov05}. We observe a small increase of the
magnetic moment with increasing mixing parameter, a general tendency noticed for the other La$M$O$_3$
compounds. A more extensive discussion of the ground state properties of LaMnO$_3$ can be found in
our previous works\cite{Franchini12, He12}.

\begin{figure}
\includegraphics[clip,width=0.5\textwidth]{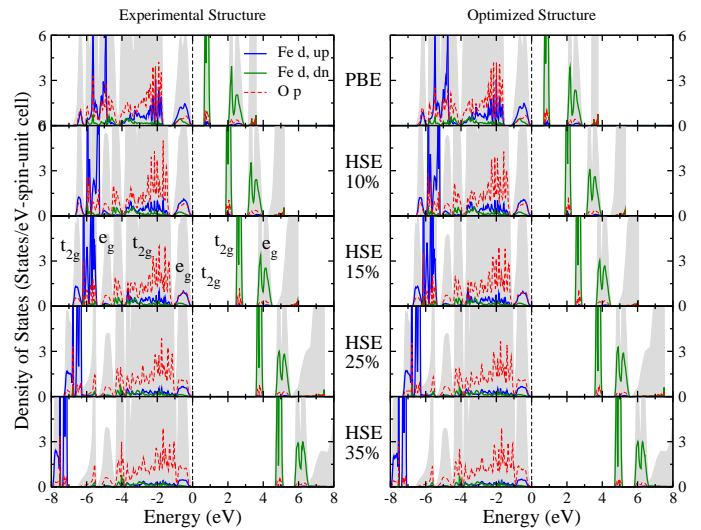}
\caption
{
(Color online) $l$-projected DOS of AFM-G ordered LaFeO$_3$ with experimental (left) and relaxed (right) structure
based on PBE and HSE (HSE-35, HSE-25, HSE-15, HSE-10) functionals. The shadow area indicates the total DOS.
}
\label{fig:12}
\end{figure}

\begin{table}
\caption{
The band gap $\Delta$ (eV), magnetic moment $m$ ($\mu_{B}$/Fe), magnetic energy
(given with respect to the FM energy, in meV) of $\rm LaFeO_3$.
calculated by PBE and HSE (HSE-35, HSE-25, HSE-15, HSE-10) using both the experimental and
relaxed structures (Table \ref{tab:6}).
Other theoretical values are also listed for comparison, along with the experimental measurements.
} \vspace{0.3cm}
\begin{ruledtabular}
\begin{tabular}{cccccc}
          & \multicolumn{5}{c}{Theory}                            \\ \hline
          & \multicolumn{5}{c}{Optimized Structure}               \\
          & HSE-35   & HSE-25   & HSE-15  & HSE-10   & PBE   \\
$\Delta$  &  4.680   &  3.570   &  2.460  &  1.875   & 0.660 \\
$m$       &  4.198   &  4.110   &  4.001  &  3.933   & 3.719 \\
A-AFM     &  -259    &   -323   &   -417  &   -487   &  -75  \\
C-AFM     &  -530    &   -653   &   -832  &   -947   & -278  \\
G-AFM     &  -760    &   -930   &  -1166  &  -1316   & -696  \\
          & \multicolumn{5}{c}{Experimental Structure}      \\
          & HSE-35   & HSE-25   & HSE-15  & HSE-10   & PBE   \\
$\Delta$  &  4.665   &  3.570   &  2.445  &  1.875   & 0.615 \\
$m$       &  4.202   &  4.111   &  3.998  &  3.927   & 3.708 \\
A-AFM     &  -251    &  -321    &   -427  &   -511   &    -9 \\
C-AFM     &  -518    &  -655    &   -854  &   -993   &  -134 \\
G-AFM     &  -742    &  -930    &  -1194  &  -1372   &  -552 \\
          &\multicolumn{5}{c}{Other works}                        \\
          &  LDA     &   LDA+U  &  GW     &  HF                  &       \\
Band gap  & 0.0 $^a$ & 0.10$^b$, 2.1$^a$ &  1.78$^b$ & 4.0$^g$   &       \\
m         & 3.5 $^a$ & 3.54$^b$, 4.1$^a$ &  3.37$^b$ & 4.6$^g$   &       \\
\hline\hline
          & \multicolumn{5}{c}{Experiment}                        \\ \hline
$\Delta$  & \multicolumn{5}{c}{2.1$^c$, 2.4$^d$}             \\
$m$       & \multicolumn{5}{c}{3.9$^e$, 4.6$^f$}             \\
\end{tabular}
\end{ruledtabular}
\label{tab:15}
\begin{flushleft}
$^a$Ref.\cite{yang99}, $^b$Ref.\cite{nohara09}, $^c$Ref.\cite{arima}, $^d$Ref.\cite{bellakki}, $^e$Ref.\cite{zhou},
$^f$Ref.\cite{koehler}, $^g$Ref.\cite{Mizokawa96}.
\end{flushleft}
\end{table}

\begin{figure}
\includegraphics[clip,width=0.45\textwidth]{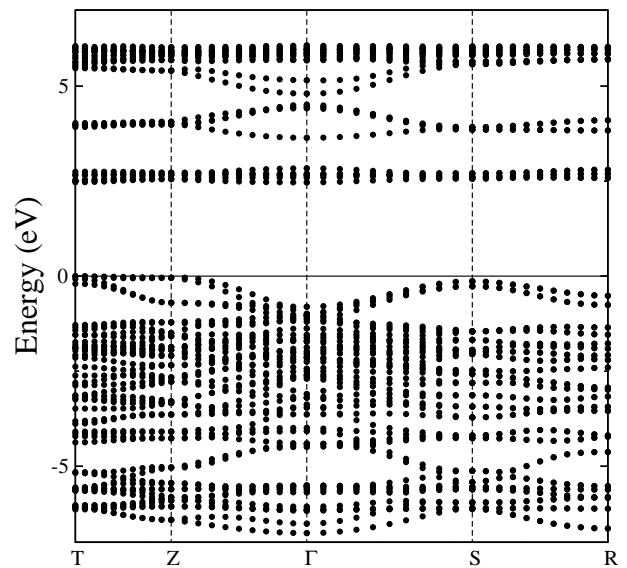}
\caption
{Bandstructure of LaFeO$_3$ computed at HSE level ($\alpha$=0.15) using the optimized structure.
}
\label{fig:13}
\end{figure}

\subsubsection{$d^5$: LaFeO$_3$}
The electronic configuration of Fe$^{3+}$ ion in LaFeO$_3$ is the high spin state
($t^{\uparrow\uparrow\uparrow}_{2g})(e^{\uparrow\uparrow}_g$). Below the rather high magnetic
ordering temperature T$_N=750$ K,\cite{white} LaFeO$_3$ displays a G-type AFM spin ordering, and
the $d^5$ spin saturation prevents the formation of orbital ordering.
Arima\cite{arima} reported that the spectrum of LaFeO$_3$ is similar to that of LaMnO$_3$, except for
an increase of the insulating gap which is found to be 2.1-2.4 eV, about 0.5 eV larger than the
LaMnO$_3$ energy gap. The bandgap is opened between the predominantly O-$p$ and Fe-$e_g$ valence band
maxima and the lowest unoccupied $t_{2g}$ band as shown in the density of states of Fig. \ref{fig:12}. As such LaFeO$_3$
should be considered an intermediate CT/MH insulator, as originally
suggested by Arima, who found almost identical CT and MH gaps\cite{arima}.
PBE does an appreciable job in predicting the correct AFM-G insulating ground state, though the value of
the band gap, $\approx$ 0.6 eV, is significantly underestimated with respect to experiment
(see the collection of electronic and magnetic data in Table \ref{tab:15}). Similarly, the PBE estimates
of the magnetic moment, 3.7 $\mu_B$, is below the observed value. However, it should be noted that
the available low temperature experimental measures of the magnetic moments are very different,
3.9 $\mu_B$\cite{zhou} and 4.6 $\mu_B$\cite{koehler}, thus a firm comparison is presently out of reach.

The best agreement with the experimental gap is obtained also in this case for $\alpha$=0.15
for which HSE gives a gap of about 2.4 eV, for both the optimized and experimental structure
(this is not surprising considering that in LaFeO$_3$ the optimized structure differences by less
than 1\% from the experimental one, as discussed previously). For this value of the mixing parameter
we achieve an excellent comparison with photoemission data of Wadati {\em et al.}\cite{Wadati05}, in
terms of the position and character of the main peaks at -0.5 eV (Fe-$e_g$, O-$p$), -2 eV (Fe-$t_{2g}$-O-$p$)
and -6 eV (Fe-$e_g$, O-$p$). These findings agree with the GW spectra computed by Nohara\cite{nohara09}.
By increasing the fraction of HF exchange the position of the lowest occupied $t_{2g}$ and $e_g$ states are
gradually pushed down in energy and become progressively more localized whereas the position and bandwidth of
the O-$p$ band remains essentially unaffected. This leads to a worsening of the comparison with the experiment
for $\alpha$ $\ge$ 0.25. The $\alpha=0.15$ bandstructure is shown in Fig.\ref{fig:13}.
Finally, we note that the energy separation between the unoccupied $t_{2g}$ and $e_g$ states
(the two lowest conduction bands, respectively, as indicated in Fig.\ref{fig:12}), about 1.3 eV, is almost
independent from $\alpha$ and in good agreement with x-ray absorption spectroscopy\cite{Wadati05} and the
GW\cite{nohara09} results.

\begin{figure}
\includegraphics[clip,width=0.5\textwidth]{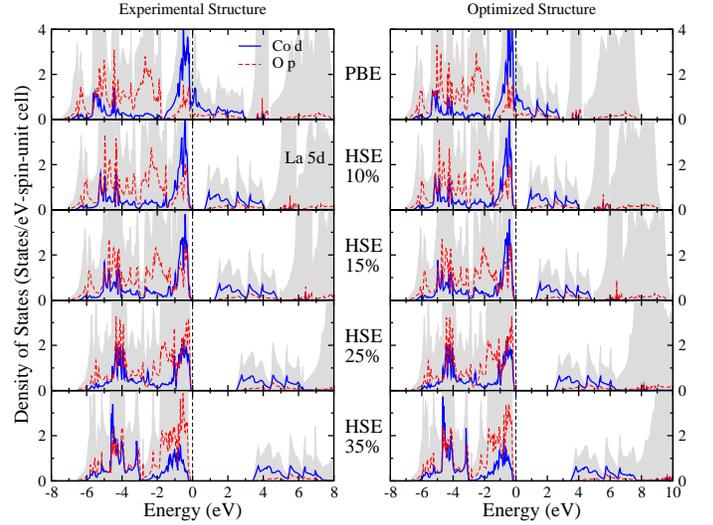}
\caption
{
(Color online) $l$-projected DOS of non-magnetic LaCoO$_3$ with experimental (left) and relaxed (right) structure
based on PBE and HSE (HSE-35, HSE-25, HSE-15, HSE-10) functionals. The shadow area indicates the total DOS.
}
\label{fig:14}
\end{figure}

\begin{table}
\caption{
The band gap $\Delta$ (eV) of non-magnetic $\rm LaCoO_3$.
calculated by PBE and HSE (HSE-35, HSE-25, HSE-15, HSE-10) using both the experimental and
relaxed structures (Table \ref{tab:7}).
Other theoretical values are also listed for comparison, along with the experimental measurements.
} \vspace{0.3cm}
\begin{ruledtabular}
\begin{tabular}{ccccccc}
          & \multicolumn{6}{c}{Theory}                        \\ \hline
          & \multicolumn{6}{c}{Optimized Structure}           \\
          & HSE-35   & HSE-25   & HSE-15   & HSE-10  &HSE-05& PBE  \\
$\Delta$  & 3.480    &  2.415   & 1.215    &  0.660  &0.165 & 0.0  \\
          & \multicolumn{6}{c}{Experimental Structure}             \\
          & HSE-35   & HSE-25   & HSE-15   & HSE-10  &HSE-05&PBE   \\
$\Delta$  &  3.390   &  2.445   &  1.200   &  0.615  &0.105 & 0.0  \\
          &\multicolumn{6}{c}{Other works}                         \\
          &   LDA             &  LDA+U                      & PBE0  &   GW     &  HF   &\\
$\Delta$  & 1.06$^a$, 0.0$^b$ & 1.0$^c$, 2.06$^d$           & 2.50$^b$, 3.14$^b$ & 1.28$^f$ & 3.5$^l$& \\
          &                   &                    1.8$^e$  &                    &          &          \\
\hline\hline
          & \multicolumn{6}{c}{Experiment}                    \\ \hline
$\Delta$  & \multicolumn{6}{c}{0.3$^g$, 0.1$^e$ }                       \\
\end{tabular}
\end{ruledtabular}
\label{tab:16}
\begin{flushleft}
$^a$Ref.\cite{sahnoun05}, $^b$Ref.\cite{Gryaznov10}, $^c$Ref.\cite{knizek09}, $^d$Ref.\cite{korotin96}, $^e$Ref.\cite{yamaguchi96}, $^f$Ref.\cite{nohara09}, $^g$Ref.\cite{arima}, $^h$Ref.\cite{chainani}, $^i$Ref.\cite{laref10},$^l$Ref.\cite{Mizokawa96}
\end{flushleft}
\end{table}

\begin{figure}
\includegraphics[clip,width=0.45\textwidth]{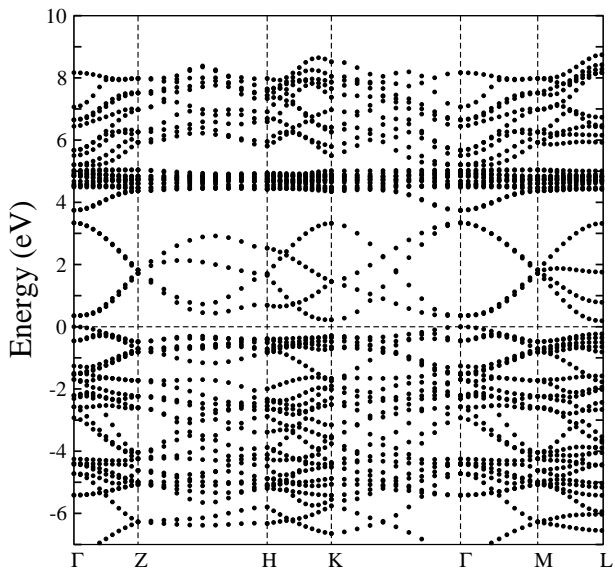}
\caption
{Bandstructure of LaCoO$_3$ computed at HSE level ($\alpha$=0.05) using the optimized structure.
}
\label{fig:15}
\end{figure}

\subsubsection{$d^6$: LaCoO$_3$}

The complex magnetic behavior of LaCoO$_3$ represents a great challenge for theory.
At low temperature LaCoO$_3$ is a diamagnetic insulator in which the Co$^{3+}$ are aligned in the low spin (LS)
state $(t^{\uparrow\downarrow\uparrow\downarrow\uparrow\downarrow}_{2g})(e^0_g)$, with a total spin S=0.
At about 100~K it undergoes a transition towards a paramagnetic state associated with magnetic excitations
involving high-spin ($(t^{4}_{2g})(e^{2}_g)$, S=2) and intermediate spin ($(t^{5}_{2g})(e^{1}_g)$, S=1)
configurations, and at high temperature (T $\approx$ 500~K) it shows a second magnetic anomaly associated with
an insulator-to-metal transition\cite{Imada98, Goodenough01, Rao04}.  These issues have been widely discuss in
literature but a general consensus is still missing and their detailed understanding remain highly
controversial\cite{Hsu10, laref10}.

Standard LDA (or GGA) predicts a metallic and magnetic ground opposite to
experiment\cite{sahnoun05,yang99,Gryaznov10}. Conversely, DFT+U can reproduce the correct non-magnetic insulating
state, but the results depends critically on the choice of U and the results seems to be strongly dependent on the
specific computational schemes adopted. Small values of U ($<$ 2 eV) lead to the erroneous
DFT-like solution. It has been shown that the correct LS insulating solution can be obtained using rather
different U, ranging from U$\approx$3 eV\cite{knizek06, knizek09} to U$\approx$8\cite{korotin96,laref10,Hsu09}.
Hsu and coworkers\cite{Hsu09} have recently performed an optimization of the value of U based on an accurate
account of the structural properties, and show that the best agreement with experiment is achieved for
a rather large U$\approx$8.2 eV\cite{Hsu09}. A similar value of U has been also found by Laref {\em et al.}
throughout the inverse response matrices\cite{laref10}.
Finally, using the unscreened hybrid functional PBE0 scheme with the standard choice of the mixing parameter (0.25)
Gryaznov {\em et al.} were able to find the correct LS state with a bandgap of 2.5 eV \cite{Gryaznov10}.
Our HSE results for $\alpha=0.25$ delivers a LS gap of 2.4 eV, in excellent agreement with these PBE0 results.

In Table \ref{tab:16} we collect the values of the bandgap for the more stable S=0
HSE solution along with available experimental and other theoretical estimations.
The best comparison with experiment is achieved for a rather small $\alpha=0.05$ for which HSE delivers a bandgap
of about 0.1 eV, in good agreement with optical measurements, 0.1-0.3 eV\cite{arima,yamaguchi96} (photoemission data
of Chainani {\em et al.}\cite{chainani} give a somehow larger gap of about 0.6 eV).
We remind that this value of $\alpha$ leads to the most accurate optimized geometry, as discussed in the previous
section (see Fig. \ref{fig:mare}). From the density of states shown in Fig.\ref{fig:14} we  evince that the gap is opened between
valence band mixed O $p$ and Co $d$ states and empty $d$-like Co states, in agreement with the DFT+U
and PBE0 results mentioned above.  The effect of the inclusion of a fraction of HF exchange is the splitting of
the occupied $t_{2g}$ manifold and the $e_g$ states (this forms a continuous band which crosses the Fermi energy
at PBE level). The valence band DOS is characterized by three main regions located at -1 eV, -3 eV and -5 eV,
reproducing well the XPS\cite{Abbate93} and GW\cite{nohara09} spectra.
Equally satisfactory is the distribution of the conduction
band states, with the Co and La $d$ states centered at $\approx$ 2 eV and 7-9 eV, respectively.

The bandstructure plotted in Fig.\ref{fig:15} provides further evidence for the large
degree of hybridization of the top of the valence band
and the rather dispersive character of the lowest
$e_g$ unoccupied states. On the basis of this
analysis LaCoO$_3$ can thus be considered to be predominantly a CT like (O $p$ $\rightarrow$ Co $d$)
insulator (in agreement with the initial assignment by Arima\cite{arima}) but a Mott mechanism is necessary to
split apart the Co $d$ band crossing $\rm E_F$ at PBE level, possibly indicating minor
$t_{2g} \rightarrow e_g$ MH like excitations, which have not been specifically
investigated so far, by both theory and experiment.

\begin{figure}
\includegraphics[clip,width=0.5\textwidth]{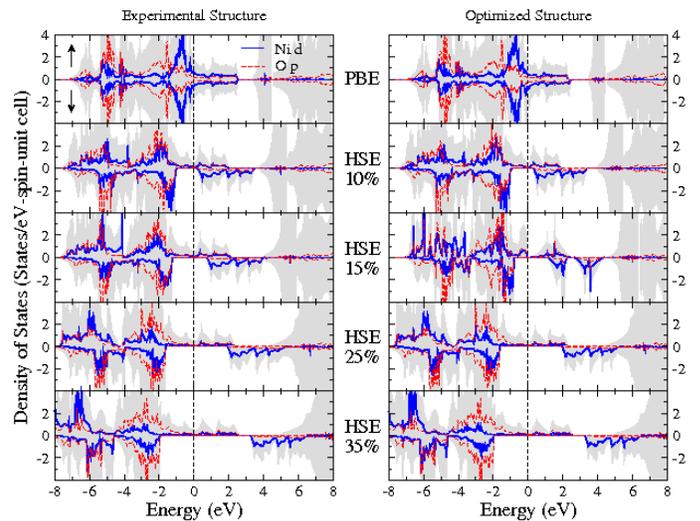}
\caption
{
(Color online) $l$-projected DOS of FM LaNiO$_3$ with experimental (left) and relaxed (right) structure
based on PBE and HSE (HSE-35, HSE-25, HSE-15, HSE-10) functionals. The shadow area indicates the total DOS.
}
\label{fig:16}
\end{figure}

\begin{table}
\caption{
The band gap $\Delta$ (eV) and magnetic moment $m$ ($\mu_{B}$/Ni) of FM ordered $\rm LaNiO_3$,
calculated by PBE and HSE (HSE-35, HSE-25, HSE-15, HSE-10) using both the experimental and
relaxed structures (Table \ref{tab:8}).
Other theoretical values are also listed for comparison, along with the experimental measurements.
} \vspace{0.3cm}
\begin{ruledtabular}
\begin{tabular}{cccccc}
          & \multicolumn{5}{c}{Theory}                            \\
          & \multicolumn{5}{c}{Optimized Structure}               \\
          & HSE-35   & HSE-25   & HSE-15  & HSE-10   & PBE   \\
$\Delta$  &   HM     &  HM      & HM      &  HM      & 0.00  \\
$m$       & 1.303    & 1.187    & 1.034   &  0.960   & 0.169 \\
          & \multicolumn{5}{c}{Experimental Structure}      \\
          & HSE-35   & HSE-25   & HSE-15  & HSE-10   & PBE   \\
$\Delta$  &   HM     &   HM     &  HM     &  HM      & 0.0   \\
$m$       &   1.308  &  1.186   & 1.039   &  0.956   & 0.002 \\
          &\multicolumn{5}{c}{Other works}                        \\
          &   LDA    &  LDA+U   & PBE0/HSE&   GW     &  HF   \\
$\Delta$  & 0.0$^{a,b}$  & 0.0$^a$, HM$^b$  &  HM$^b$ &  0.0$^d$ & 0.3$^g$ \\
$m$       & 0.0$^{a,b}$  & 1.1$^a$, 1.0$^b$  &        &          &       \\
\hline\hline
          & \multicolumn{5}{c}{Experiment}                  \\
$\Delta$  & \multicolumn{5}{c}{0.0$^e$}                     \\
$m$       & \multicolumn{5}{c}{0.0$^f$ (PM)}                     \\
\end{tabular}
\end{ruledtabular}
\label{tab:17}
\begin{flushleft}
$^a$Ref.\cite{solovyev96}, $^b$Ref.\cite{guo11}, $^d$Ref.\cite{nohara09}, $^e$Ref.\cite{arima}, $^f$Ref.\cite{sreedhar},
$^g$Ref.\cite{Mizokawa96}.
\end{flushleft}
\end{table}

\begin{figure}
\includegraphics[clip,width=0.45\textwidth]{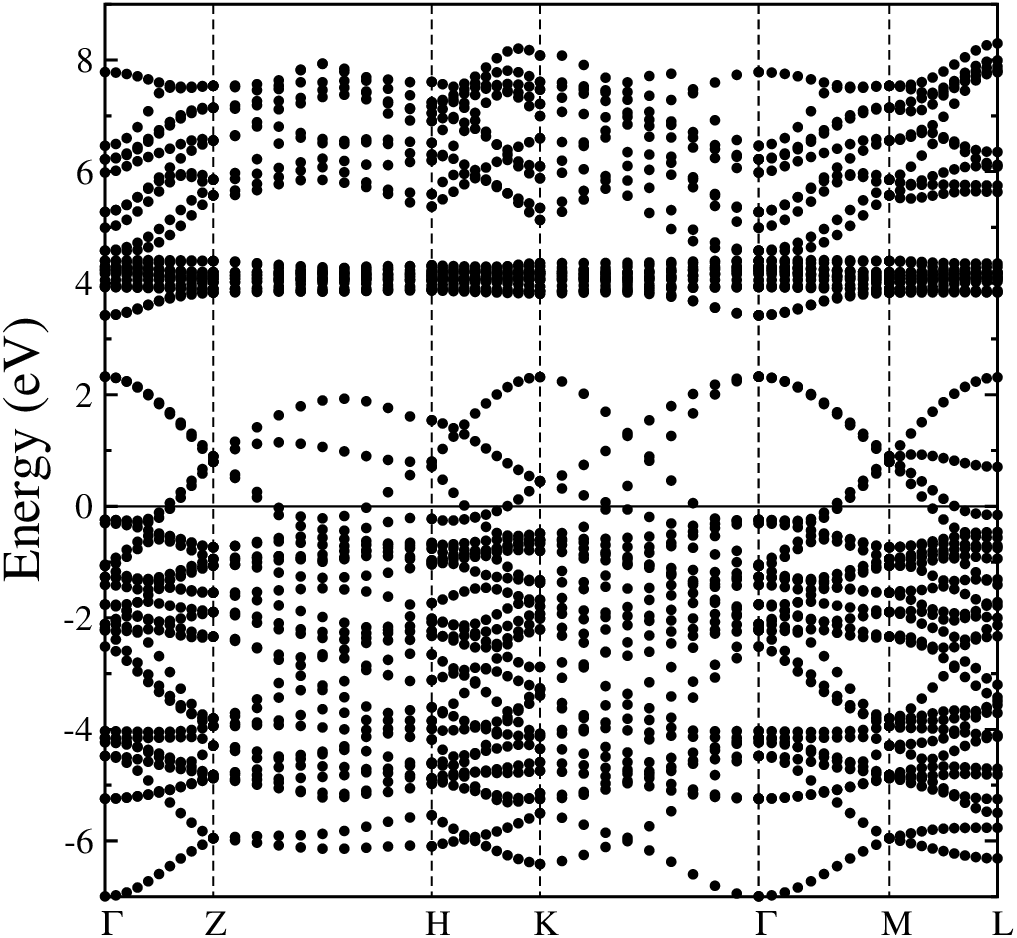}
\caption
{Bandstructure of LaNiO$_3$ computed at PBE level using the optimized structure.
}
\label{fig:17}
\end{figure}

\subsubsection{$d^7$: LaNiO$_3$}

LaNiO$_3$ is a weakly correlated PM metal in which the Ni$^{+3}$ ion possesses the low-spin 3d$^7$ configuration
$(t^{\uparrow\downarrow\uparrow\downarrow\uparrow\downarrow}_{2g})(e^\uparrow_g)$. The electron-electron
correlation associated with the partially filled Ni 3$d^7$ shell is inhibited by an efficient
electrostatic screening, originated by the strong Ni 3$d$-O 2$p$ hybridization (relatively small Ni-O distance),
and $d$-$d$ hybridization (large valence $d$-bandwidth)\cite{arima, sarma95, guo11}.
The electronic structure of LaNiO$_3$ has been recently extensively investigated and thoughtfully discussed
by the group of J. M. Rondinelli\cite{guo11} using an array of several above-standard first principles methods including
LDA+U, PBE0 and HSE.

Our HSE results (summarized in Fig.\ref{fig:16} and Table \ref{tab:17}) reproduce the
trends observed by Rondinelli and coworkers and support their conclusions:

(i) Conventional DFT works fairly well as it provides a correct non-magnetic metallic solution.
This was already pointed out in precedent works\cite{Sarma94b,Anisimov99}.
We should however note that for this specific case structural effects are extremely important in the
determination of the relative stability between the non-magnetic ground state and the competing FM
solution. Using the experimental structure PBE favors the non-magnetic
case by about 130 meV/f.u., but adopting the PBE-optimized structure the FM ordering become the most stable
solution by about 110 meV/f.u.. This should be attributable to the PBE overestimation of the volume (+2.3 \%),
as all other structural properties are described by PBE with an error smaller than 1\% (see Table \ref{tab:8}).
The comparison with PES data gives further support to the quality of the DFT
performance, as discussed in Ref.\onlinecite{guo11}. Minor differences have been observed between LSDA and PBE,
relative to the width of the valence band which is better described at LSDA level.
The bandstructure computed at PBE level for the most stable non-magnetic solution
given in Fig.\ref{fig:17} show evident similarities with the LaCoO$_3$ bands. The major difference is the downward shift
of the $e_g$ manifold at the bottom of the conduction band which now crosses the Fermi level and get mixed
with the lower laying occupied $t_{2g}$ orbitals. The strong $d$-$p$ and $d$-$d$ hybridization is reflected in the
highly dispersive character of the valence bands, in accordance with the DOS.

(ii) HSE, similarly to PBE0 and DFT+U\cite{guo11}, delivers a very lacking picture:
LaNiO$_3$ is described as a FM half-metal with a magnetic moment $m$ of about 1 $\mu_B$. $m$ increases gradually as a
function of $\alpha$ and reaches the value 1.3 $\mu_B$ for $\alpha$=0.35. The deficient HSE results is mostly due to
an excessive downward shift of the $t_{2g}$ manifold (this increases the overall bandwidth with respect to PBE),
a strong depletion of Ni $d$ states on top of the valence band, and a much too large exchange splitting.
Clearly, the relative strength of these effects increases with increasing $\alpha$, as clarified in Fig.\ref{fig:16}.
This makes the comparison with the PES data much worse, in terms of both bandwidth, and number and positions
of the main peaks.\cite{guo11}. The fundamental failure of hybrid functional for itinerant magnets was already
reported by Paier {\em et al.} for bulk Fe, Co and Ni\cite{paier06}.

Though conventional DFT leads to a decent account of the ground state of LaNiO$_3$ it should be emphasized that
LaNiO$_3$ is experimentally recognized as being a correlated metals, with important dynamical correlation effects
associated with the Ni $e_g$ orbitals\cite{Stewart11} which cannot be captured at DFT level.
More suitable methodologies such as dynamical mean-filed theory are need to appreciate the fundamental
nature of LaNiO$_3$, as recently demonstrated\cite{Hansmann10, Stewart11, Deng12}.

\begin{figure}
\includegraphics[clip,width=0.5\textwidth]{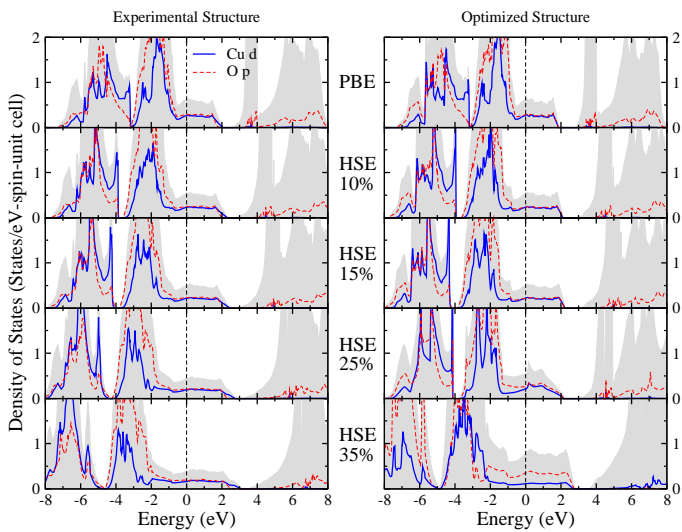}
\caption
{
(Color online) $l$-projected DOS of non-magnetic LaCuO$_3$ with experimental (left) and relaxed (right) structure
based on PBE and HSE (HSE-25, HSE-15, HSE-10) functionals.  The shadow area indicates the total DOS.
}
\label{fig:18}
\end{figure}

\begin{table}
\caption{
The band gap $\Delta$ (eV) of non-magnetic $\rm LaCuO_3$,
calculated by PBE and HSE (HSE-35, HSE-25, HSE-15, HSE-10) using both the experimental and
relaxed structures (Table \ref{tab:9}). HSE-35 favors an FM-ordered ground state with $m$=1.197$\mu_{B}$,
similarly to the LDA+U calculation of Ref. \onlinecite{czyzyk94}.
Other theoretical values are also listed for comparison, along with the experimental measurements.
} \vspace{0.3cm}
\begin{ruledtabular}
\begin{tabular}{cccccc}
          & \multicolumn{5}{c}{Theory}                            \\ \hline
          & \multicolumn{5}{c}{Optimized Structure}               \\
          & HSE-35   & HSE-25   & HSE-15   & HSE-10   & PBE         \\
$\Delta$  &  0.00    &  0.000   &  0.000   &  0.00    &  0.00       \\
$m$       &  0.0     &  0.000   &  0.000   &  0.00    &  0.00       \\
          & \multicolumn{5}{c}{Experimental Structure}      \\
          & HSE-35   & HSE-25   & HSE-15   & HSE-10   & PBE         \\
$\Delta$  &  0.000   &  0.000   &  0.000   &  0.000   &  0.000      \\
$m$       &  0.000   &  0.000   &  0.000   &  0.00    &  0.00       \\
          &\multicolumn{5}{c}{Other works}                        \\
          &   LDA    &  LDA+U             &  GW      &  HF      &       \\
$\Delta$  & 0.0$^a$  &  0.0$^b$, 0.95$^a$ & 0.0$^c$  & 2.2$^f$  &       \\
$m$       & 0.0$^a$  & 0.01$^b$, 0.98$^a$ &          &          &       \\
\hline
\hline
          & \multicolumn{5}{c}{Experiment}                        \\ \hline
$\Delta$  &  \multicolumn{5}{c}{0.0$^d$}                   \\
$m$       &  \multicolumn{5}{c}{0.0$^e$ (PM)}                      \\
\end{tabular}
\end{ruledtabular}
\label{tab:18}
\begin{flushleft}
$^a$Ref.\cite{czyzyk94}, $^b$Ref.\cite{solovyev96}, $^c$Ref.\cite{nohara09}, $^d$Ref.\cite{arima}, $^e$Ref.\cite{bringley},
$^f$Ref.\cite{Mizokawa96}.
\end{flushleft}
\end{table}

\begin{figure}
\includegraphics[clip,width=0.45\textwidth]{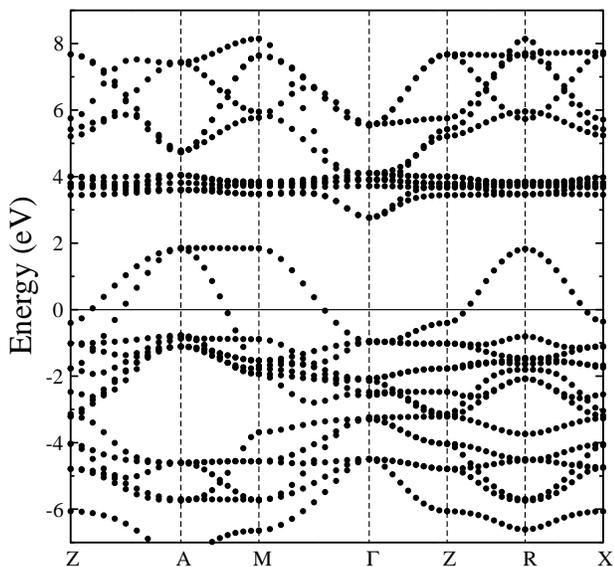}
\caption
{Bandstructure of LaCuO$_3$ computed at PBE level using the optimized structure.
}
\label{fig:19}
\end{figure}

\subsubsection{$d^8$: LaCuO$_3$}

LaCuO$_3$ is a PM metal\cite{arima}. Cu$^{3+}$ ions are formally in the low spin configuration
$(t^{\uparrow\downarrow\uparrow\downarrow\uparrow\downarrow}_{2g})(e^{\uparrow\downarrow}_g)$
(the $t_{2g}$ shell is fully occupied and the $e_g$ orbitals are half-filled), but it has been argued that
this $d^8$ state is strongly hybridized with the $d^9{\underline{L}}$ configuration in which ${\underline{L}}$
denotes a ligand hole, thus suggesting the existence of orbital degeneracy associated with significant
Cu-Cu many-body excitations\cite{Mizokawa98,Okada98,Yalovega00}. This is another challenging case both for
theory (orbital degeneracy and dynamical correlation) and experiment (it is very difficult to to prepare a
stoichiometric tetragonal phase of LaCuO$_3$ without oxygen vacancies).
Thus, the final methodological comments given for LaNiO$_3$ on the necessity to employ many-body schemes
to achieve a fundamentally more accurate theoretical description remain valid for LaCuO$_3$ as well.

Our PBE and HSE results are collected in Table \ref{tab:18} and Figs. \ref{fig:18} and \ref{fig:19}.
In agreement with the results of M.T. Czy\.zyk and G.A. Sawatzky\cite{czyzyk94} we find that standard DFT
finds the correct metallic non-magnetic ground state. The DOS (Fig. \ref{fig:18}) is characterized by a
wide band crossing the Fermi level formed by Cu $d$ (primarily $e_g$) and O $p$ states, associated with an
highly dispersive bands (Fig. \ref{fig:19}).
In analogy with  HF\cite{Mizokawa96,Mizokawa98} and LDA+U\cite{czyzyk94} calculations also within HSE
the G-type AFM insulating state is lower in energy than the non magnetic metallic state, in contradiction with experiment.
Here we only report the results for the non-magnetic solution. From the DOS shown in Fig.\ref{fig:18} we infer that the
electronic structure stays almost unchanged with respect to the non-magnetic PBE case. The only notable difference
is a progressive downward shift of the $t_{2g}$ Cu states with increasing $\alpha$ and a gradual further
broadening of the Cu $d$/O $p$ band crossing the $\rm E_F$.

\vspace{5mm}

\begin{figure*}
\includegraphics[clip,width=1.00\textwidth]{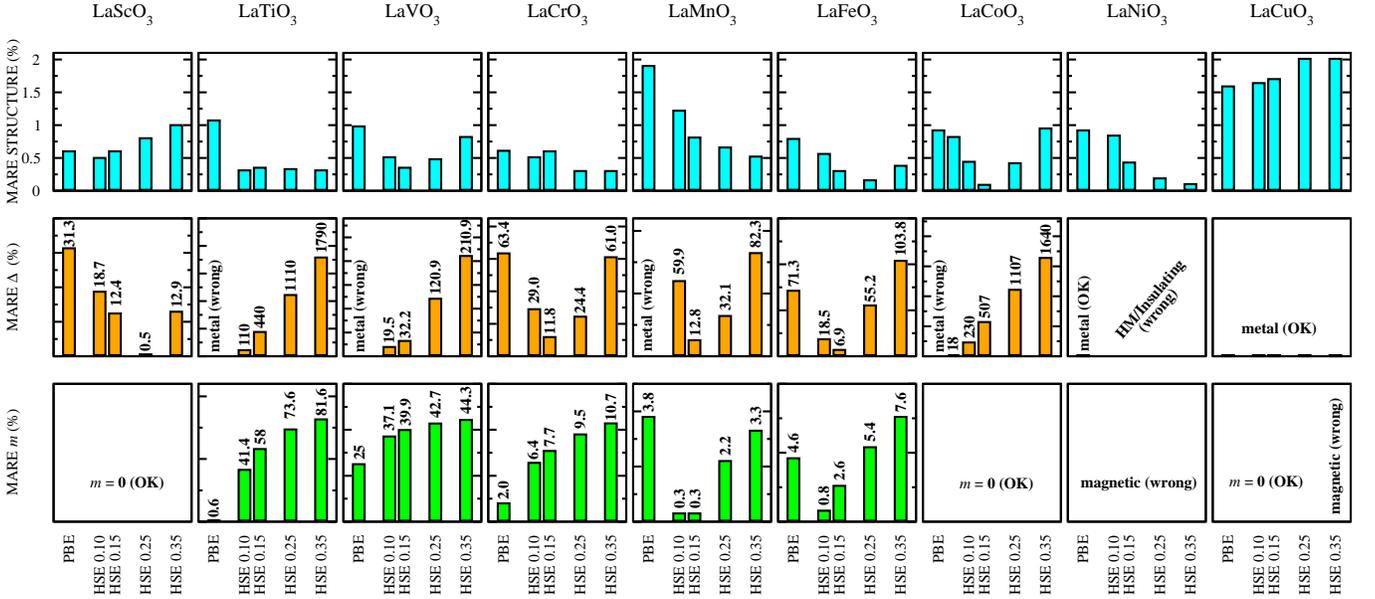}
\caption
{(Color online) Summary of the MARE for the structural properties (top panel), band gap $\Delta$ (middle panel)
and magnetic moment $m$ (lower panel), at PBE and HSE level. For the band gap $\Delta$ and the magnetic moment $m$
the MARE are indicated by the numbers associated to each bar.
A few specifications for the labels 'OK' and 'wrong':
(i)  LaScO$_3$, $m$: all methods correctly predicts a non-magnetic ground state;
(ii) LaCoO$_3$, $m$: all methods correctly predicts a zero magnetic moment;
(iii) LaNiO$_3$, $\Delta$: PBE is the only approach which correctly find a metallic solution;
(iv) LaNiO$_3$, $m$: all methods wrongly predict a magnetic ground state;
(v) LaCuO$_3$, $\Delta$: all methods correctly predicts a metallic solution;
(vi) LaCuO$_3$, $m$: PBE and HSE (0.05, 0.15, and 0.25) correctly predicts a zero magnetic moment, whereas HSE 0.35
wrongly stabilized a magnetic ground state.
}
\label{fig:mare}
\end{figure*}

We conclude this section by providing in Fig. \ref{fig:mare} a schematic graphical interpetation of the
comparison between computed and measured structural, electronic and magnetic properties, given in terms
of the obtained MARE. A more elaborated discussion will be developed in the next section.

\section{DISCUSSION}
\label{sec:discussion}

From the analysis of the structural, electronic and magnetic properties developed in the previous section
we have derived a set of 'optimum' values for the mixing parameter ($\alpha_{opt}^{\rm HSE}$) for which HSE (and in two
cases PBE, i.e. $\alpha$=0)
delivers a substantially correct and quantitatively satisfying description of the La$M$O$_3$ series (within the limits
discussed previously). This set of $\alpha_{opt}^{\rm HSE}$ parameters, collected in Table \ref{tab:19}, includes
$\alpha$=0.25 (for the wide band gap insulator LaScO$_3$), $\alpha$=0.1-0.15 (for the MH and intermediate MH/CT
insulators LaTiO$_3$-LaFeO$_3$), $\alpha$=0.05 (for the small band gap CT insulator LaCoO$_3$), and
$\alpha$=0 (for metallic LaNiO$_3$ and LaCuO$_3$). As already reported in Sec.\ref{sec:elecTi},
it is important to underline that for LaTiO$_3$ the overall best quantitative agreement with experiment is
achieved for $\alpha$=0.1 (the computed band gap is $\approx$ 0.2 eV, almost identical
to the measured value), but the incorporation of this fraction of exact exchange in HSE leads to the stabilization of the wrong
magnetic ordering, AFM-A instead of AFM-G. 

\begin{table*}[t]
\caption{Comparison between the set of optimum mixing factors $\alpha$ for the 3$d$ perovskite series La$M$O$_3$
($M$=Sc-Cu) computed throughout the HSE fitting procedure developed in Sec. \ref{sec:results} and those obtained
using the relations $\alpha_{opt}^{\epsilon_\infty}=1/\epsilon_\infty$ (Eq. \ref{eq:alpha}, with $\epsilon_\infty$ taken from experiment) and
$\alpha_{opt}^{\rm {\Delta}}$=$\frac{{\Delta}^{\rm Expt}-{\Delta}^{\rm semilocal}}{k}$  (Eq. \ref{eq:gap}).
The experimental values of the dielectric constant taken from Ref.\onlinecite{Arima95} are compared with the HSE
values obtained using the optimum value of $\alpha$ ($\alpha_{opt}^{\rm HSE}$).}
\begin{ruledtabular}
\begin{tabular}{cccccccccc}
                         & LaScO$_3$ & LaTiO$_3$ & LaVO$_3$ & LaCrO$_3$ & LaMnO$_3$ & LaFeO$_3$ & LaCoO$_3$ & LaNiO$_3$ & LaCuO$_3$  \\
& \multicolumn{9}{c}{Optimum $\alpha$}                        \\ 
$\alpha_{opt}^{\rm HSE}$       &  0.25     & 0.10 (0.15) & 0.10-0.15&   0.15    &  0.15     &    0.15   &     0.05  &    0      &    0       \\
$\alpha_{opt}^{\epsilon_\infty}$      &  0.323    &   0.125   &    0.192 &   0.250   &  0.204    &    0.200  &     0.105 &    0      &   0       \\
$\alpha_{opt}^{\rm {\Delta}}$  &0.245, 0.283&0.050, 0.087&0.115 &0.184 & 0.102, 0.173, 0.190, 0.201&0.117, 0.144&0.029, 0.065  &    0     &   0      \\
& \multicolumn{9}{c}{Dielectric constant $\epsilon_\infty$}                        \\ 
Expt.                    &  3.1      &   8.0     &    5.2   &   4.0     &  4.9      &    5.0    &     9.5   & $\infty$  &  $\infty$ \\
HSE                      &  4.4      &   8.3     &    5.9   &   5.5     &  5.8      &    5.7    &    10.7   & $\infty$  &  $\infty$ 
\end{tabular}
\end{ruledtabular}
\label{tab:19}
\end{table*}

\begin{figure}
\includegraphics[clip,width=0.48\textwidth]{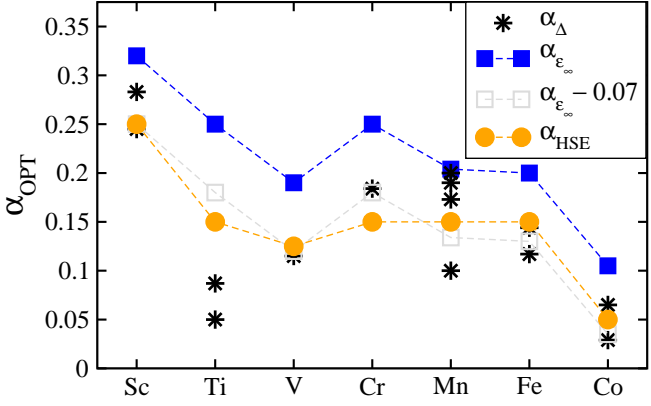}
\caption
{Graphical interpretation of the optimum values of $\alpha$ listed in Table \ref{tab:19} showing the
correlation between the HSE fitted parameters (HSE fit), the inverse dielectric constant relation ($1/\epsilon_\infty$),
and $\alpha_{opt}^{\rm {\Delta}}$ (Eq. \ref{eq:gap}).
The light-gray squares represent the $1/\epsilon_\infty$ values shifted by 0.07. This shift
roughly reflects the amount of screening incorporated in HSE via the screening factor $\mu$ as compared to
fully unscreened PBE0 (see text).
}
\label{fig:gapeps}
\end{figure}

It is instructive at this point to compare the set of parameters $\alpha_{opt}^{\rm HSE}$ with the optimum
values of $\alpha$ obtained throughout the inverse dielectric constant relation
$\alpha_{opt}^{\epsilon_\infty} \approx \frac{1}{\epsilon_\infty}$
introduced in the computational section (Eq. \ref{eq:alpha}) and derived by mapping hybrid DFT with GW.
The measured dielectric constant $\epsilon_\infty$\cite{Arima95} and the corresponding $\frac{1}{\epsilon_\infty}$
values are also listed in Table \ref{tab:19}, along with the PEAD HSE values of $\epsilon_\infty$ obtained for $\alpha$=$\alpha_{opt}^{\rm HSE}$.
Remarkably, the agreement between the measured and calculated $\epsilon_\infty$ is very good.
The nice correlation between $\alpha_{opt}^{\rm HSE}$ and $\alpha_{opt}^{\epsilon_\infty}$ can be appreciated graphically in Fig. \ref{fig:gapeps}.                 
Theses two curves follow a very similar behavior characterized by an initial large value of $\alpha$ for the poorly screened d$^0$ band insulator LaScO$_3$
followed by a plateau of similar values in the range d$^1$ (LaTiO$_3$) $\rightarrow$ d$^5$ (LaFeO$_3$) and finally
a sharp decrease towards the more strongly screened metallic compounds characterized by a completely filled $t_{2g}$
manifold. For LaNiO$_3$ and LaCuO$_3$ the optimum value of $\alpha$ is zero (not shown).
Thus, the $\alpha_{\rm opt}$ curve derived from the HSE fitting procedure conducted by computing a wide set of structural
(volume, cell shape, JT distortions, atomic positions) electronic (band gap and DOS) and magnetic
(spin ordering, magnetic moment) properties as a function of $\alpha$ reflects well the evolution of the screening
properties across the La$M$O$_3$ series represented by the dielectric function ${\epsilon_\infty}$.

However, from a quantitative point
of view the two sets of value differ by about 0.07, as clarified graphically by the open squares
in Fig. \ref{fig:gapeps}.
In order to achieve a good quantitative match between the $\alpha_{opt}^{\rm HSE}$ and
$\alpha_{opt}^{\epsilon_\infty}$ curves it is necessary to shift downwards the latter by about 0.07.
This behavior is attributable to two main reasons:
(i) The relation $\alpha_{opt}^{\epsilon_\infty} \approx
\frac{1}{\epsilon_\infty}$ identifies a proportionality between $\alpha_{\rm opt}^{\epsilon_\infty}$
and $\frac{1}{\epsilon_\infty}$,
not an exact equality (the factor of proportionality is not exactly 1, as discussed in Refs.
\onlinecite{Gygi89, Clark10, Alkauskas10});
(ii) As already mentioned before, Eq. \ref{eq:alpha} holds for standard {\em unscreened} hybrid functionals such as PBE0.
HSE is a range-separated screened hybrid functional which contains already a certain degree of screening (controlled
by the screening factor $\mu$). Therefore it is expected that the optimum $\alpha$ derived for PBE0
($\alpha_{opt}^{\rm PBE0}$) will be smaller than the corresponding $\mu$-dependent HSE value ($\alpha_{opt}^{\rm HSE}$)
\cite{Alkauskas10}. Needless to say, that in absence of a systematic study of the role of $\mu$ it is very difficult
to quantify its effect on $\alpha_{opt}^{\rm HSE}$. We leave this issue open for future studies.

Recently, Alkauskas {\em et al.} has proposed that an optimal mixing coefficient can generally be
found for any material using the formula\cite{Alkauskas10}:

\begin{equation}
\alpha_{opt}^{\Delta} = \frac{{\Delta}^{\rm Expt}-{\Delta}^{\rm semilocal}}{k}
\label{eq:gap}
\end{equation}

where ${\Delta}^{\rm Expt}$ and ${\Delta}^{\rm semilocal}$ indicate the experimental and semilocal (GGA/LDA)
band gap, and $k={\rm d}{\Delta}(\alpha)/{\rm d}\alpha$ ($\Delta$($\alpha$) represents the variation of the band gap as
a function of $\alpha$)\cite{Alkauskas10}. This practical relation takes advantage of the linear relation between
$\Delta$ and $\alpha$, which holds true as long as the valence band maxima and conduction band minimum (and their
associated wave functions) do not change much by changing $\alpha$\cite{Alkauskas10}.
In practice, if the experimental bang gap is known, it is sufficient to perform only one
hybrid functional calculation for an arbitrary value of $\alpha$ plus one semilocal calculation to derive the
optimum value of $\alpha$. Clearly, this empirical procedure does not guarantee that the so obtained optimum
$\alpha$ is the best choice for what concerns the other ground state properties. We have already addressed this
issue for LaTiO$_3$ for which $\alpha$=0.1 give the best band gap but leads to the incorrect magnetic ordering.

\begin{figure}
\includegraphics[clip,width=0.4\textwidth]{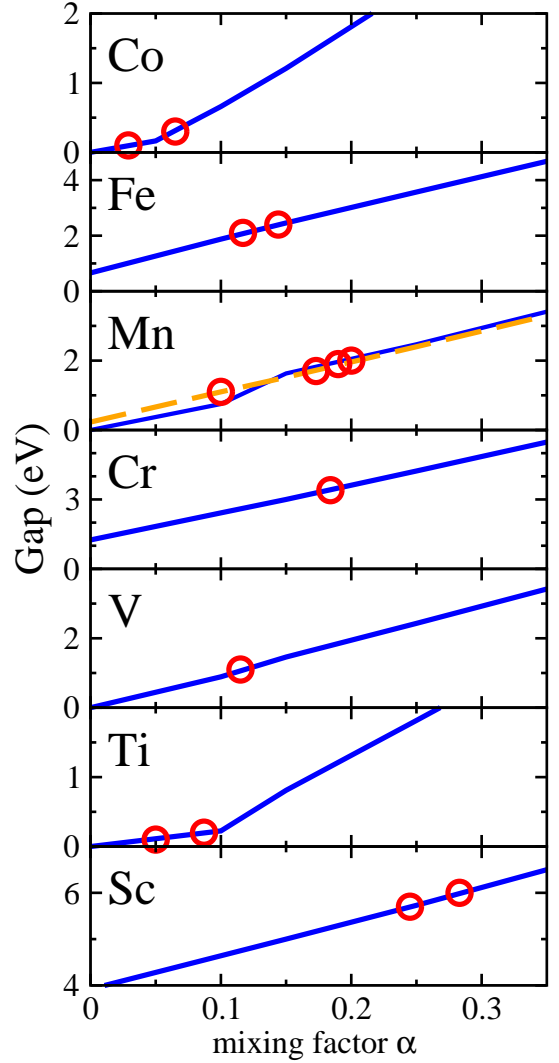}
\caption
{Change of the band gap as a function of $\alpha$ (think lines) and optimum
values of $\alpha$ (circles) obtained throughout the practical formula
$\alpha_{opt}^{\rm {\Delta}}$=$\frac{{\Delta}^{\rm Expt}-{\Delta}^{\rm semilocal}}{k}$  (Eq. \ref{eq:gap}).
}
\label{fig:gap}
\end{figure}

The changes of the bang gap as a function of $\alpha$ for the series LaScO$_3$-LaCoO$_3$ are reported in
Fig. \ref{fig:gap}. The linearity is well preserved for most of the materials with the exception
of the small band gap compounds LaTiO$_3$ and LaCoO$_3$ for which a sudden change of
$k={\rm d}{\Delta}(\alpha)/{\rm d}\alpha$ is observed for a critical value of $\alpha$.
A departure from linearity is also found for the JT/MH insulator LaMnO$_3$ if we adopt the fully relaxed structure (full line).
This is due to the fact that the cooperative JT distortions, which contribute to the opening of the band gap,
do not change linearly with $\alpha$ (as such, the associated wave function will change with $\alpha$).
Indeed, by keeping the atomic coordinates fixed to the experimental positions
the gap grows linearly by increasing $\alpha$ (dashed line).

The values of $\alpha_{opt}^{\Delta}$ obtained from this
set of curves are indicated with empty (red) circles and included in Table \ref{tab:19}. For some materials we provide more than
one optimum mixing parameter since different experimental gaps are reported in literature (see previous
section). Not surprisingly, we find that the values of $\alpha_{opt}^{\Delta}$ are very similar to the corresponding
$\alpha_{opt}^{\rm HSE}$ data, with the exception of LaTiO$_3$ and to a lesser extent LaMnO$_3$, and correlates well
with $\alpha_{opt}^{\epsilon_\infty}$, as visualized in Fig. \ref{fig:gapeps}.

\begin{figure}
\includegraphics[clip,width=0.50\textwidth]{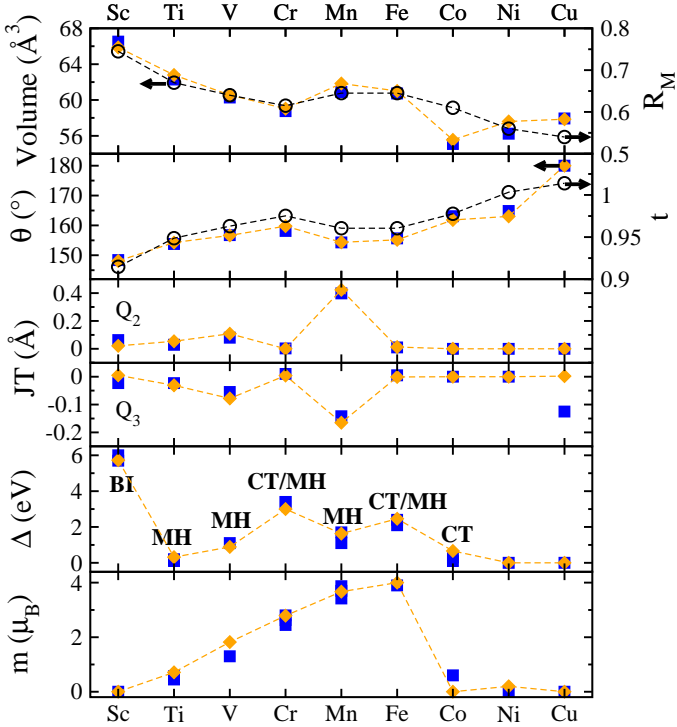}
\caption
{Trend of selected structural (Volume V, tilting angle $\theta$, and JT distortions Q$_2$ and  Q$_3$),
electronic (bandgap $\Delta$),  magnetic (magnetic moment $m$), and dielectric constant ($\epsilon_\infty$) quantities along the La$R$O$_3$ series
from $M$=Sc to $M$=Cu. We also show the trend of the tolerance factor
$t$=($\rm R_A$+$\rm R_0$)/$\sqrt{2}$($\rm R_M$+$\rm R_O$), where $\rm R_A$, $\rm R_M$ and $\rm R_O$
indicate the ionic radius for La, $M$=Sc-Cu and O, respectively, as well as $\rm R_M$. For LaTiO$_3$
we used $\alpha$=0.1. The character of the insulating gap is also indicated (BI = band insulator,
CT = charge transfer, MH = Mott-Hubbard, CT/MH = mixed CT and MH character).
}
\label{fig:trend}
\end{figure}

Now, with the rough-and-ready set of optimum HSE-fitted values $\alpha_{opt}^{\rm HSE}$ we
conclude this paper by providing a general picture of the variation of the properties of La$M$O$_3$ from $M$=Sc to $M$=Cu,
by comparing our computed results with the available experimental data. 
Figure \ref{fig:trend} shows the remarkably good agreement between the calculated and measured values of the Volume (V), tilting angle
($\theta$), JT distortion, band gap ($\Delta$), magnetic moment ($m$) and dielectric constant ($\epsilon_\infty$). The correlation between V and $\rm R_M$,
as well as between $\theta$ and $t$ was already discussed at the beginning of Sec. \ref{sec:struc}.
The variation of the magnetic moment as a function of $M$ can be easily understood in terms of the progressive
$t_{2g}$ and $e_g$ band filling in the high-spin compounds LaTiO$_3$ (${t_{2g}}^{\uparrow}$, $m$=0.51 $\mu_{\rm B}$),
LaVO$_3$ (${t_{2g}}^{\uparrow\uparrow}$, $m$=1.3 $\mu_{\rm B}$),
LaCrO$_3$ (${t_{2g}}^{\uparrow\uparrow\uparrow}$, $m$=2.63 $\mu_{\rm B}$),
LaMnO$_3$ (${t_{2g}}^{\uparrow\uparrow\uparrow}{e_{g}}^{\uparrow}$, $m$=3.66 $\mu_{\rm B}$), and
LaFeO$_3$ (${t_{2g}}^{\uparrow\uparrow\uparrow}{e_{g}}^{\uparrow\uparrow}$, $m$=3.9-4.6 $\mu{\rm B}$).
As already specified the experimental and computed magnetic moments of LaCoO$_3$ should be taken with
a certain caution. LaNiO$_3$ and LaCuO$_3$ shown a non-magnetic behavior at PBE level only (we have already
reported that the a small magnetic moment of 0.169 $\mu_{\rm B}$ for LaNiO$_3$ is found for the fully relaxed
structure).

The variation of the band gaps with the $M$ species shown in Fig. \ref{fig:trend} is rather complex and
in line with the earlier observation of Arima\cite{arima,Arima95}: we observe a general tendency of the MH gap
to increase as the TM atomic number increases,
whereas the CT gaps follow an opposite behavior. As expected, there is an apparent correlation between the
trend of the band gaps and the optimum fraction of exact exchange displayed in Fig. \ref{fig:gapeps},
especially $\alpha_{opt}^{\Delta}$. In LDA+U language this behavior is interpreted as a correlation between the strength
of the effective Coulomb interaction U and the band gap (this become particularly evident by comparing the
$\Delta$ curve in Fig.\ref{fig:trend} with Fig. 2 in Ref. \onlinecite{solovyev96} showing the changes of the effective
U with respect to $M$).

\begin{figure*}
\includegraphics[clip,width=1.00\textwidth]{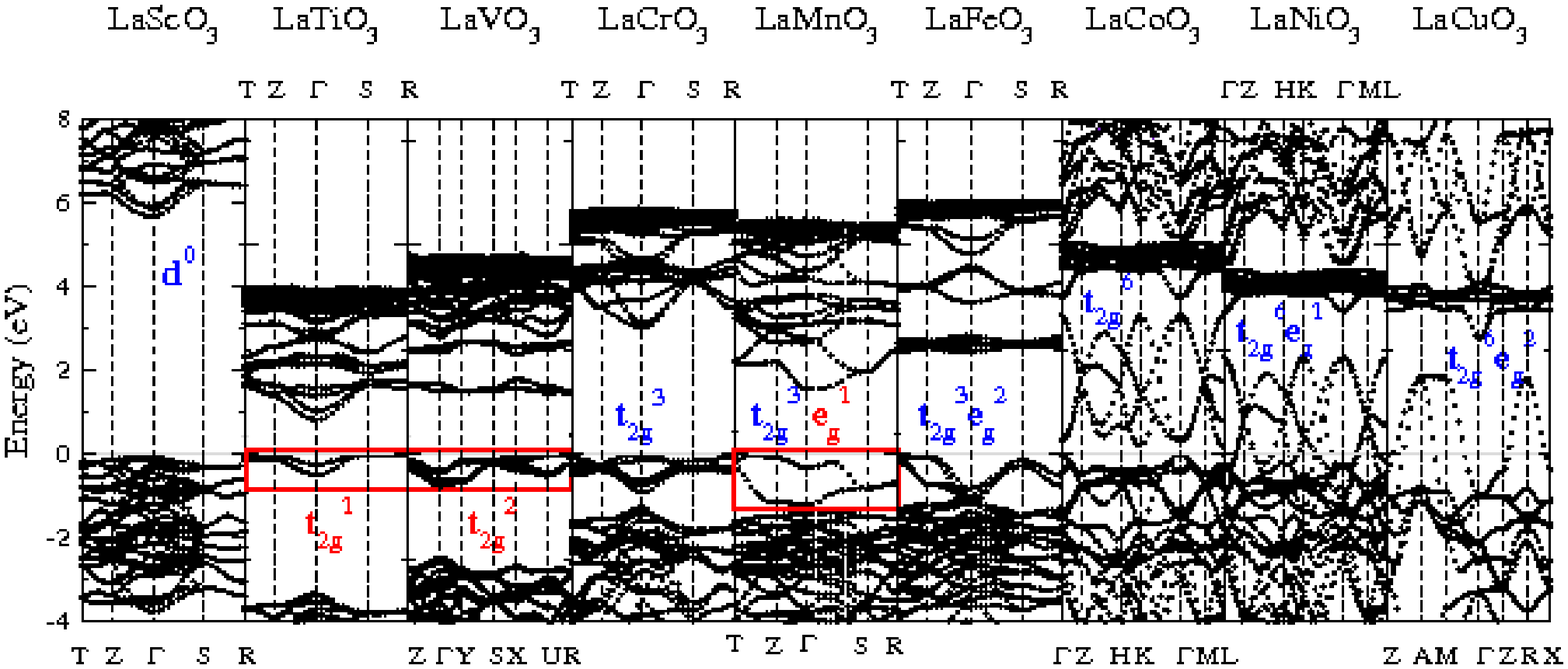}
\caption
{(Color online) Summary of the HSE electronic dispersion relations showing the complete
trend from the band insulator LaScO$_3$ to metallic LaCuO$_3$. The thik (red) lines
demarcate the $d$ bands responsible for the observed orbital-ordering in
LaTiO$_3$ ($t_{2g}$), LaVO$_3$ ($t_{2g}$) and LaMnO$_3$ ($e_{g}$).
}
\label{fig:bands}
\end{figure*}

In Fig. \ref{fig:bands} we collect the bandstructures of La$M$O$_3$ obtained using the optimum
$\alpha_{opt}^{\rm HSE}$ values showing the variation of the electronic dispersion across the whole series.
Starting from the $d^0$ band insulator LaScO$_3$ the addition of one $d$ electron creates an highly localized
$t_{2g}$ state right below $E_{\rm F}$ in LaTiO$_3$. The gradual filling of this $t_{2g}$ manifold
leads to a continuous increase of the band width from ${t_{2g}}^1$ (LaTiO$_3$) to ${t_{2g}}^3$ (LaCrO$_3$),
connected with a gradual increase of the crystal field splitting.
In LaMnO$_3$ the fully occupied $t_{2g}$ band is pushed down in energy and the valence band maxima is
dominated by the half-filled ${e_g}^1$ subbands. The ${e_g}$ orbital gets completely filled in LaFeO$_3$ which is the last
member of the series having a predominantly MH gap. The inclusion of one additional electron yields a sudden
change of the bandstructure manifested by a high increase of $p$-$d$ hybridization and bandwidth around E$_{\rm F}$.
which finally lead to the onset of a metallic state in LaNiO$_3$ and LaCuO$_3$.

\begin{figure*}
\includegraphics[clip,width=1.00\textwidth]{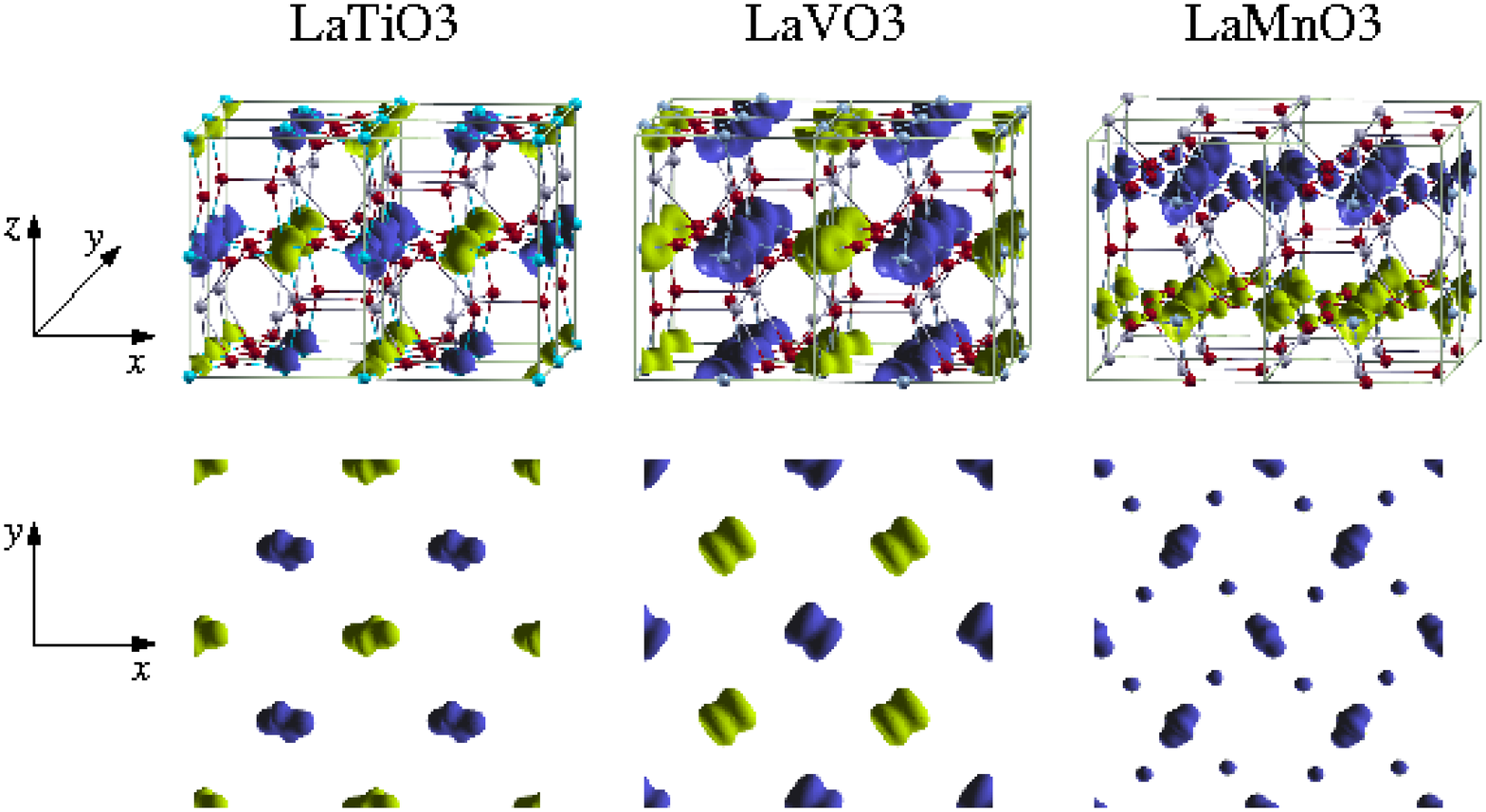}
\caption
{(Color online) Isosurface of the magnetic orbitally ordered charge density
for LaTiO$_3$, LaVO$_3$, and LaMnO$_3$
associated with the topmost occupied bands highlighted in the insets of Fig.\ref{fig:bands}.
Light (yellow) and dark (blue) areas represent spin-down and spin-up, respectively, indicating
the different types of spin orderings in LaTiO$_3$ (G-type), LaVO$_3$ (C-type), and LaMnO$_3$ (A-type).
Top panel: three dimensional view; Bottom: projection onto the $xy$ plane.
}
\label{fig:oo}
\end{figure*}

Three members of the La$M$O$_3$ family (LaTiO$_3$, LaVO$_3$, and LaMnO$_3$) are known to display orbital-ordering (OO)
associated with the partially filled $t_{2g}$ and $e_{g}$ orbitals located on top of the valence band (these states are
demarcated by thick lines in Fig. \ref{fig:bands}).
A visual representation of the OO states derived from the 'optimum' HSE wavefunctions is shown in
Fig.\ref{fig:oo} in terms of charge density isosurfaces of the highest occupied $d$ states.
In the following we describe briefly the most important characteristics of the observed OO states.

(i) In LaTiO$_3$, where the OO originates from the single $t_{2g}$ electron, the lobes
have a quasi cigar-like shape with asymmetric contributions along the two main directions,
indicating an almost identical occupation of the three $xy$, $xz$, and $yz$ $t_{2g}$ shells.
Co-planar lobes are arranged in a chessboard-like way with a sign alternation along z, in good agreement
with previously reported theoretical\cite{Pavarini04, Filippetti11,Mochizuki03} and experimental works
\cite{cwik, kiyama03}.
There is a clear connection between this chessboard-like Ti $d^1$ ordering and the JT structural instability,
which is manifested by the tendency of the occupied $t_{2g}$ state to lie along the longer Ti-O bond.
This also explains why the chessboard-like OO in LaTiO$_3$ is not as much evident as in LaMnO$_3$:
in LaTiO$_3$ the difference between the distinct Ti-O bondlengths Ti-O$_s$ Ti-O$_m$ Ti-O$_l$, quantified by the JT parameters Q$_2$ and
Q$_3$, is about one order of magnitude smaller than in LaMnO$_3$ (see Table \ref{tab:2} and Table \ref{tab:5}).

(ii) The V$^{3+}$ ions in LaVO$_3$ can accommodate two electrons in the three possible orbital states
$d_{xy}$, $d_{xz}$, and $d_{yz}$. The spins are arranged according to the C-type ordering, whereas the
OO state is found to be G-type, in accordance with the Goodenough-Kanamori rules\cite{Goodenough63} and
in agreement with x-ray diffraction\cite{Ren03} and previous GGA\cite{sawada96} and HF\cite{Solovyev06} calculations.
The distribution of the $t_{2g}$ orbitals in the G-type OO state follows the cooperative JT-induced V-O
bond-alternation in the xy plane and along the z axis, i.e. the $t_{2g}$ charge density in one specific V
site is rotated by 90$^{\circ}$  with respect to that in the 6 neighboring V sites (four in-plane and two
in the adjacent vertical planes). As already observed for LaTiO$_3$, the $t_{2g}$ orbitals are preferentially
occupied along the long-bond direction.

(iii) The C-type OO in LaMnO$_3$, originating from the singly occupied $e_g$ state of the Mn$^{+3}$
3$d$ electrons in the high-spin configuration ${t_{2g}}^{3}{e_{g}}^{1}$ has been extensively studied
both experimentally\cite{Murakami98,kovaleva,Kruger04}, and theoretically\cite{Yin06, Pavarini10}.
We have also recently addressed this issue throughout a maximally localized Wannier functions representation
of the $e_g$ states\cite{Franchini12}. This C-type OO state can be written in the form
$|\theta\rangle = \mathrm{cos}\frac{\theta}{2}|3z^2-r^2\rangle + \mathrm{sin}\frac{\theta}{2}|x^2 - y^2\rangle$
with the sign of $\theta\sim 108^\circ$ alternating along $x$ and $y$ and repeating along $z$, as correctly
represented by our HSE charge density plots.

\begin{figure}
\includegraphics[clip,width=0.50\textwidth]{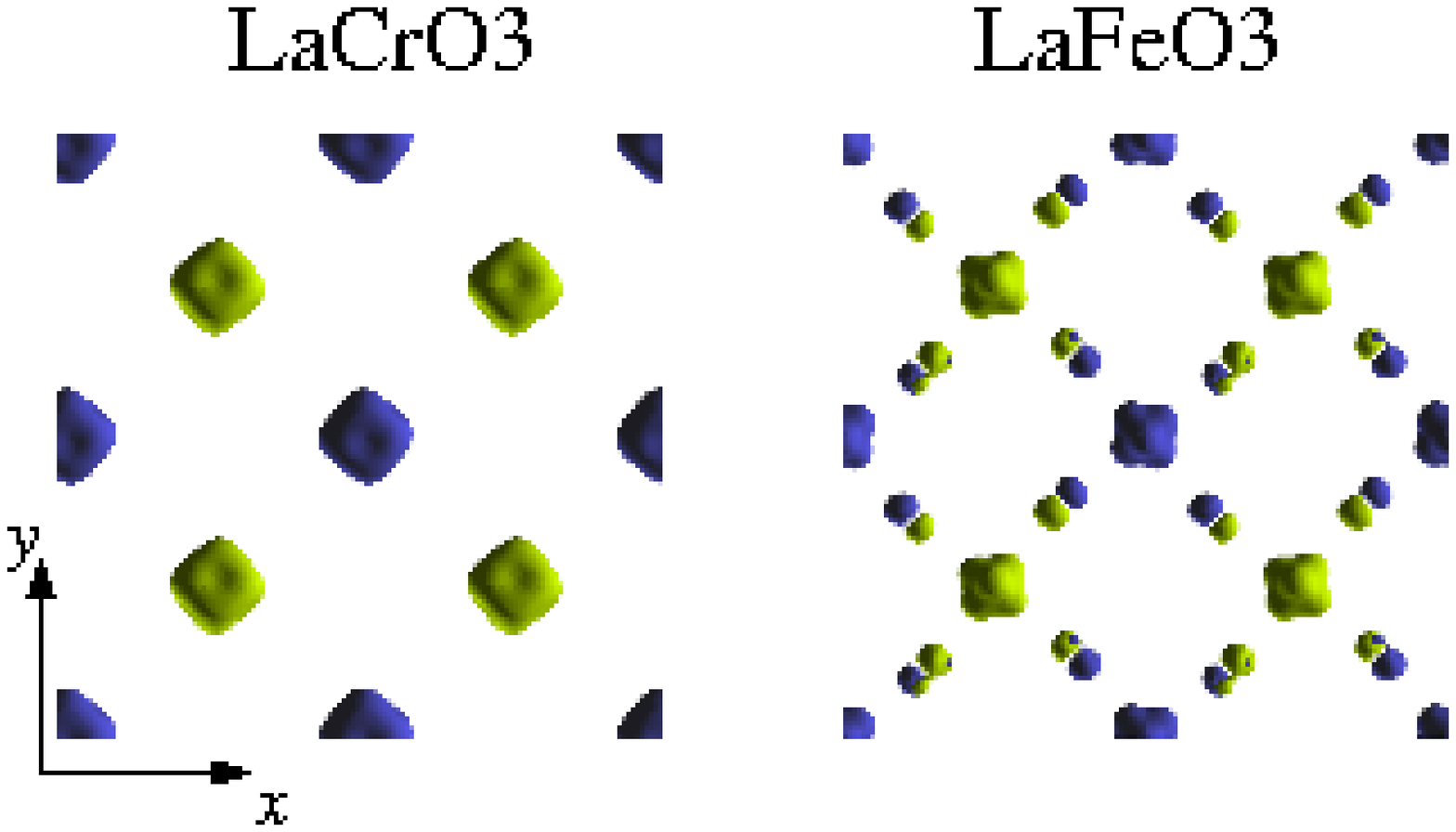}
\caption
{(Color online) Isosurface of the {\em non} orbitally ordered magnetic charge density
for LaCrO$_3$ (G-type) and LaFeO$_3$ (A-type) associated with the topmost occupied bands (See Fig.\ref{fig:bands}).
Light (yellow) and dark (blue) areas indicate spin-down and spin-up, respectively.
}
\label{fig:oofecr}
\end{figure}

For comparison we show in Fig. \ref{fig:oofecr} the similar charge density isosurfaces
calculated for LaCrO$_3$ and LaFeO$_3$, in which the half-filling of the
$t_{2g}$ and $e_g$ orbitals inhibit any orbital flexibility. As expected there is no sign of orbital ordering.

\begin{figure}[hb]
\includegraphics[clip,width=0.4\textwidth]{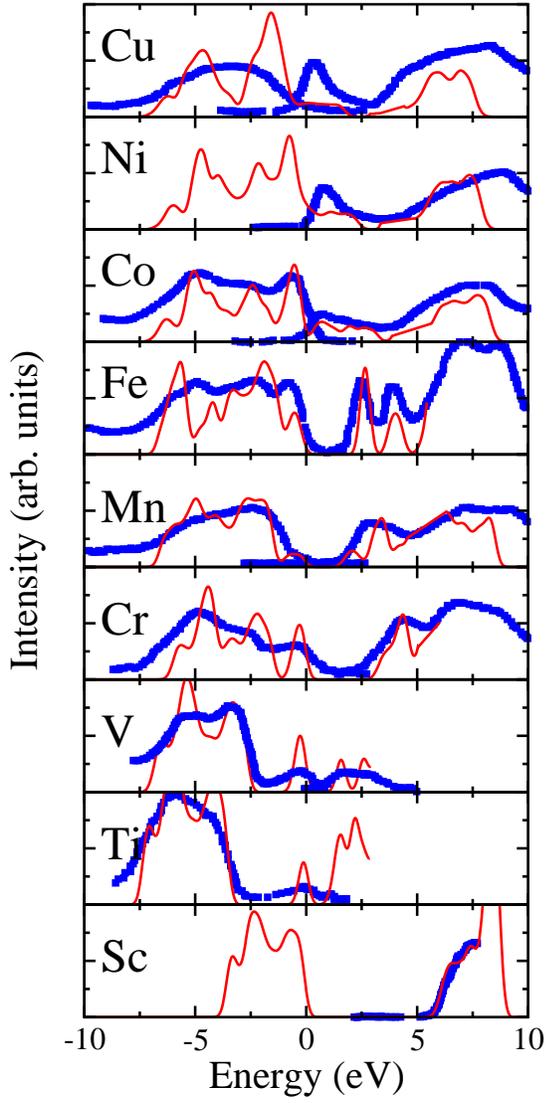}
\caption
{
Comparison between experimental (blue squares) and calculated (red full lines) valence and
conduction band spectra at the 'optimum' value of the $\alpha$ parameter:
(i) LaScO$_3$: $\alpha=0.25$;
(ii) LaTiO$_3$: $\alpha=0.15$;
(iii) LaVO$_3$: $\alpha=0.15$;
(iv) LaCrO$_3$: $\alpha=0.15$;
(v) LaMnO$_3$: $\alpha=0.15$;
(vi) LaFeO$_3$: $\alpha=0.15$;
(vii) LaCoO$_3$: $\alpha=0.05$;
(viii) LaNiO$_3$: $\alpha=0$;
(ix) LaCuO$_3$: $\alpha=0$.
The calculated and measured spectra have been aligned by overlapping the
valence band maxima and conduction band minima.
The experimental data are taken from the collection of spectra presented in Ref.\onlinecite{nohara09},
originally published in separated articles:
(i) LaScO$_3$: Ref.\onlinecite{arima};
(ii) LaTiO$_3$: Ref.\onlinecite{Nakamura99};
(iii) LaVO$_3$: Ref.\onlinecite{Maiti00};
(iv) LaCrO$_3$: Ref.\onlinecite{Sarma96};
(v) LaMnO$_3$: Ref.\onlinecite{saitoh};
(vi) LaFeO$_3$: Ref.\onlinecite{Wadati05};
(vii) LaCoO$_3$: Ref.\onlinecite{Abbate93};
(viii) LaNiO$_3$: Ref.\onlinecite{Abbate02};
(ix) LaCuO$_3$: Ref.\onlinecite{Mizokawa98}.
}
\label{fig:spectra}
\end{figure}

We conclude this section with a comparison of the calculated density of valence and conduction states
with available x-ray photoemission and x-ray adsorption spectra. This is summarized in Fig. \ref{fig:spectra}.
The overall picture is satisfactory in terms of bandwidth, intensity and peaks position, though
some sizable deviations are visible for LaCrO$_3$, LaFeO$_3$ and for the metallic compounds
LaCuO$_3$ and LaNiO$_3$.
These differences between theory and experiment should be attributable to the approximations included in the adopted
computational scheme, as mentioned in the previous section.

\section{SUMMARY AND CONCLUSIONS}
\label{sec:con}

In summary, we have studied the ground state properties of the perovskite series La$M$O$_3$ by
means of screened hybrid DFT following the HSE formulation, based on the inclusion of a fraction
of exact HF exchange in the short range Coulomb kernel of the underlying semilocal PBE exchange-correlation
functional. In particular, we have investigated the role of the HSE mixing parameter $\alpha$ on the
(i) structural parameters (Volume, JT/GFO distortions, lattice parameters and unit cell symmetry)
(ii) electronic character (band gap, DOS, and bandstructure), and
(iii) spin orderings and magnetic moment, as a function of the gradual filling of the $d$ manifold from
LaScO$_3$ ($d^0$) to LaCuO$_3$ ($d^8$: ${t_{2g}}^{6}{e_{g}}^{2}$).

On the basis of a computational fitting of the most relevant experimentally available data
we have derived a set of mixing factors which leads to an accurate qualitative and quantitative description of
the physical behavior of all members of the representative La$M$O$_3$ family.
It is found that, apart from LaScO$_3$, the 'optimum' values of $\alpha$ ($\alpha_{opt}^{\rm HSE}$) are significantly smaller
than the routinely used standard choice $\alpha$=0.25, especially for the end members
(
LaScO$_3$: $\alpha_{opt}^{\rm HSE}$=0.25;
LaTiO$_3$ \& LaVO$_3$: $\alpha_{opt}^{\rm HSE}$=0.10-0.15;
LaCrO$_3$, LaMnO$_3$, and LaFeO$_3$: $\alpha_{opt}^{\rm HSE}$=0.15;
LaCoO$_3$: $\alpha_{opt}^{\rm HSE}$=0.05;
LaNiO$_3$ \& LaCuO$_3$: $\alpha_{opt}^{\rm HSE}$=0.0, i.e. for these two cases PBE is better than HSE
).
This can be understood by correlating the changes of $\alpha_{opt}^{\rm HSE}$
from Sc to Cu with the corresponding values of the inverse dielectric constant $1/\epsilon_\infty$, and by considering
that a certain degree of screening is already included by construction in the HSE functional throughout the
screening length $\mu$, at variance with the unscreened parent hybrid functional PBE0 (for which $\mu=0$).
This suggests that the 'optimum' value of $\alpha$ in HSE should be smaller than the corresponding PBE0 one:
in our specific case it is proposed that the difference between $\alpha_{opt}^{\rm HSE}$ and $\alpha_{opt}^{\rm PBE0}$
should about 0.05-0.07, but a more detailed analysis on the influence of $\mu$ is required to achieve more accurate
and comprehensive conclusions.

An alternative way to obtain a set of optimum $\alpha$ is the fitting
of the bandgap only, by applying the practical recipe represented by Eq. \ref{eq:gap}. However, though
this procedure has the clear advantage of reducing considerably the computational cost, it can lead to
an erroneous description of other properties (for example the best-bandgap $\alpha$ in LaTiO$_3$
stabilizes the wrong spin ordering) and can only be applied under the assumption that the wavefunction does not
change with $\alpha$.

For what concerns the description of the modulation of the electronic and magnetic properties across the La$M$O$_3$
series, we found that for all insulating compounds ($M$=Sc to Co) HSE is capable to capture the localization of
the $t_{eg}$/$e_g$ orbitals and to provide a consistent and quantitatively satisfactory description of all
considered quantities, thereby improving the deficient DFT based predictions.

For the structural properties, on the other hand, PBE performs rather well, delivering optimized geometry
within 1\%. The only exception are the JT parameters in LaMnO$_3$, which PBE finds 60\% smaller than experiment.
HSE cures this limitation, reproducing quite well the critical JT and GFO structural instabilities, and, in general,
its application improves even further the overall agreement with experiment as compared to PBE.

The complex nature of the PM correlated metals LaNiO$_3$ and LaCuO$_3$, end members of the La$M$O$_3$ series,
is only marginally accounted for by PBE and rather poorly treated at HSE level. This is mostly due to
underlying dynamical correlation effects which cannot be easily treated at DFT/HF level. For these compounds, PBE
might be considered to be a good starting point for more elaborated many-body approaches.

\section{ACKNOWLEDGMENTS}
This research was sponsored by the FP7 European Community grant ATHENA.
All calculations have been performed on the Vienna Scientific Cluster (VSC).

\end{document}